\pdfoutput=1
\documentclass[%
 twocolumn,
 aip, apl, 
 amsmath,amssymb,
 reprint,%
]{revtex4-1}

\usepackage{graphicx}% Include figure files
\usepackage{dcolumn}% Align table columns on decimal point
\usepackage{bm}% bold math
\usepackage[caption=false,justification=centering,font=small]{subfig}
\usepackage[colorlinks=true,linkcolor = blue,urlcolor=blue,citecolor=blue]{hyperref}
\usepackage[utf8]{inputenc}
\usepackage[T1]{fontenc}
\usepackage{mathptmx}
\usepackage{etoolbox}

\makeatletter
\newcommand{\nocaptionlabel}[1]{%
    \long\def\@makecaption##1##2{}% Redefine to suppress "Fig.", number, and dot
    \caption{\phantom{#1}}% Invisible caption
    \long\def\@makecaption##1##2{\ifx\@captype\@typeset@table TABLE \else FIG. \fi\kern1em ##1: ##2}% Restore original
}
\makeatother
\makeatletter
\def\@email#1#2{%
 \endgroup
 \patchcmd{\titleblock@produce}
  {\frontmatter@RRAPformat}
  {\frontmatter@RRAPformat{\produce@RRAP{*#1\href{mailto:#2}{#2}}}\frontmatter@RRAPformat}
  {}{}
}%
\makeatother

\makeatletter
\renewcommand{\p@subfigure}[1]{\thefigure(#1)} 
\makeatother

\begin{document}

\preprint{AIP/123-QED}

\title[Carrier density and mobility interplay in (Al,Ga)N]{Interplay of carrier density and mobility in Al-Rich (Al,Ga)N-Channel HEMTs: Impact on high-power device performance potential}
% Force line breaks with \\
\author{Badal Mondal}
\author{Pietro Pampili}%
\affiliation{Tyndall National Institute, University College Cork, Cork T12 R5CP, Ireland}%

\author{Jayjit Mukherjee}
\author{David Moran}
\affiliation{James Watt School of Engineering, University of Glasgow, Glasgow G12 8LT, UK}%

\author{Peter James Parbrook}%
\affiliation{Tyndall National Institute, University College Cork, Cork T12 R5CP, Ireland}%
\affiliation{School of Engineering, University College Cork, Western Road, Cork, Ireland}

\author{Stefan Schulz}
\email{stefan.schulz@tyndall.ie}
\affiliation{Tyndall National Institute, University College Cork, Cork T12 R5CP, Ireland}%
\affiliation{School of Physics, University College Cork, Cork T12 YN60, Ireland}%

\date{19 February 2025}%

\begin{abstract}
Despite considerable advancements, high electron mobility transistors (HEMTs) based on gallium nitride (GaN) channels remain largely limited to power applications below 650 V. For higher power demands, the ultra-wide bandgap semiconductor alloy aluminium gallium nitride, (Al,Ga)N, has emerged as a key contender for next-generation HEMTs. In this theoretical study, we show that Al-rich Al$_x$Ga$_{1-x}$N-channel HEMTs (with $x \geq 0.5$) outperform the GaN-channel counterparts at and above room temperature, across all Al compositions, $x$. This contrasts with recent theory reports which suggest that only Al$_x$Ga$_{1-x}$N HEMTs with high Al content ($x \geq 0.85$) offer comparable performance to GaN-channel devices. Unlike previous assumptions of a constant two-dimensional electron gas (2DEG) density across the entire composition range $x$, we show that the 2DEG density is highly sensitive to both the Al content and thickness of the individual layers in a HEMT structure. We demonstrate that the superior performance of Al-rich (Al,Ga)N-channel HEMTs is driven by a competing effect between 2DEG density and electron mobility. This work challenges the assumptions of prior studies, which can result in a significant under or overestimation of the potential of high Al content HEMTs. The insights gained from our work provide a comprehensive understanding of the trade-offs between device and material parameters, thus help to guide the design of future Al-rich ($x = 0.5 - 1.0$) Al$_x$Ga$_{1-x}$N-channel HEMTs for high-power applications.
\end{abstract}

\maketitle

The ultra-wide bandgap (UWBG) semiconductor alloy aluminium gallium nitride, (Al,Ga)N, exhibits immense potential for high-frequency and high-power applications, positioning it as a key contender for next generation of mobile-communication base-stations, satellite systems, voltage converters in electric vehicles, and power supply devices.\cite{Baca2020Al-richTransistors, Tsao2018UltrawideBandgapChallenges,Wong2021Ultrawide-bandgapOverview,Kaplar2017ReviewUltra-Wide-BandgapDevices} Over the past decades substantial research efforts have been focused on establishing and improving the performance of (Al,Ga)N/GaN-based high electron mobility transistors (HEMTs).\cite{He2021RecentDevices} Despite significant advancements, the use of (Al,Ga)N/GaN HEMTs in the high-power market is largely limited to applications below 650 V.\cite{Ma2019ReviewW} However, introducing aluminium (Al) into the channel of a GaN-based HEMT structure allows to produce (Al,Ga)N alloys which exhibit significantly larger bandgaps when compared to pure GaN. As a consequence, the critical breakdown field of a HEMT device can be increased.\cite{Hudgins2003AnDevices} Recent experiments have shown superior performance of high Al content (Al,Ga)N/(Al,Ga)N HEMTs with increased breakdown voltage and high-temperature operation stability, thus demonstrating their suitability in high-power and high-frequency device application domains.\cite{Kaplar2017ReviewUltra-Wide-BandgapDevices,Hussain2023HighSubstrates,Khachariya2022RecordSubstrates,Baca2016AnContact,Carey2019OperationTransistors,Yafune2014AlN/AlGaNOperation,Armstrong2018Ultra-wideTransistor}

Nevertheless, the intrinsic performance of (Al,Ga)N/(Al,Ga)N HEMTs also critically depends on the density and mobility of the two-dimensional electron gas (2DEG) formed at the channel-barrier interface,\cite{Jogai2003InfluenceTransistors,Ibbetson2000PolarizationTransistors,Zhang2008TheContent,Singhal2022TowardHeterostructures} which in general varies significantly with Al composition and layer thicknesses of a HEMT heterostructure. Previous studies often simplified these dependencies, assuming a constant 2DEG density ($n_{\text{2D}}$) of $1 \times 10^{13}$ cm$^{-2}$ \cite{Bassaler2024AlRichMobility,Coltrin2017AnalysisAlloys,Bajaj2014ModelingVoltage} --- a typical target for GaN-channel HEMTs --- across the entire composition range and layer thickness. We demonstrate here through device-scale simulations --- which explicitly account for individual layer thicknesses, alloy compositions, and strain fields --- that assuming a constant 2DEG density imposes experimentally challenging device design requirements, especially with high Al content (Al,Ga)N HEMTs. Moreover, earlier studies,\cite{Zhang2008TheContent,Bassaler2024AlRichMobility,Coltrin2017AnalysisAlloys,Ambacher1999Two-dimensionalHeterostructures} based on the constant $n_{\text{2D}}$ assumption, concluded that Al$_x$Ga$_{1-x}$N-channel HEMTs only outperform GaN-channel HEMTs for $x \geq 0.85$. In contrast, our simulations show that this performance parity occurs at a much lower Al composition $x \geq 0.5$, highlighting the potential benefit of Al-rich (Al,Ga)N-channel HEMTs. Our simulations reveal a critical interplay between 2DEG density and mobility, and we find that this interplay must be considered carefully when designing (Al,Ga)N/(Al,Ga)N HEMT devices. Our work, thereby, addresses prior simplified assumptions and underscores the importance of detailed device simulations to fully exploit the potential of (Al,Ga)N-based HEMTs. 

\begin{figure}%[!htbp]
    \centering
    \subfloat[]{\label{fig:fig1a}\includegraphics[width=0.75\columnwidth]{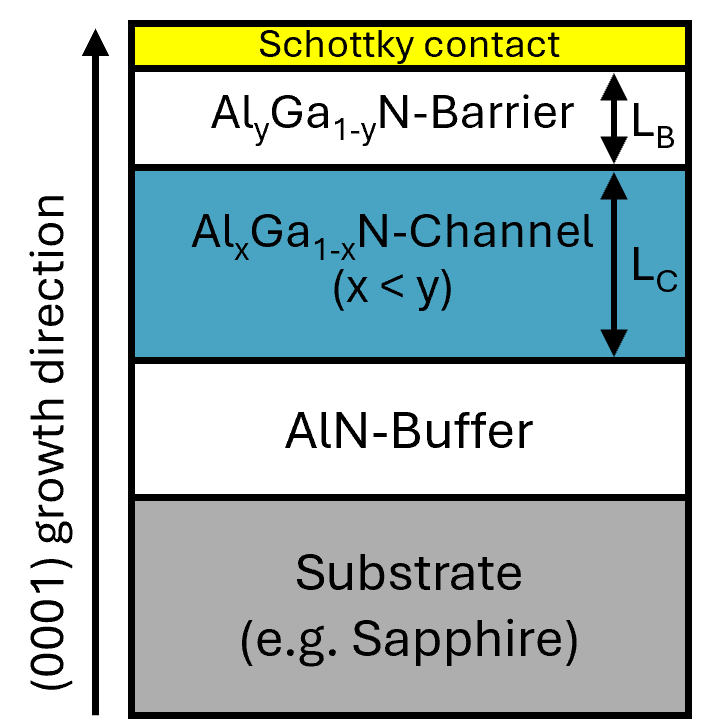}}\\
    \subfloat[]{\label{fig:fig1b}\includegraphics[width=0.9\columnwidth]{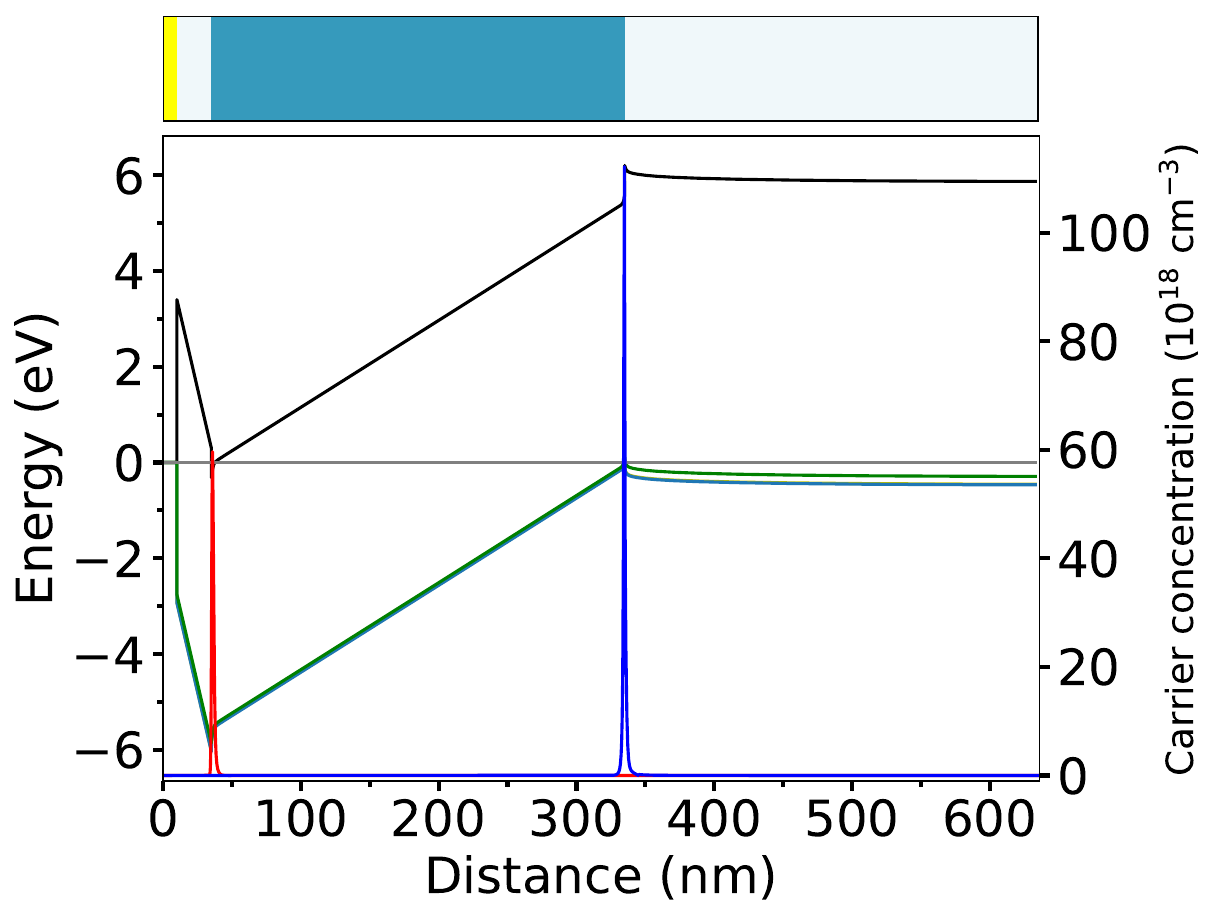}}
    \caption{(a) 2D schematic of the composition profile for a (Al,Ga)N/(Al,Ga)N/AlN HEMT. The substrate is not included in the simulations since its contribution is of secondary importance in this study. (b) Band diagram of an AlN/Al$_{0.75}$Ga$_{0.25}$N/AlN HEMT, showing the conduction band (black), three valence bands --- heavy hole, light hole, and split-off --- in green, and the Fermi level (gray). The 2DEG (red) and 2DHG (blue) density distributions are shown on the right axis; the composition profile is indicated above the band diagram.}
    \label{fig:fig1}
\end{figure}

Figure~\ref{fig:fig1a} shows the schematic of an undoped (Al,Ga)N/(Al,Ga)N/AlN double heterostructure HEMT device used in this study. We assume in the following abrupt interfaces and conventional (0001) cation faced growth direction. Recent experiments have demonstrated pseudomorphic epitaxial growth of Al rich (Al,Ga)N/(Al,Ga)N-layers on thick AlN buffer layer.\cite{Hussain2023HighSubstrates,Singhal2022TowardHeterostructures,Singhal2023AlN/AlGaN/AlNHEMTs,Mehta2023HighSubstrate,Maeda2022AlN/AlSputtering,Kozaka2024PulsedStructures,Abid2021AlGaNContacts,Kometani2024AlN/AlGaNDeposition} Here, we specifically focus on Al-rich Al$_y$Ga$_{1-y}$N/Al$_x$Ga$_{1-x}$N systems with $x, y \geq 0.5$. A representative band diagram for an AlN/Al$_{0.75}$Ga$_{0.25}$N/AlN HEMT obtained within our simulation framework is presented in Fig.~\ref{fig:fig1b}. The composition contrast at the top (Al,Ga)N/(Al,Ga)N barrier-channel interface induces a polarization field discontinuity, which leads to the formation of a 2DEG, when the Al composition in the barrier is higher than the channel (i.e., $y > x$). Similarly, the polarization field discontinuity of opposite sign at the bottom (Al,Ga)N/AlN channel-buffer interface results in the formation of a two-dimensional hole gas (2DHG).\cite{Jogai2003InfluenceTransistors,Ibbetson2000PolarizationTransistors,Singhal2022TowardHeterostructures,Ambacher1999Two-dimensionalHeterostructures}

To study the impact of the varied Al content and thickness of (Al,Ga)N layers on the 2DEG and 2DHG densities, we employ a one-dimensional (1D) self-consistent 2+6-band \textbf{k}$\cdot$\textbf{p} Schr{\"o}dinger-Poisson solver, as implemented in Nextnano++ (v-1.21.24).\cite{Birner2007Nextnano:Simulations,Trellakis2006TheResults} Further details on the Nextnano++ simulations, including discussions on the band diagram and the Fermi level, are provided in Secs.~S1--S3 of the Supplementary Information (SI).

Figure~\ref{fig:fig2a} shows the variation of $n_{\text{2D}}$ with Al composition in both the channel, $x$, and barrier, $y$, for a fixed barrier thickness (L$_\text{B}$) of 50 nm. Similar map of $n_{\text{2D}}$ for other L$_\text{B}$ values can be found in Sec.~S4 of the SI. The highest $n_{\text{2D}}$ is achieved with an AlN barrier, $y = 1$, and an Al$_{0.5}$Ga$_{0.5}$N channel, $x = 0.5$, i.e.  an AlN/Al$_{0.5}$Ga$_{0.5}$N/AlN HEMT structure. This configuration exhibits the largest composition contrast, $\Delta_{yx} = y-x$, between the channel and barrier, resulting in the highest polarization field discontinuity at the interfaces and thereby maximizing $n_{\text{2D}}$.
Despite challenges posed by the large lattice mismatch strain at high $\Delta_{yx}$, recent studies have demonstrated successful epitaxial growth and promising HEMT performance in structures such as  AlN/Al$_{0.5}$Ga$_{0.5}$N/AlN.\cite{Kozaka2024PulsedStructures,Abid2021AlGaNContacts,Mehta2023HighSubstrate,Maeda2022AlN/AlSputtering} This study investigates a broader yet practical $\Delta_{yx}$ range to guide future (Al,Ga)N-based HEMT technologies, while covering commonly studied $\Delta_{yx}$ values ($\sim 0.2 - 0.3$).\cite{Hussain2023HighSubstrates,Singhal2022TowardHeterostructures,Singhal2023AlN/AlGaN/AlNHEMTs,Baca2020Al-richTransistors}

Figure~\ref{fig:fig2b} presents $n_{\text{2D}}$ as a function of $\Delta_{yx}$ for different L$_\text{B}$ values, where each value of $\Delta_{yx}$ (e.g., $\Delta_{yx} = c$) represents all $y$ and $x$ combinations satisfying $y-x = c$ [all $n_{\text{2D}}$ values along the same diagonal in Fig.~\ref{fig:fig2a}]. We find that $n_{\text{2D}}$ consistently increases with L$_\text{B}$, reaching its highest value at 50 nm. We note that, although a thicker barrier enhances $n_{\text{2D}}$ further, $n_{\text{2D}}$ plateaus soon after 50 nm L$_\text{B}$ [Fig.~S6(a)]. Additionally, thicker barriers, exceeding 50 nm, present practical challenges for realizing electrical contacts on Al-rich (Al,Ga)N barriers.\cite{Baca2020Al-richTransistors} While recessed contacts offer a viable solution for thick barriers,\cite{Abid2021AlGaNContacts,Baca2020Al-richTransistors,Sharbati2021AnalyticalTransistors} the effective reduction in barrier thickness also affects $n_{\text{2D}}$ in the recessed regions.\cite{Sharbati2021AnalyticalTransistors} Due to the additional complexities involved, we do not explore recess contacts further in this study and restrict our analysis to HEMT structures with L$_\text{B} \leq 50$ nm. In all cases, we use a 300 nm channel thickness (L$_\text{C}$). The $n_{\text{2D}}$ is found to be increased with L$_\text{C}$ and saturates at around 200 nm [Fig.~S6(b)]. The choice of an optimum 300 nm L$_\text{C}$ strikes a balance between the channel being thin enough to minimize strain relaxation and defect formation while being thick enough to ensure minimal influence of the 2DHG from the bottom (Al,Ga)N/AlN heterojunction (e.g, Coulomb drag effect).\cite{Li2014Two-dimensionalAlN} 

\begin{figure}%[!htbp]
    \centering
    \subfloat[]{\label{fig:fig2a}\includegraphics[width=0.85\columnwidth]{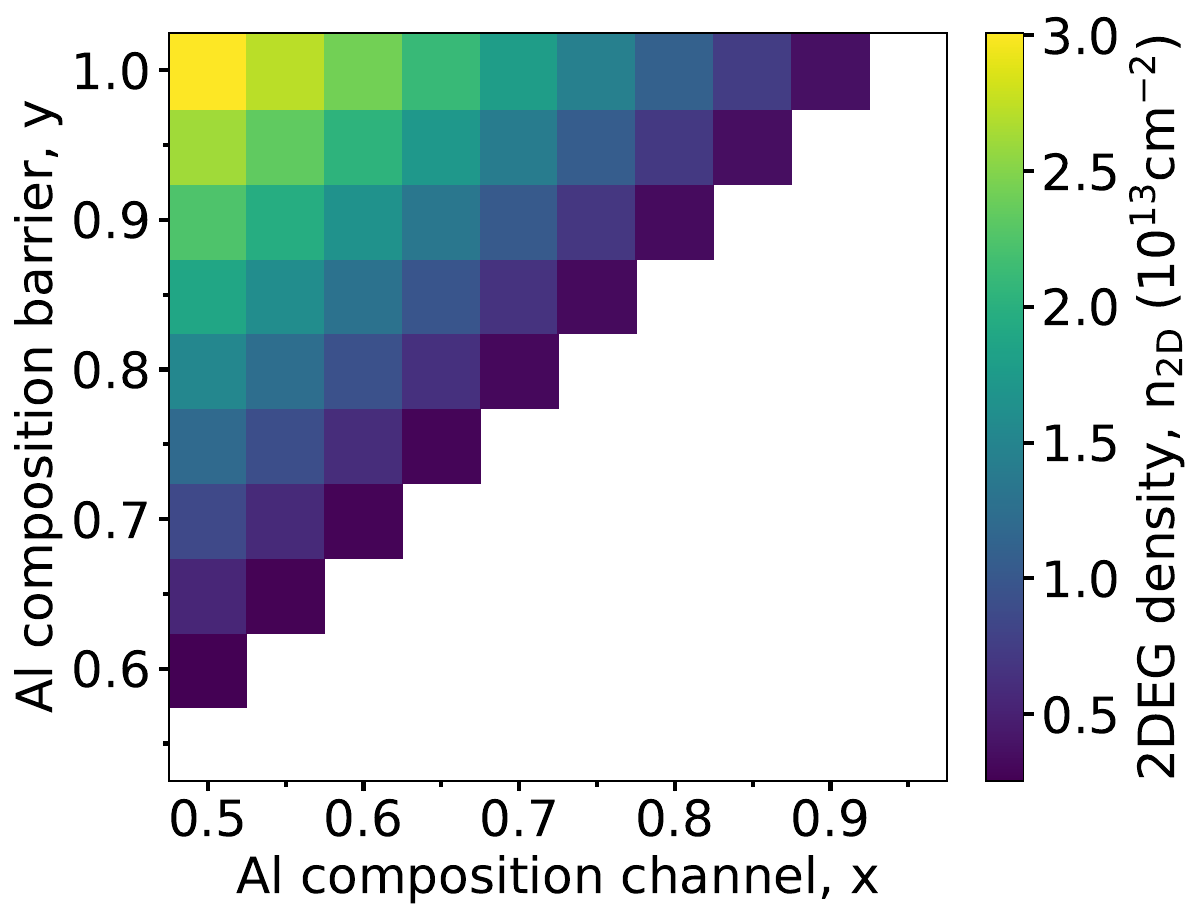}}\\
    \subfloat[]{\label{fig:fig2b}\includegraphics[width=0.85\columnwidth]{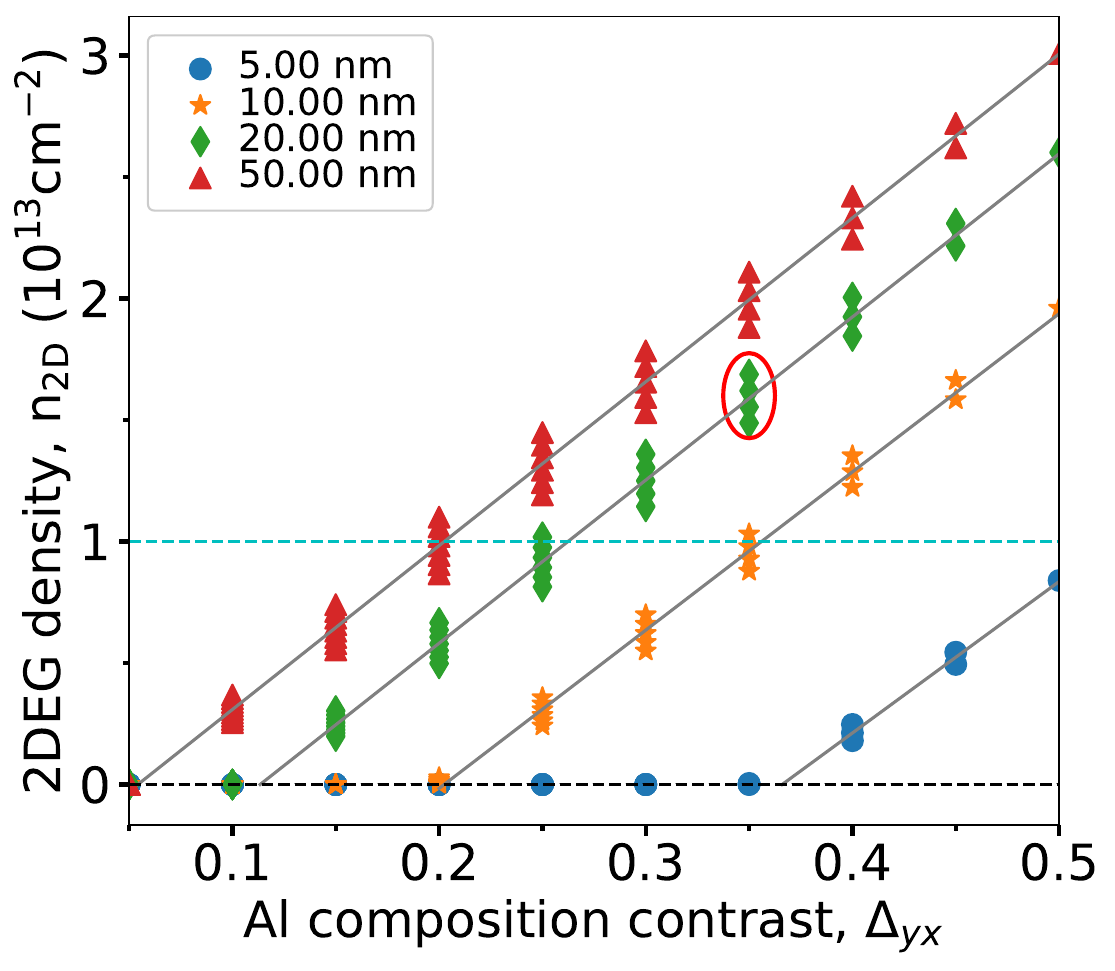}}\\
    \subfloat[]{\label{fig:fig2c}\includegraphics[width=0.85\columnwidth]{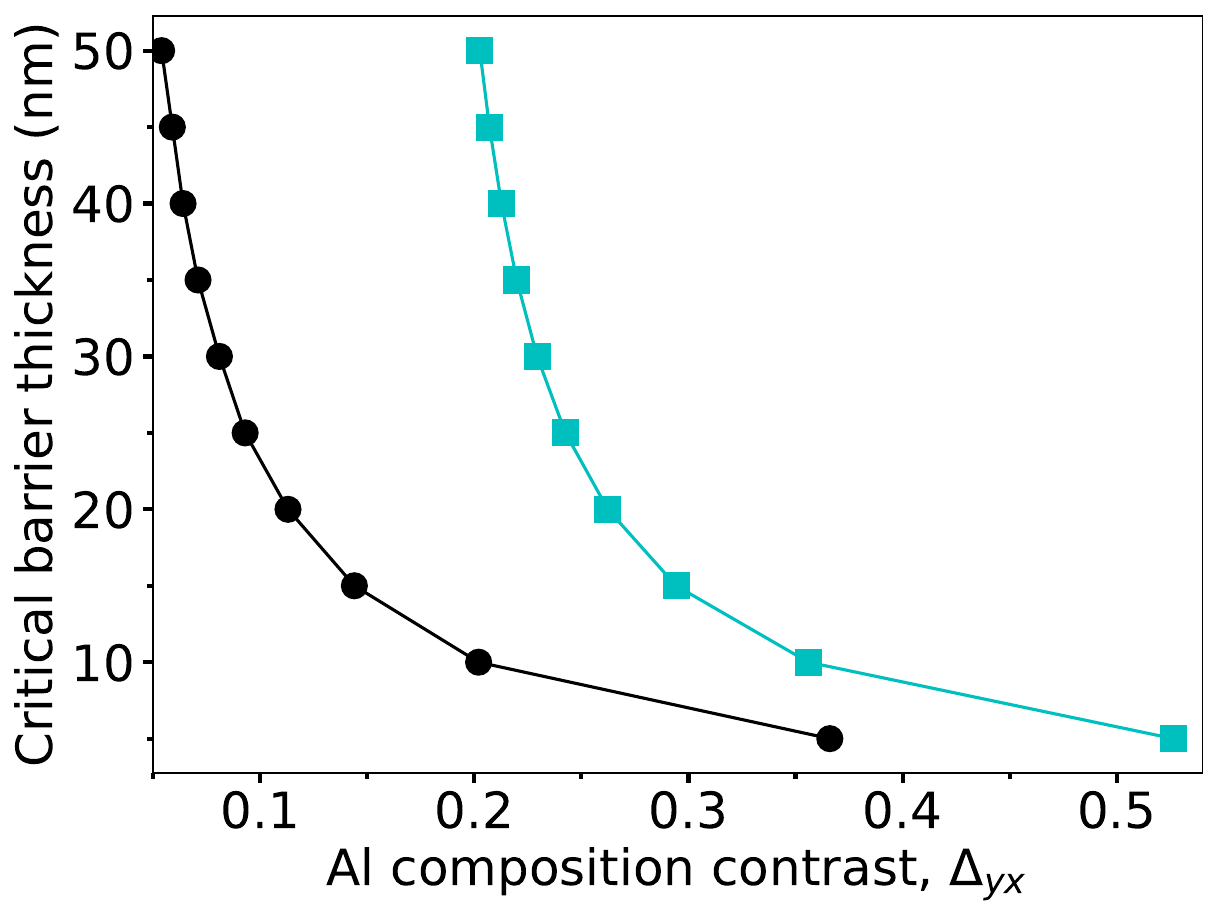}}\\
    \caption{(a) Map of 2DEG density, $n_{\text{2D}}$, as a function of Al composition in the (Al,Ga)N channel, $x$, and barrier layers, $y$, for a barrier thickness, L$_\text{B}$ = 50 nm. (b) $n_{\text{2D}}$ as a function of Al composition difference between the barrier and channel, $\Delta_{yx} = y-x$, plotted for several L$_\text{B}$ values as indicated in the legend. Horizontal dashed lines represent $n_{\text{2D}} = 0$ cm$^{-2}$ (no 2DEG, black) and $n_{\text{2D}}$ = $1 \times 10^{13}$ cm$^{-2}$ (cyan).  (c) Critical barrier thickness required to achieve $n_{\text{2D}} = 0$ cm$^{-2}$ (black) and $n_{\text{2D}}$ = $1 \times 10^{13}$ cm$^{-2}$ (cyan) as a function of composition contrast, $\Delta_{yx}$, determined from the intercepts of the linear fits in (b).  Only $n_{\text{2D}}$ $> 0$ cm$^{-2}$ are fitted.}
    \label{fig:fig2}
\end{figure}

Figure~\ref{fig:fig2b} further indicates that even for the same value of composition contrast, $\Delta_{yx}$, slight variation in $n_{\text{2D}}$ can occur (red circle in figure). Since the 2DEG originates from polarization field discontinuity at the interface, and the magnitude of this discontinuity can differ slightly depending on the absolute Al content in the individual layers, $n_{\text{2D}}$ depends not only on the composition contrast but also on the absolute Al content in the layers.

Moreover, generating a 2DEG requires not only a sufficient composition contrast but also a minimum barrier thickness.\cite{Coltrin2017AnalysisAlloys,Jogai2003InfluenceTransistors,Ibbetson2000PolarizationTransistors} This is demonstrated in Fig.~\ref{fig:fig2c}, which shows the critical barrier thickness required for different composition contrasts to realize a non-zero $n_{\text{2D}}$ ($> 0$ cm$^{-2}$). As the composition contrast increases, the required critical barrier thickness rapidly decreases. 

Figure~\ref{fig:fig2c} also highlights the shortcomings of assuming a constant $n_{\text{2D}}$ across the full composition range, as done in previous studies.\cite{Bassaler2024AlRichMobility,Coltrin2017AnalysisAlloys,Bajaj2014ModelingVoltage} As shown, while $n_{\text{2D}}$ = $1 \times 10^{13}$ cm$^{-2}$ can be achieved with a relatively thin barrier in the high composition contrast structures (e.g., $\sim 5$ nm AlN barrier in AlN/Al$_{0.5}$Ga$_{0.5}$N), realizing the same $n_{\text{2D}}$ in low Al contrast structures would require impractically thick barriers (e.g., over 300 nm in AlN/Al$_{0.85}$Ga$_{0.15}$N) (see also Fig.~S7). This demonstrates the importance of our device-level simulations, which allow us to explicitly reflect on the structural properties required to achieve target $n_{\text{2D}}$ values in (Al,Ga)N/(Al,Ga)N HEMTs.

Next, we explore the low-field 2DEG mobility in these devices.\cite{Coltrin2017AnalysisAlloys} High electron mobility is crucial for achieving fast switching speeds and high on-state current densities.\cite{Bassaler2024AlRichMobility} Here, the total mobility is analytically modelled as the cumulative effects of various scattering mechanisms, including acoustic phonon, deformation-potential, piezoelectric effect, polar optical phonon, alloy disorder, interface roughness, and dislocation mediated scattering.\cite{Zhang2008TheContent,Bassaler2024AlRichMobility} For the 2DEG densities, we use  $n_{\text{2D}}$ values obtained from our device  simulations. Further details on the mobility models are provided in the SI, Sec.~S6.

\begin{figure}
    \centering
    \includegraphics[width=0.85\columnwidth]{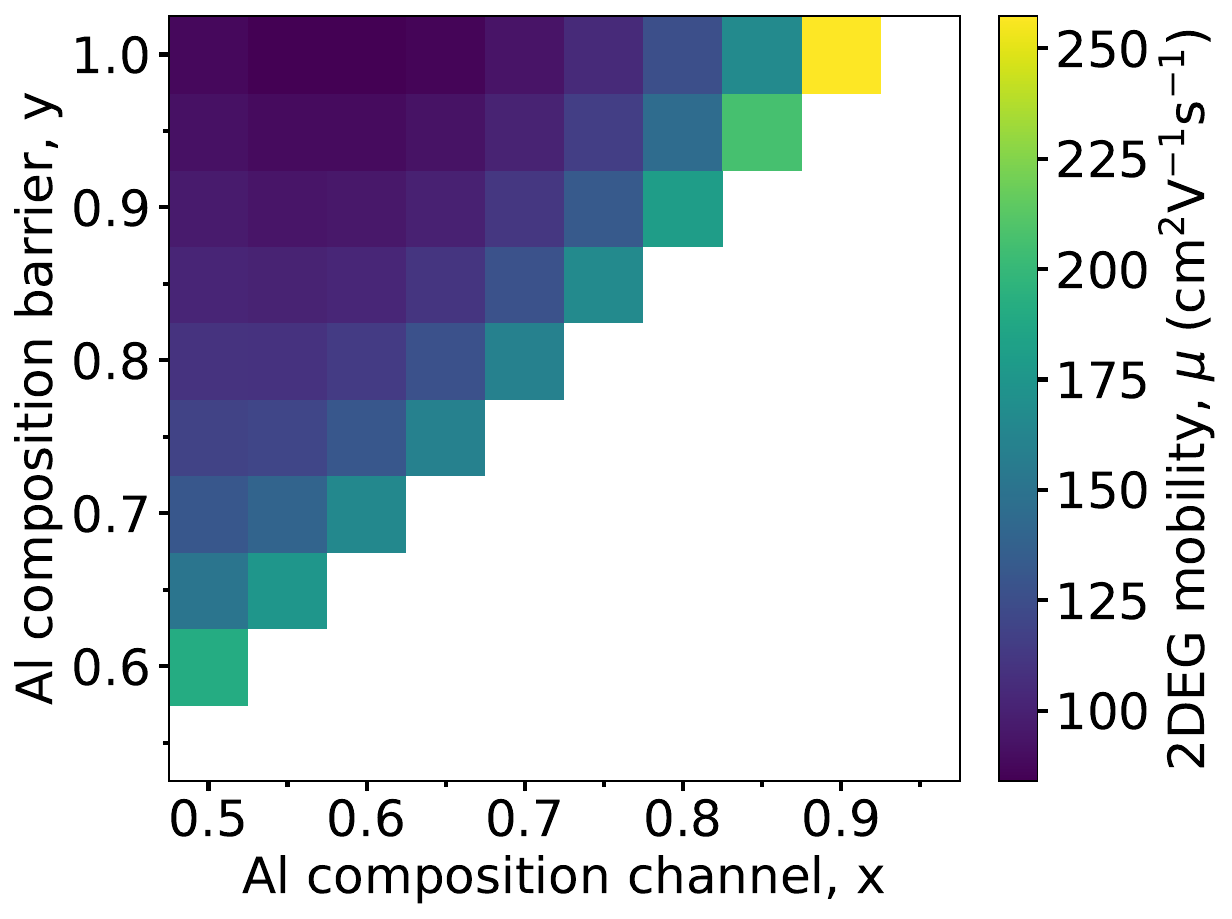}
    \caption{Maps showing the low-field 2DEG mobility, $\mu$, as a function of Al composition in the channel, $x$, and barrier, $y$, for a barrier thickness, L$_\text{B}$ = 50 nm, at 300 K temperature.}
    \label{fig:fig3}
\end{figure}

Figure~\ref{fig:fig3} presents the calculated low-field 2DEG mobility, $\mu$, at 300 K as a function of Al content in the channel, $x$, and barrier, $y$, with L$_\text{B}$ = 50 nm. When decomposing the total mobility into individual scattering mechanisms, we find that alloy disorder scattering is the dominant contributor limiting the mobility (Fig.~S8). This means that the Al contrast, $\Delta_{yx}$, should be small, i.e., higher $x$ values are required to achieve high $\mu$ values for a given y in Al$_y$Ga$_{1-y}$N/Al$_x$Ga$_{1-x}$N HEMTs (Fig.~\ref{fig:fig3}). This contrasts with the trend observed for $n_{\text{2D}}$ in Fig.~\ref{fig:fig2a}. Comparing Figs.~\ref{fig:fig2a} and \ref{fig:fig3}, clearly, while $n_{\text{2D}}$ increases with increasing $\Delta_{yx}$, $\mu$ decreases due to enhanced alloy disorder: the highest $n_{\text{2D}}$ value is found for an AlN/Al$_{0.5}$Ga$_{0.5}$N structure, whereas the AlN/Al$_{0.9}$Ga$_{0.1}$N system exhibits the largest $\mu$ value. 

Similar mobility maps for other L$_\text{B}$ values ($5 - 50$ nm) are provided in the SI. Across all barrier thicknesses, the highest $n_{\text{2D}}$ and $\mu$ for (Al,Ga)N-channel HEMTs are found when using AlN as the barrier. Therefore, in the following, we focus on AlN/Al$_x$Ga$_{1-x}$N HEMT configurations.

\begin{figure}
    \centering
    \includegraphics[width=0.85\columnwidth]{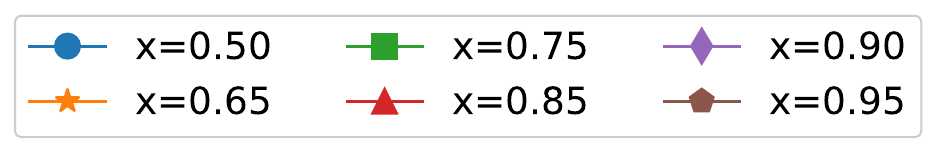}\\
    \subfloat[]{\label{fig:fig4a}\includegraphics[width=0.85\columnwidth]{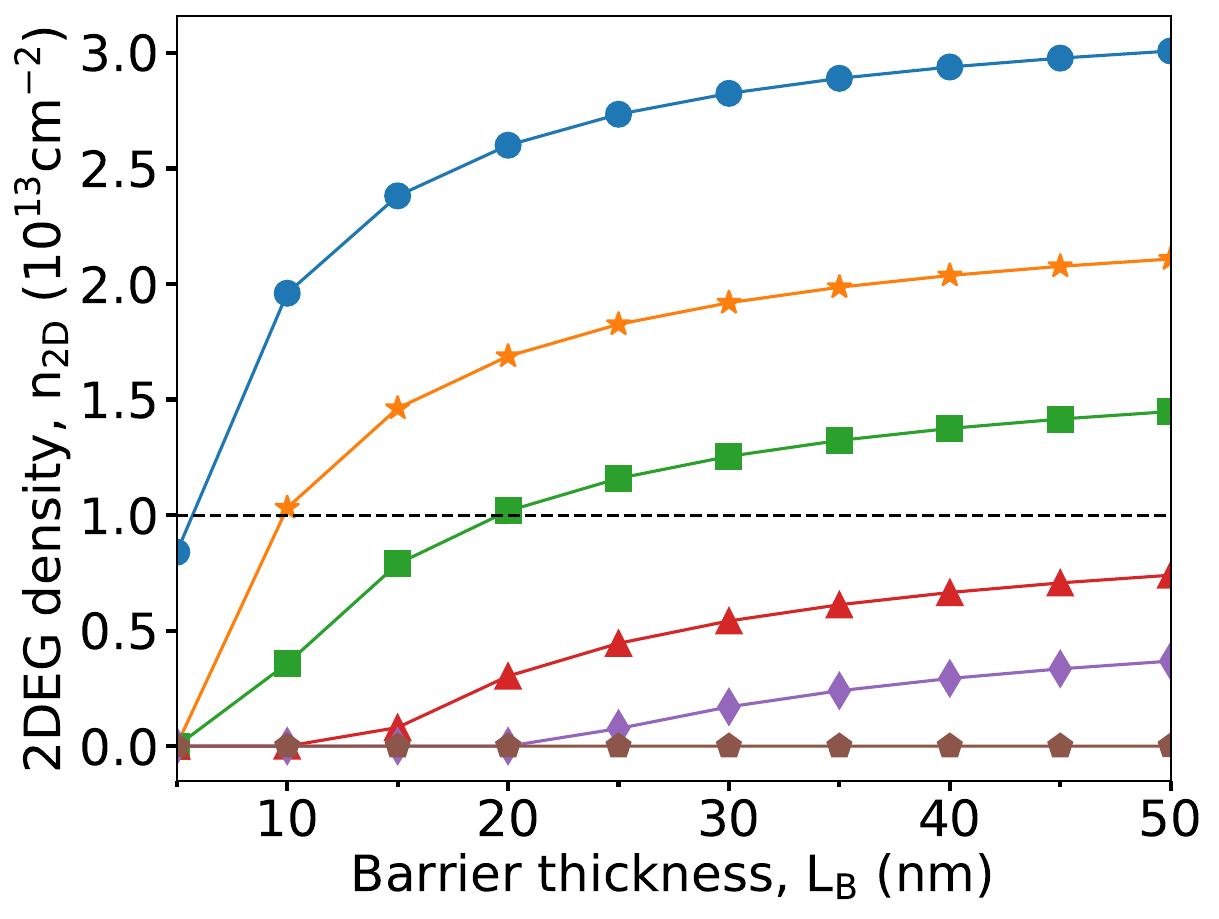}}\\
    \subfloat[]{\label{fig:fig4b}\includegraphics[width=0.85\columnwidth]{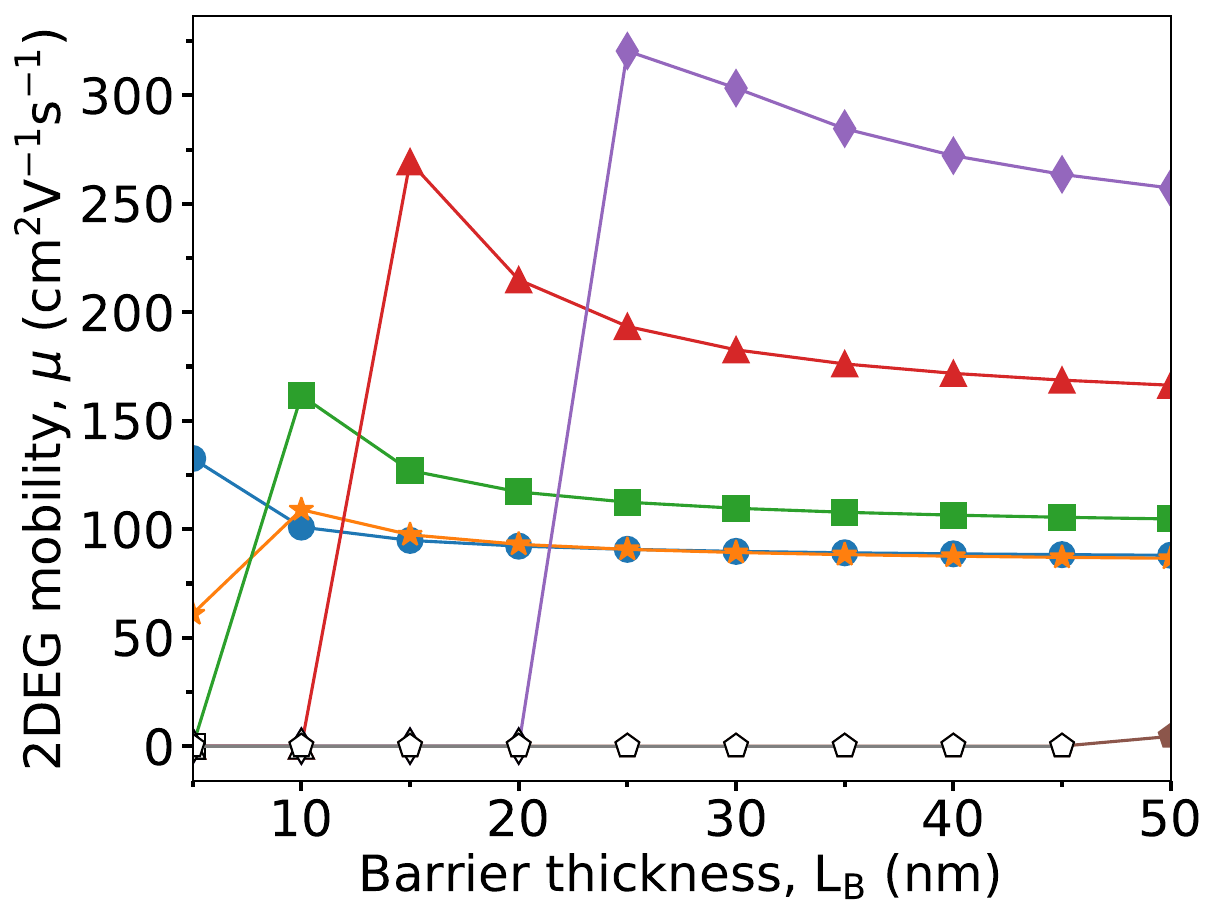}}\\
    \subfloat[]{\label{fig:fig4c}\includegraphics[width=0.85\columnwidth]{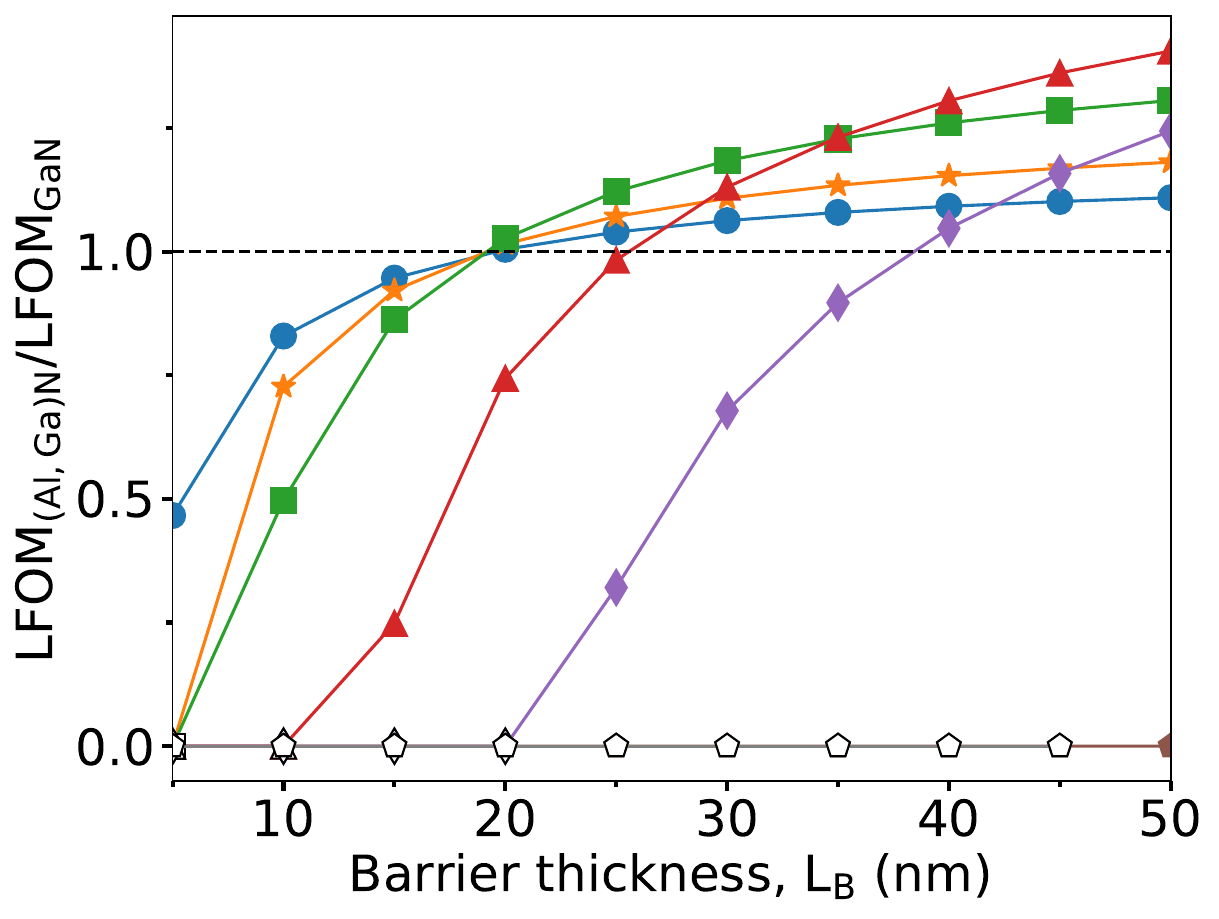}}\\
    \caption{Maps depicting (a) 2DEG density, $n_{\text{2D}}$, (b) electron mobility, $\mu$, and (c) normalized lateral figure-of-merit, LFOM$_{\text{norm}}$ (= LFOM$_{\text{(Al,Ga)N}}$ / LFOM$_{\text{GaN}}$), as functions of barrier thickness, L$_\text{B}$, at 300 K and for different Al mole fractions in the channel, $x$, of AlN(L$_\text{B}$)/Al$_x$Ga$_{1-x}$N HEMT structures. The $x$ legends are displayed at the top. Empty markers in (b) and (c) represent undefined $\mu$ and LFOM$_{\text{norm}}$ values due to the absence of 2DEG.}
    \label{fig:fig4}
\end{figure}

Figures~\ref{fig:fig4a} and \ref{fig:fig4b} illustrate $n_{\text{2D}}$ and $\mu$ for AlN/Al$_x$Ga$_{1-x}$N HEMTs and their dependence on barrier thicknesses L$_\text{B}$ and for different $x$. As $x$ increases, the composition contrast between the AlN barrier and Al$_x$Ga$_{1-x}$N channel decreases, reducing $n_{\text{2D}}$ [Fig.~\ref{fig:fig4a}]. Conversely, higher $x$ reduces alloy-disorder scattering, enhancing $\mu$ [Fig.~\ref{fig:fig4b}]. Furthermore, while increasing L$_\text{B}$ raises the 2DEG density, our calculations also show that the 2DEG wave functions extend further into the channel material with increasing L$_\text{B}$ (Fig.~S9). As a consequence, alloy-disorder-mediated scattering processes are increased, resulting in a slight decrease in $\mu$.\cite{Bassaler2024AlRichMobility} Overall, these competing effects result in the highest $n_{\text{2D}}$ for AlN(50nm)/Al$_{0.5}$Ga$_{0.5}$N, while highest $\mu$ is found in an AlN(25nm)/Al$_{0.9}$Ga$_{0.1}$N structure. 

Finally, the power-switching performance of the HEMT devices is analysed using the lateral figure-of-merit (LFOM).\cite{Coltrin2017AnalysisAlloys} While we focus here on higher power devices, we note that, (Al,Ga)N-channel HEMTs are also promising for high-frequency RF applications,\cite{Baca2020Al-richTransistors} where their performance is often evaluated using Johnson figure-of-merit (JFOM).\cite{Coltrin2017AnalysisAlloys,Johnson2005PhysicalTransistors} However, since JFOM depends on carrier saturation velocity, which requires extensive computational techniques, such as Monte Carlo simulations\cite{Farahmand2001MonteTernaries} and phonon-lasing models,\cite{Khurgin2016AmplifiedTransistors} this aspect is beyond the scope of this paper. 

Notably, LFOM is an intrinsic material metric that reflects potential but does not directly translate to HEMT device performance, as it does not account for parasitics or explicit device geometry.\cite{Coltrin2017AnalysisAlloys} Thus, the LFOM-based analysis presented in the following indicates device performance limits determined by the limitations posed by fundamental material properties but excludes factors such as contact resistance, device scaling, and geometry.

Figure~\ref{fig:fig4c} shows the LFOM for the (Al,Ga)N-channel HEMTs [LFOM$_{\text{(Al,Ga)N}}$] at 300 K, normalized to the LFOM of a widely discussed reference GaN-channel HEMT, namely an Al$_{0.25}$Ga$_{0.75}$N(25nm)/GaN system (LFOM$_{\text{GaN}}$).\cite{Ma2019ReviewW,Udrea2024GaNTechnology} Further details on the LFOM calculations and the selection of the reference GaN-channel HEMT are discussed in SI, Sec.~S9.

As shown in Fig.~\ref{fig:fig4c}, above minimum barrier thicknesses, the normalized LFOM (LFOM$_{\text{norm}}$ = LFOM$_{\text{(Al,Ga)N}}$ / LFOM$_{\text{GaN}}$) surpasses unity, demonstrating the superior performance of Al-rich (Al,Ga)N-channel devices over GaN-channel devices performance across the entire Al composition range considered (i.e., $0.5-0.9$ Al mole fraction). Despite the highest mobility being found in the AlN(25nm)/Al$_{0.9}$Ga$_{0.1}$N structure, the low $n_{\text{2D}}$ value of this structure results in a low LFOM$_{\text{(Al,Ga)N}}$, and thus, a lower LFOM$_{\text{norm}}$ [Figs.~\ref{fig:fig4a} -- \ref{fig:fig4c}]; LFOM$_{\text{GaN}}$ in Fig.~\ref{fig:fig4c} is a single number and serves as a constant rescaling factor. In contrast, while the AlN(50nm)/Al$_{0.85}$Ga$_{0.15}$N structure has relatively lower $\mu$, the higher $n_{\text{2D}}$ value leads to a superior LFOM$_{\text{(Al,Ga)N}}$ and LFOM$_{\text{norm}}$. Furthermore, the AlN/Al$_{0.75}$Ga$_{0.25}$N structure achieves the highest LFOM$_{\text{norm}}$ for L$_\text{B}$ values between 20 and 35 nm (commonly used in GaN HEMT devices), while the AlN/Al$_{0.85}$Ga$_{0.15}$N structure exhibits the largest LFOM$_{\text{norm}}$ for L$_\text{B}$ values in the $35 - 50$ nm range.

\begin{figure}
    \centering
    \subfloat[]{\label{fig:fig5a}\includegraphics[width=0.85\columnwidth]{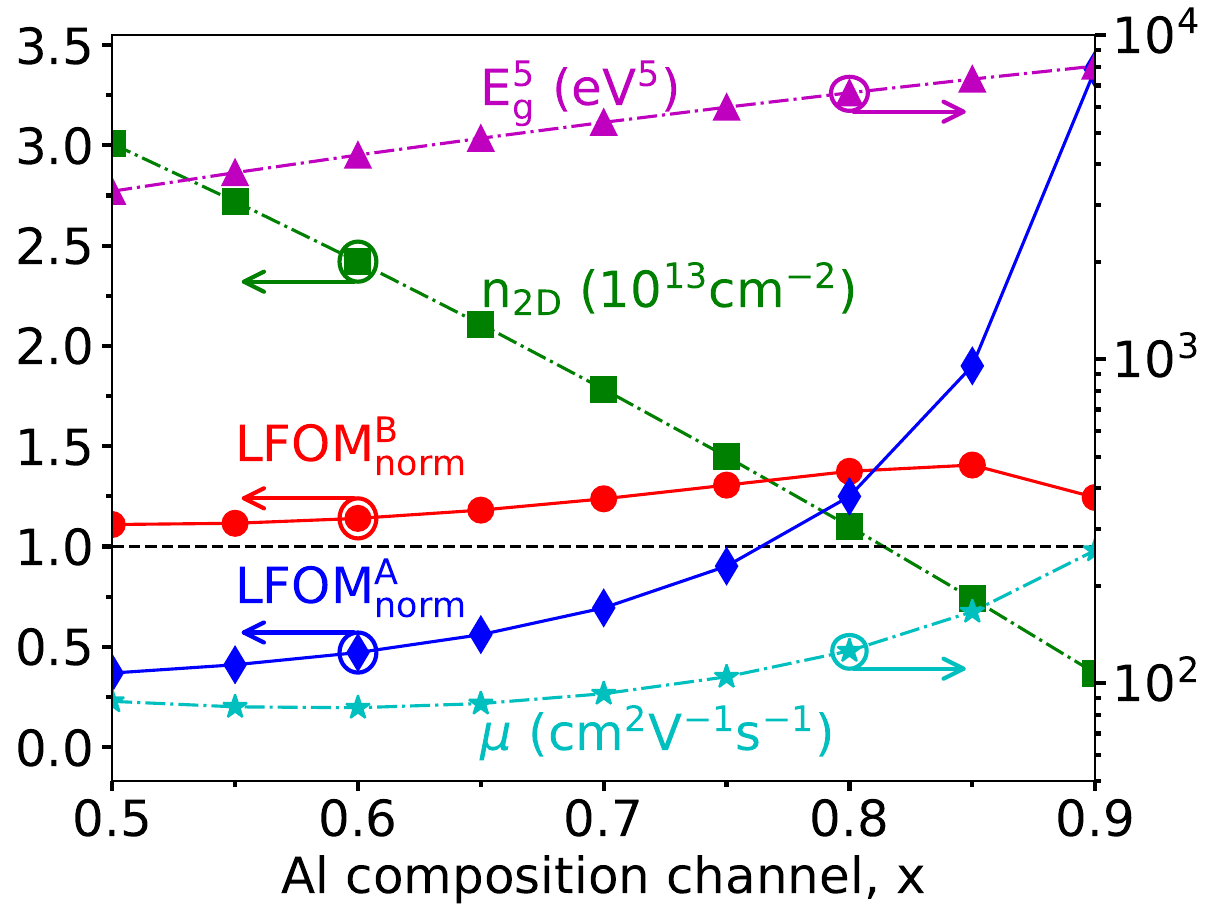}}\\
    \includegraphics[width=0.8\columnwidth]{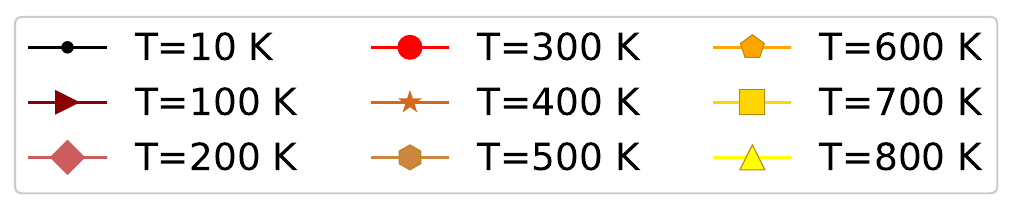}\\
    \vspace*{-0.4cm}
    \subfloat[]{\label{fig:fig5b}\includegraphics[width=0.85\columnwidth]{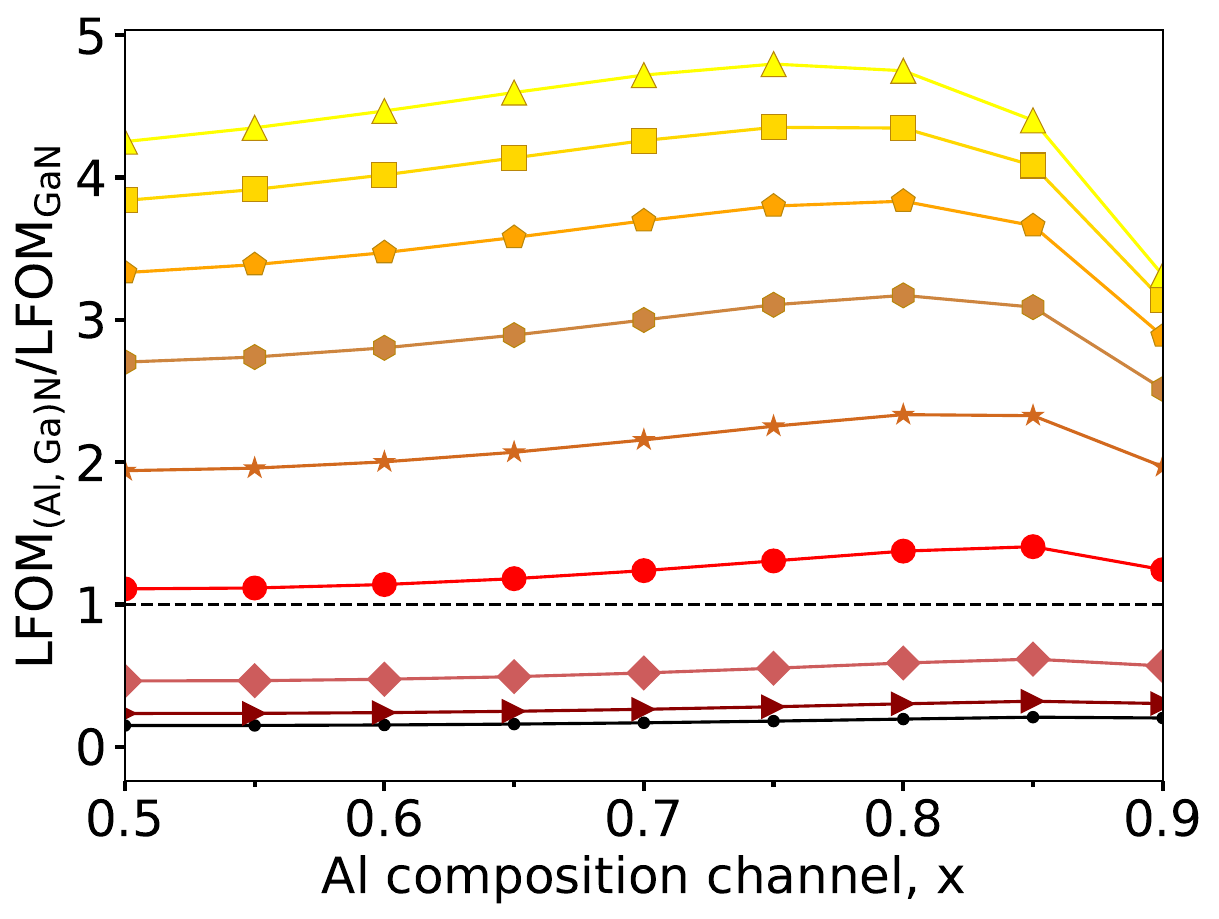}}
    \caption{(a) Normalized lateral figure-of-merit, LFOM$_{\text{norm}}$ (= LFOM$_{\text{(Al,Ga)N}}$ / LFOM$_{\text{GaN}}$) as a function of channel Al composition, $x$, for AlN(50nm)/Al$_x$Ga$_{1-x}$N HEMTs at 300 K temperature. The red solid curve shows LFOM$_{\text{norm}}$ using our calculated LFOM$_{\text{(Al,Ga)N}}$ (LFOM$^{\text{B}}_{\text{norm}}$), while the blue solid curve assumes a constant $n_{\text{2D}}$ = $1 \times 10^{13}$ cm$^{-2}$ (LFOM$^{\text{A}}_{\text{norm}}$). Dotted dashed lines show individual contributions to LFOM$_{\text{norm}}$. $n_{\text{2D}}$ and LFOMs are plotted on the left y-axis, with $\mu$ and E$_\text{g}^5$ on the right y-axis. (b) LFOM$_{\text{norm}}$ at various temperatures, T, with T (in K) values indicated in the legend. LFOM$_{\text{(Al,Ga)N}}$ is normalized against LFOM$_{\text{GaN}}$ at each T. }
    \label{fig:fig5}
\end{figure}

It is important to note that the LFOMs discussed in this study (LFOM $\propto n_{\text{2D}}\,\mu \,\text{E}^5_g$) are built on the $n_{\text{2D}}$ and $\mu$ values obtained from our device simulations. Figure~\ref{fig:fig5a} illustrates the variation of LFOM$_{\text{norm}}$ with Al composition in the channel, $x$, for an example AlN/Al$_x$Ga$_{1-x}$N HEMT structure, when (i) assuming constant $n_{\text{2D}}$ = $1 \times 10^{13}$ cm$^{-2}$, as done in previous studies (LFOM$^{\text{A}}_{\text{norm}}$)\cite{Bassaler2024AlRichMobility,Coltrin2017AnalysisAlloys,Bajaj2014ModelingVoltage} and (ii) using our directly calculated and thus x dependent $n_{\text{2D}}$ values (LFOM$^{\text{B}}_{\text{norm}}$). 

In the constant $n_{\text{2D}}$ case, LFOM is determined solely by $\mu$ and the alloy bandgap (E$_\text{g}$). For Al$_x$Ga$_{1-x}$N both $\mu$ and E$_\text{g}$ increase monotonically with Al composition, $x$, resulting in a monotonically increasing LFOM$_{\text{(Al,Ga)N}}$. Since, the LFOM$_{\text{GaN}}$ is a $x$ independent constant normalization factor here, LFOM$^{\text{A}}_{\text{norm}}$ shows a steady increase with $x$.\cite{Zhang2008TheContent,Bassaler2024AlRichMobility,Coltrin2017AnalysisAlloys,Bajaj2014ModelingVoltage,Ambacher1999Two-dimensionalHeterostructures} However, when the variation in $n_{\text{2D}}$ with $x$ is included --- which exhibits an opposite trend to $\mu$ and E$_\text{g}$ --- a non-monotonic trend with composition emerges in LFOM$^{\text{B}}_{\text{norm}}$. This behaviour highlights the interplay of opposing contributions in LFOM. 

A comparison between LFOM$^{\text{A}}_{\text{norm}}$ and LFOM$^{\text{B}}_{\text{norm}}$ in Fig.~\ref{fig:fig5a} further reveals that assuming a constant $n_{\text{2D}}$ underestimates LFOM for $x < 0.80$ and overestimates it for $x > 0.80$. These findings again emphasize the critical importance of explicit simulations for an accurate description of the performance of (Al,Ga)N-based HEMT structures. Notably, a theory-experiment comparison in Ref.~\citenum{Coltrin2017AnalysisAlloys} also hinted a similar performance disparity of AlGaN-channel HEMTs, potentially stemming from the assumption of a composition-independent $n_{\text{2D}}$.

Following the promises demonstrated for high-temperature operation stability of Al-rich (Al,Ga)N/(Al,Ga)N HEMTs in previous studies,\cite{Kaplar2017ReviewUltra-Wide-BandgapDevices,Hussain2023HighSubstrates,Khachariya2022RecordSubstrates,Baca2016AnContact,Carey2019OperationTransistors,Yafune2014AlN/AlGaNOperation} next, we focus on the impact of temperature on the HEMT performance when directly calculating $n_{\text{2D}}$ instead of assuming it to be constant. Figure~\ref{fig:fig5b} compares the performance of AlN(50nm)/(Al,Ga)N HEMTs across a temperature range of $10 - 800$ K. However, noting the practical challenges of realizing electrical contacts on a 50 nm thick AlN barrier,\cite{Baca2020Al-richTransistors} similar temperature dependence data for a relatively thinner and more practical AlN(25nm)/(Al,Ga)N structure is provided in Fig.~S10. While the quantitative values slightly change with thinner barriers, these differences do not impact the overall conclusions of this study.

From Fig.~\ref{fig:fig5b}, it is evident that the LFOM$_{\text{norm}}$ surpasses unity at and above room temperature, clearly showing that Al-rich (Al,Ga)N-channel devices outperform GaN-channel counterparts at elevated temperatures. We find that although both (Al,Ga)N- and GaN-channel HEMTs experience a decrease in absolute LFOMs with temperature, the GaN-channel devices exhibit a comparatively more pronounced decline (Fig.~S11).  In both cases, the 2DEG density is predominantly confined in the channel region, and we observe that the mobility in (Al,Ga)N-channel HEMTs is limited by alloy-disorder-induced scattering effects, whereas in GaN-channel devices, which lack alloy disorder, phonon scattering is the dominant mechanism affecting $\mu$. Given the strong temperature dependence of phonon-mediated scattering processes, especially when contrasted to alloy effects, a greater temperature sensitivity of GaN-channel HEMTs in comparison to the (Al,Ga)N-channel device is expected.\cite{Singhal2022TowardHeterostructures,Coltrin2017AnalysisAlloys}

Notably, the Al$_{0.85}$Ga$_{0.15}$N-channel exhibited the highest LFOM$_{\text{norm}}$ up to 300 K, with the peak shifting towards lower Al content, reaching Al$_{0.75}$Ga$_{0.25}$N-channel at 800 K [Figs.~\ref{fig:fig5b} and S12]. This shift highlights the importance of selecting appropriate Al compositions to optimize the HEMT device performance for applications across a diverse temperature range such as in jet engines ($\sim 700-1300$ K), Venus's surface exploration ($\sim 750$ K), and electric vehicles ($\sim 1100$ K).\cite{Neudeck2002High-temperatureSemiconductors,Johnson2004TheElectronics} To ensure optimal performance, the Al content in HEMT structures must be tailored to the operating temperatures of specific applications. 

%\pagebreak
In summary, we address the limitations of assuming a constant 2DEG density in previous studies and demonstrate that such simplification can lead to impractical device design choices. Our results show that while 2DEG density increases with higher Al composition contrast between the barrier and channel, it negatively impacts mobility due to alloy disorder scattering. Additionally, individual layer thicknesses in HEMT heterostructures significantly affect the 2DEG density. We demonstrate a complex interplay between 2DEG density and mobility, revealing a greater potential of Al-rich (Al,Ga)N-channel HEMTs than previously predicted. We show that Al$_x$Ga$_{1-x}$N-channel HEMTs outperform GaN-channel HEMTs in lateral figure-of-merit at and above room temperature, across all Al compositions $x \geq 0.5$. This contrasts with previous studies that suggested parity between (Al,Ga)N- and GaN-channel HEMTs at room temperature only at high Al content ($x \geq 0.85$). Our work emphasizes the need for detailed device simulations that explicitly account for layer thickness and alloy composition to fully exploit the potential of (Al,Ga)N-channel HEMTs. The insights gained from this study provide an in-depth understanding of the trade-offs between device and material parameters, offering valuable guidance for the design of next-generation Al-rich ($x = 0.5 - 1.0$) Al$_x$Ga$_{1-x}$N-channel HEMTs for high-power applications.

%\clearpage
\begin{acknowledgments}
The authors thank Andrew M. Armstrong from Sandia National Laboratories, Albuquerque, NM, USA for valuable discussions. This research is supported by
Taighde \'Eireann - Research Ireland, formerly Science
Foundation Ireland (SFI) under Grant No. 23/EPSRC/3888. The computing resources are provided by Research Ireland to the Tyndall National Institute. 
\end{acknowledgments}

\section*{Author declarations}
\subsection*{Conflict of interest} 
The authors have no conflicts to disclose.
\subsection*{Author contributions}
B. Mondal: Conceptualization (equal); Data curation (lead); Formal analysis (lead); Methodology (lead); Resources (equal); Software (lead); Validation (equal); Visualization (lead); Writing – original draft (lead). P. Pampili: Conceptualization (support); Formal analysis (support); Writing – review \& editing (support). J. Mukherjee: Conceptualization (support); Formal analysis (support); Writing – review \& editing (support). D. Moran: Conceptualization (support); Formal analysis (support); Funding acquisition (equal); Project administration (equal); Writing – review \& editing (support). P. J. Parbrook: Conceptualization (support); Formal analysis (support); Funding acquisition (equal); Project administration (equal); Writing – review \& editing (support). S. Schulz: Conceptualization (equal); Formal analysis (support); Funding acquisition (equal); Methodology (support); Project administration (equal); Resources (equal); Software (support); Supervision (lead); Validation (equal); Visualization (support); Writing – review \& editing (lead).
\section*{Data availability} 
Detailed descriptions of the Nextnano++ simulations and mobility models are provided in the Supplementary Information (Supplementary\_Information.pdf). Refs.~\citenum{Chuang1996,*Schulz2015ElectronicSystems,*Romanov2006Strain-inducedLayers,*Vurgaftman2001BandAlloys,*Vurgaftman2003BandSemiconductors,*Pant2020HighStructure,*Miao2010OxidationHeterostructures,*Chakraborty2022ComprehensiveTransistors,*Hahn2013AlNProposal,*Armstrong2016Polarization-inducedAlloys,*Arehart2006EffectCharacteristics,*Monroy2002ThermalAl0.31Ga0.69N,*Ofuonye2014ElectricalHeterostructures,*Miyoshi2015NumericalHeterostructures,*Fang1966NegativeSurfaces,*Qiao2000DependenceFraction,*NSMSemiconductors,*Chaudhuri2022InSituTransistors,*Chen2022EffectBarrier,*Zervos2024AHEMTs,*Wei1998CalculatedOrbitals,*Chaudhuri2022InSituTransistors,*Riddet20073-DMOSFETs,*Alexander2011StatisticalScattering,*Vyas2018QuantumFETs}  are additionally referenced within the SI. The raw Nextnano++ simulations data are openly available in NOMAD at doi:XXX. The attached zip file contains the post-processed data (Excel sheets) for carrier densities, mobility, and figure-of-merit from the simulations. The corresponding figures, showing band diagrams, carrier densities, mobility, and figure-of-merit as a function of Al compositions and layer thicknesses, are also included within the zip file. The Python Jupyter notebooks used for automatizing Nextnano++ simulations and data post-processing are available on GitHub.\cite{Mondal2024SchrodingerPoissonSimulations} Additionally, we have implemented the mobility models as an open-source Python package `mobilitypy', available on GitHub.\cite{Mondal2024Mobilitypy} The scripts used for mobility calculations in this paper can be found in the tutorial folder within the `mobilitypy' repository.

%%%%%%%%%%%%%%%%%%%%%%%%%%%%%%%%%%%%%%%%%%%%%%%%%%%%%%%%%%%%%%%%%%%%%%%%
\bibliography{ms}% Produces the bibliography via BibTeX.

\end{document}

%\preprint{AIP/123-QED}

\title[Carrier density and mobility interplay in (Al,Ga)N]{Supplementary Information \\ \vspace*{0.3cm} Interplay of carrier density and mobility in Al-Rich (Al,Ga)N-Channel HEMTs: Impact on high-power device performance potential}
% Force line breaks with \\
\author{Badal Mondal}
\author{Pietro Pampili}%
\affiliation{Tyndall National Institute, University College Cork, Cork T12 R5CP, Ireland}%

\author{Jayjit Mukherjee}
\author{David Moran}
\affiliation{James Watt School of Engineering, University of Glasgow, Glasgow G12 8LT, UK}%

\author{Peter James Parbrook}%
\affiliation{Tyndall National Institute, University College Cork, Cork T12 R5CP, Ireland}%
\affiliation{School of Engineering, University College Cork, Western Road, Cork, Ireland}

\author{Stefan Schulz}
\email{stefan.schulz@tyndall.ie}
\affiliation{Tyndall National Institute, University College Cork, Cork T12 R5CP, Ireland}%
\affiliation{School of Physics, University College Cork, Cork T12 YN60, Ireland}%

\date{19 February 2025}%

\maketitle

\onecolumngrid

\tableofcontents

\pagebreak
%%%%%%%%%%%%%%%%%%%%%%%%%%%%%%%%%%%%%%%%%%%%%%%%%%%%%%%%%%%%%%%%%%%%%%%%%%%%%%
\section{\label{sec:secS1}Nextnano++ simulation details}
One-dimensional (1D) Schr{\"o}dinger-Poisson simulations are performed with Nextnano++ (v-1.21.24, RHEL compilation) software.\cite{Birner2007Nextnano:Simulations,Trellakis2006TheResults} The simulation model involves a self-consistent solution of the Schr{\"o}dinger, Poisson, and charge balance equations, combined with a $6 \times 6$ \textbf{k}$\cdot$\textbf{p} Hamiltonian\cite{Chuang1996} for the hole states and 2-band effective mass model for electron eigenstates. The exchange-correlation correction to the Coulomb interaction is included in the 2-band calculations. 

Although an $8 \times 8$ \textbf{k}$\cdot$\textbf{p} model could provide further refined results, it significantly increases computational costs. However, for large bandgap materials like AlN (6.25 eV) and GaN (3.51 eV),\cite{Vurgaftman2003BandSemiconductors} the $\Gamma_{7c}$ conduction band can be decoupled from the three valence bands ($\Gamma_{7v}$ heavy hole and light holes, and the $\Gamma_{9v}$ split-off band) with minimal error.\cite{Jogai2003InfluenceTransistors} This allows the transformation of the $8\times 8$ \textbf{k}$\cdot$\textbf{p} Hamiltonian into two equivalent $1\times 1$ Hamiltonians for the $\Gamma_{7c}$ conduction band, and a $6 \times 6$ Hamiltonian for the valence bands. The two $1\times 1$ Hamiltonians (one for spin up and one for spin down) essentially correspond to the standard Schr{\"o}dinger equation with a parabolic conduction band approximation model in k-space. The s-p interaction that introduces nonparabolicity is weaker in large bandgap materials.\cite{Jogai2003InfluenceTransistors}

To further reduce the computational cost, we restrict the quantum calculations, i.e., solving the Schr{\"o}dinger equation, to two 40 nm-wide quantum regions (QRs) positioned around the buffer-channel and channel-barrier interfaces. The width of the QR is adjusted to ensure minimal leakage of two-dimensional carrier gases (2DCGs) [i.e., two-dimensional electron gas (2DEG) and two-dimensional hole gas (2DHG)] wave function tail beyond the QR boundaries. A smaller QR fails to fully confine the majority of the 2DCG wave functions, while a larger QR leads to a significant increase in computational cost. 

The total electron and hole densities are obtained by summing contributions from 50 electron and 100 hole sub-bands. These values are optimized to ensure that the higher energy states have vanishing occupations of the respective carriers (electrons and holes). Due to the larger effective mass of holes compared to electrons and the small energy separation between heavy-hole and light-hole states in (Al,Ga)N, our calculations require approximately twice as many hole eigenstates as electron eigenstates to achieve convergence.

The undoped (Al,Ga)N/(Al,Ga)N/AlN double heterostructure HEMT device with abrupt interfaces and conventional (0001) cation-faced growth direction is used in this study. The spontaneous and piezoelectric polarizations are included in the Poisson equation to solve for the electrostatic potential with charge neutrality enforced throughout the whole device.\cite{Schulz2015ElectronicSystems,Romanov2006Strain-inducedLayers} Homogeneous (pseudomorphic) strain with respect to an AlN substrate is assumed in all layers. The strain effect is included in the eigenstate calculations via deformation potential theory. The material parameters used in the simulations are given below (Table~\ref{tab:tableS1}). We note that device scaling and geometry additionally influence 2DCG densities in realistic devices --- effects that cannot be captured by 1D simulations and require 2D or 3D modeling.\cite{Riddet20073-DMOSFETs,Alexander2011StatisticalScattering,Vyas2018QuantumFETs} However, these do not affect the overall conclusions of this study. 

\begin{figure}%[!htbp]
    \centering
    \subfloat[]{\label{fig:figS1a}\includegraphics[width=0.45\textwidth]{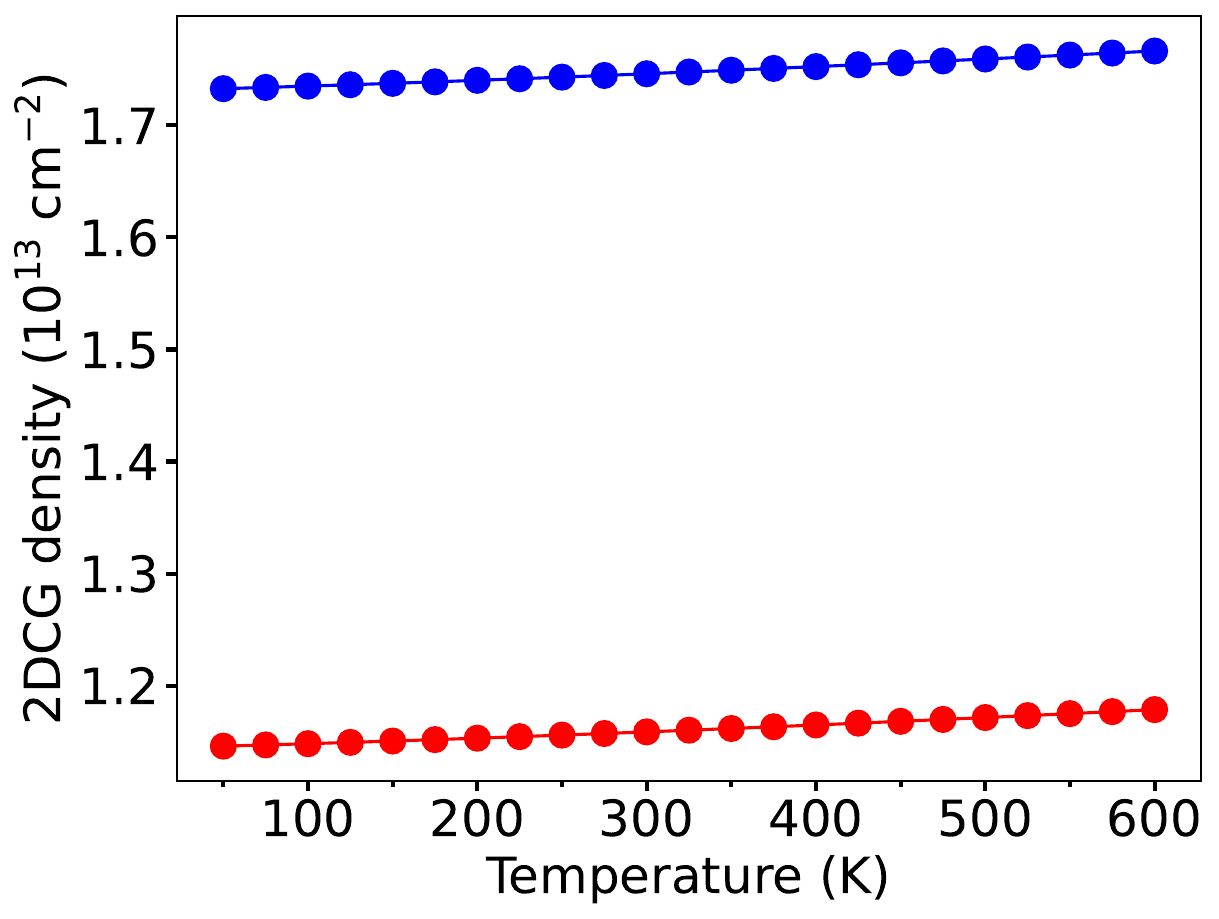}}
    \hspace{.5cm}
    \subfloat[]{\label{fig:figS1b}\includegraphics[width=0.45\textwidth]{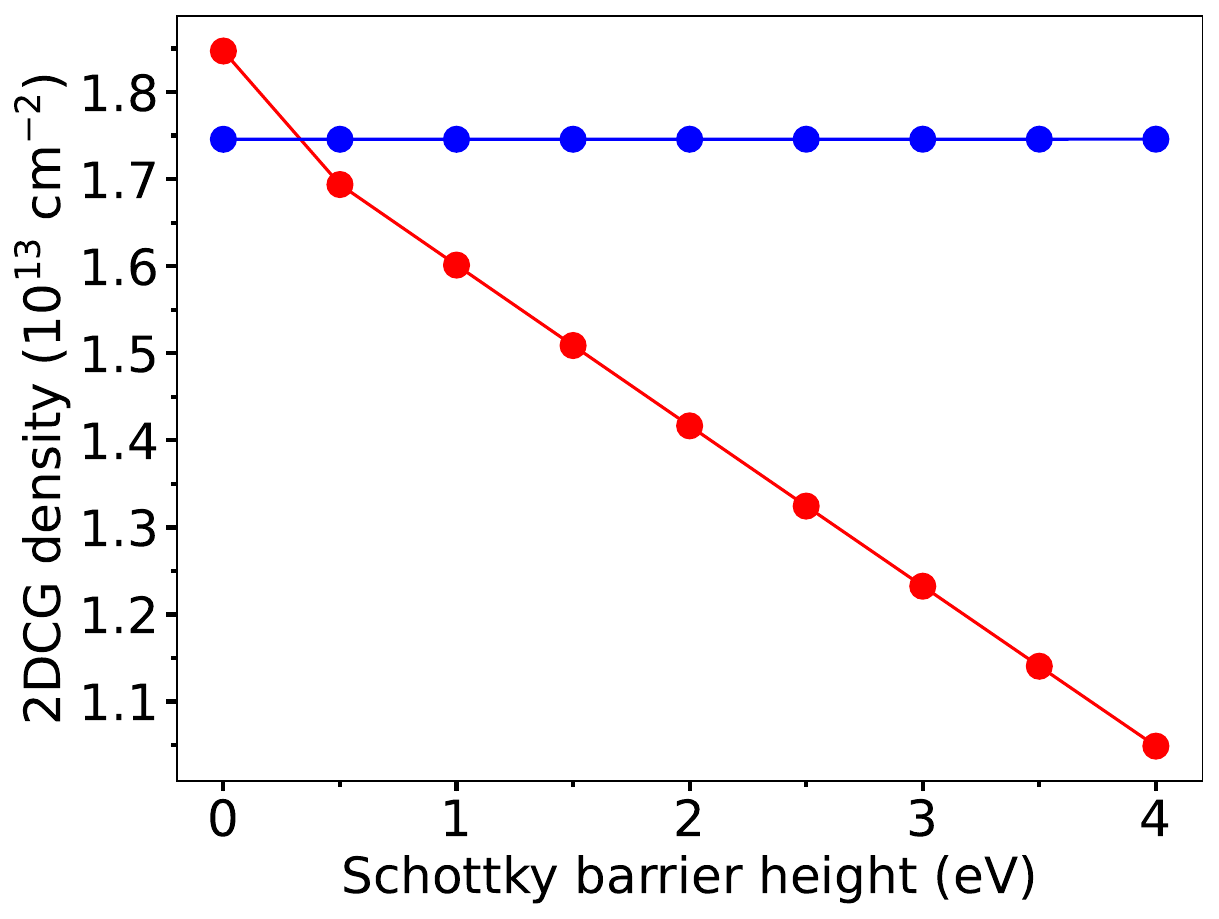}}
    \caption{Dependence of 2D carrier gas (2DCG) densities on (a) temperature and (b) Schottky barrier height for the AlN(25nm)/Al$_{0.75}$Ga$_{0.25}$N(300nm) HEMT structure. The 2D electron gas (2DEG) is shown in red, and the 2D hole gas (2DHG) is shown in blue.}
    \label{fig:figS1}
\end{figure}

Simulations are conducted at 300 K, with temperature effects incorporated into the simulation by accounting for the linear thermal expansion of lattice parameters and the temperature dependency of bandgap. However, we highlight that simulations with varied temperatures show that carrier densities, both the 2DEG and 2DHG, are largely unaffected by temperature, as shown in Fig.~\ref{fig:figS1a} for an example HEMT structure. Since 2DCG formation is primarily driven by polarization effects, it depends electrostatically on the layer thickness and Al composition, with little influence from temperature.\cite{Singhal2022TowardHeterostructures} The small temperature dependency in 2D carrier densities arises from slight changes in the bandgap (Varshni's formula) and piezoelectric response due to the thermal expansion of lattice parameters. 

A 10 nm wide Schottky contact is assumed at the top of the device, with the Schottky barrier height for (Al,Ga)N alloy in the barrier linearly interpolated between those of AlN (3.4 eV)\cite{Armstrong2016Polarization-inducedAlloys} and GaN (1.11 eV)\cite{Arehart2006EffectCharacteristics}. Experimental studies have shown a linear relationship between Schottky barrier height and Al content in (Al,Ga)N within the small strain regime.\cite{Qiao2000DependenceFraction,Hahn2013AlNProposal} For our HEMT structures, with an AlN buffer and a high Al content Al$_x$Ga$_{1-x}$N channel (x $\geq$ 0.5), the strain in the (Al,Ga)N barrier is relatively small, supporting the applicability of this linear approximation. 

Moreover, it is important to emphasize that the Schottky barrier height can also strongly depend on several other factors, such as the metal stack used in the contact (e.g., Ni vs Ni/Au vs Pt/Au, etc.),\cite{Monroy2002ThermalAl0.31Ga0.69N} temperature,\cite{Monroy2002ThermalAl0.31Ga0.69N,Ofuonye2014ElectricalHeterostructures} defect density,\cite{Arehart2006EffectCharacteristics,Chakraborty2022ComprehensiveTransistors} and surface passivation.\cite{Miao2010OxidationHeterostructures} The growth method (e.g., MBE vs. MOVPE) and strain state, additionally, may affect the Schottky barrier height, although the exact nature of this dependency remains somewhat unclear.\cite{Armstrong2016Polarization-inducedAlloys} Nevertheless, for simplicity, we assume a linear dependence of Schottky barrier height on the Al content in our all calculations. It should, however, be noted that the choice of Schottky barrier height can slightly affect the carrier densities, as demonstrated in Fig.~\ref{fig:figS1b} for an example HEMT structure. Therefore, variation of Schottky barrier height with Al content in (Al,Ga)N for high Al content ($\geq 50\%$ Al) needs further investigation. 

Finally, Neumann boundary conditions are applied to the potential at the device's end (i.e. rightmost AlN buffer–air interface in Fig.~\ref{fig:figS2}). We note that although no doping is used in any region of the device, the Neumann boundary condition causes the valence bands in the rightmost AlN buffer region to approach the Fermi level [Fig.~1(b) in main text and Fig.~\ref{fig:figS2a}]. Further details on this issue are discussed in the next section (Sec.~\ref{sec:secS2}).

%%%%%%%%%%%%%%%%%%%%%%%%%%%%%%%%%%%%%%%%%%%%%%%%%%%%%%%%%%%%%%%%%%%%%%%%%%%%%%
\section{\label{sec:secS2}Impact of boundary conditions in Nextnano++ simulations} 
\begin{figure}[!htbp]
    \centering
    \subfloat[]{\label{fig:figS2a}\includegraphics[width=0.45\textwidth]{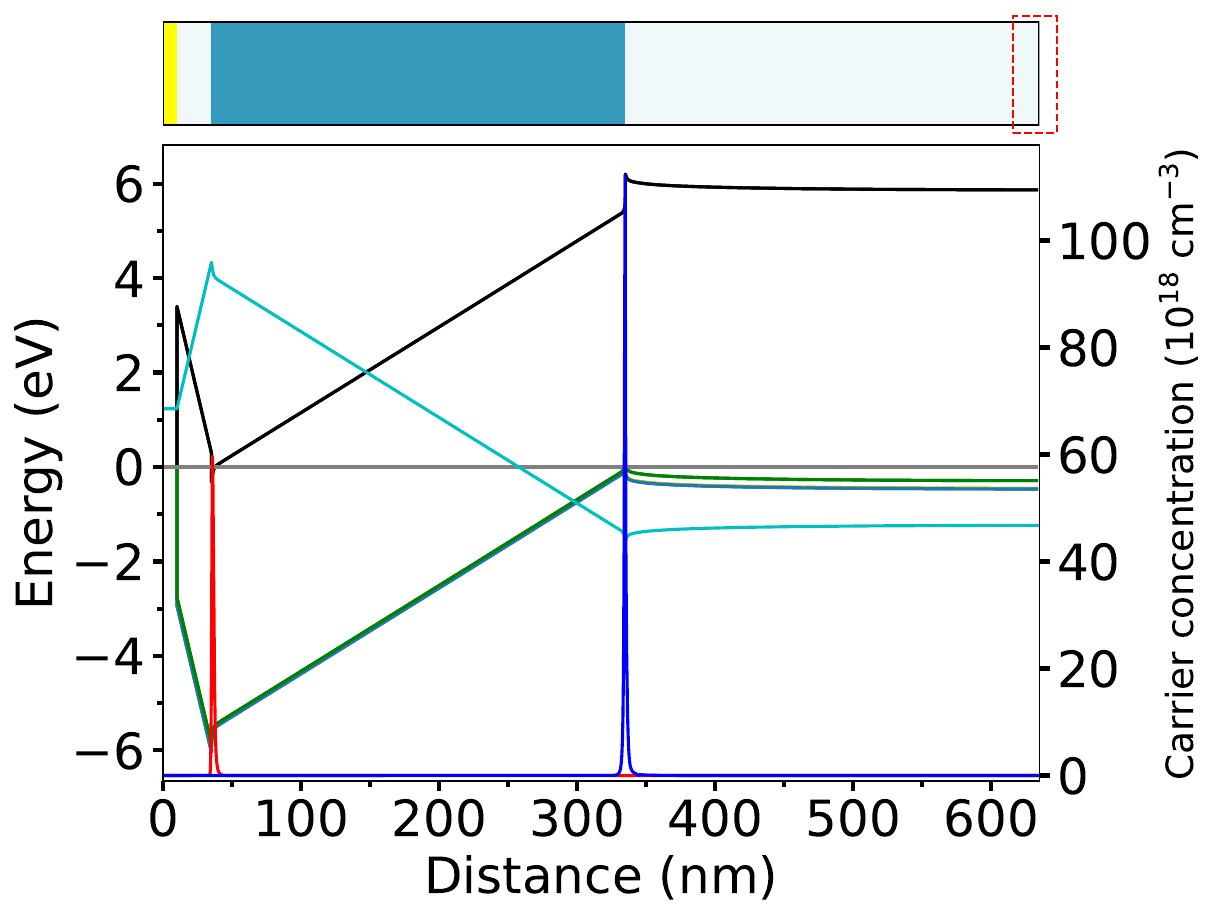}}
    \hspace{.5cm}
    \subfloat[]{\label{fig:figS2b}\includegraphics[width=0.45\textwidth]{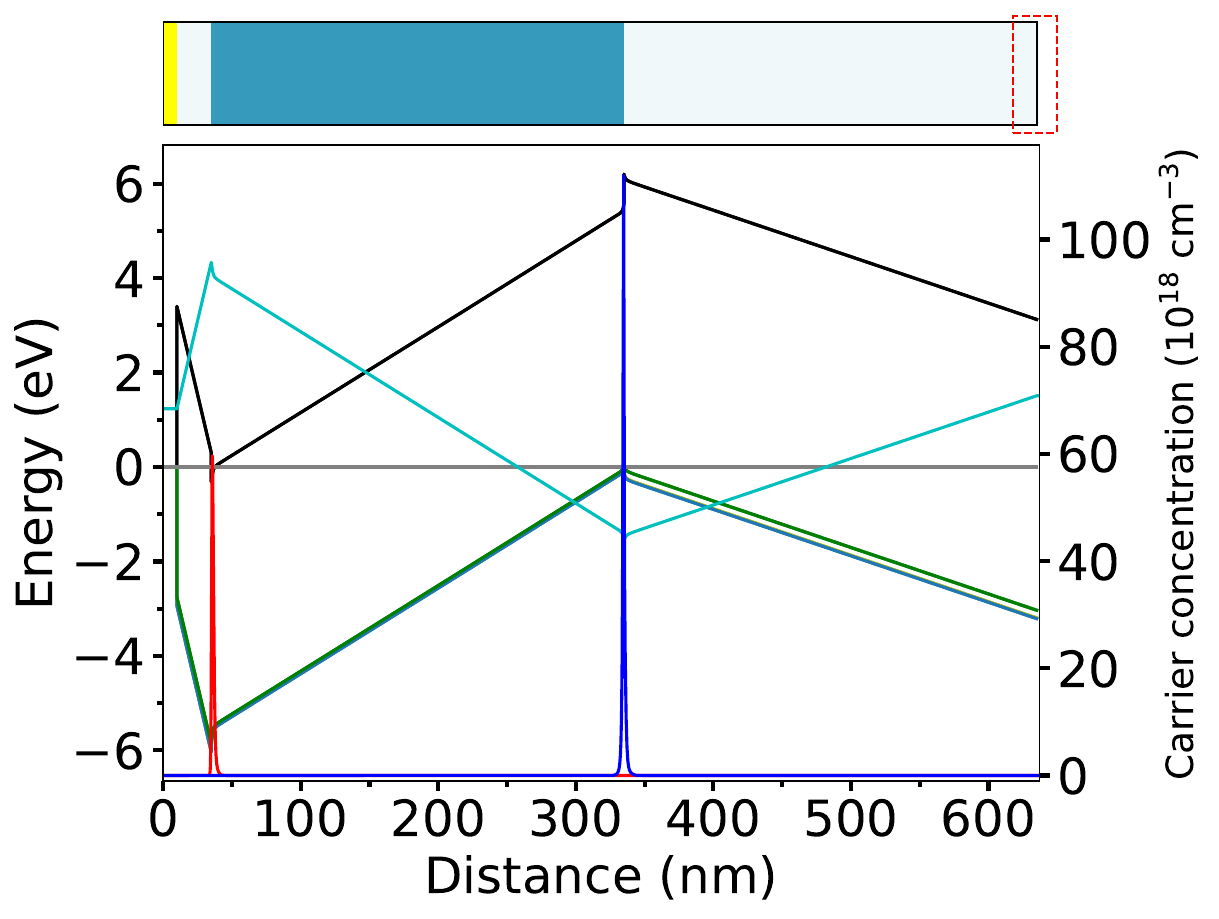}}
    \caption{Band diagram of AlN(25nm)/Al$_{0.75}$Ga$_{0.25}$N(300nm)/AlN(300nm) HEMT with (a) Neumann boundary condition and (b) Schottky contacts at the device rightmost end, highlighted in dotted red boxes. The conduction band (black), valence bands --- heavy hole, light hole, and split-off --- in green, the Fermi level (gray), and potential energy (cyan) across the device are shown. The 2D representation of the 1D composition profile along the device length is displayed above the band diagram. The x-axis (left to right) represents the top to bottom of the device, opposite to the growth direction. The 2DEG (in red) and 2DHG (in blue) distribution are plotted on the right axis.}
    \label{fig:figS2}
\end{figure}

Figure~\ref{fig:figS2} shows the band diagram for the AlN(25nm)/Al$_{0.75}$Ga$_{0.25}$N(300nm)/AlN(300nm). In both cases, a Schottky barrier is applied at the top (Al,Ga)N barrier-air interface (leftmost boundary in Fig.~\ref{fig:figS2}), with the Schottky barrier height determined by the Al composition in the (Al,Ga)N barrier, as described in Sec.~\ref{sec:secS1}. The key difference between Figs.~\ref{fig:figS2a} and ~\ref{fig:figS2b} are, in the boundary condition applied to the potential at the bottom AlN buffer-air interface (right-most boundary in Fig.~\ref{fig:figS2}). In Fig.~\ref{fig:figS2a}, a Neumann boundary condition is used, whereas in Fig.~\ref{fig:figS2b}, a Schottky contact is applied, with the Schottky barrier height set to the mid-bandgap value of AlN. 
 	 
We observe that although no doping is used to any regions of the device, the Neumann boundary condition causes the valence bands in the rightmost AlN buffer region to approach the Fermi-level [Fig.~\ref{fig:figS2a}]. In contrast, using a Schottky contact yields a more realistic band diagram, with the Fermi level positioned at the mid-bandgap, consistent with bulk semiconductor properties [Fig.~\ref{fig:figS2b}]. When the Neumann boundary condition is applied, the absence of an additional charge source at the device's end causes the electric potential [cyan line Fig.~\ref{fig:figS2a}] to stabilize at a constant value, resulting in the bands becoming flat and cannot be bent. In Fig.~\ref{fig:figS2b}, this additional charge source is introduced through a `fictitious' end Schottky contact with a finite Schottky barrier height, which alters the band structure. 

\begin{figure}
    \centering
    \includegraphics[width=0.45\textwidth]{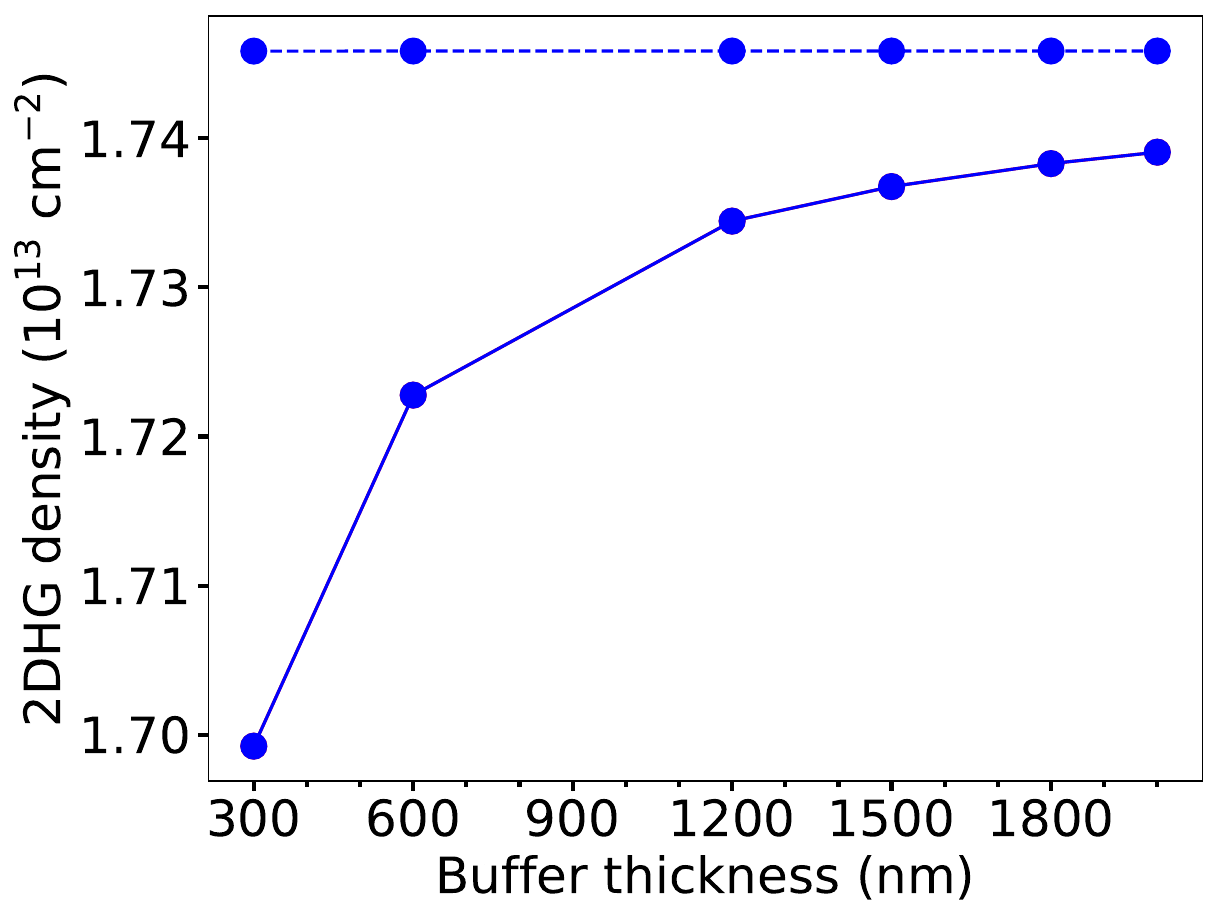}
    \caption{2DHG density as a function of AlN buffer thickness. Solid line represents the simulation with an end Schottky contact, while the dashed line corresponds to the same when using the Neumann boundary condition at the rightmost device-end AlN buffer-air interface.}
    \label{fig:figS3}
\end{figure}

The setup in Fig.~\ref{fig:figS2b}, however, as expected, significantly impacts the 2DHG density formed at the buffer-channel interface due to the extra charge supplied by the fictitious end Schottky barrier. This effect is illustrated in Fig.~\ref{fig:figS3}, where the 2DHG density decreases as the buffer thickness decreases, indicating an increasing influence of the Schottky contact on the 2DHG at the buffer-channel interface when a shorter buffer is used. However, as the buffer thickness increases, the influence of the fictitious end-device Schottky contact diminishes. For a sufficiently long buffer region ($\sim 2$ $\mu m$), the results converge to those obtained using the Neumann boundary condition. In practical HEMT structures, buffer regions on the order of $\sim \mu m$ are typically used. While a Schottky contact with a long buffer width more accurately models the end device-air interface, it also significantly increases computational cost. In contrast, using Neumann boundary conditions allows for mimicking the effects of a large (infinite) buffer but even with a shorter AlN buffer region, thereby reducing computational demands considerably. Therefore, we employ the Neumann boundary condition at the device-end AlN buffer-air interface. Notably, the 300 nm long channel ensures that the barrier-channel interface remains distant, minimizing any impact on the 2DEG density at the barrier-channel interface.

Alternatively, we could incorporate `realistic' (intentional or unintentional) doping in our calculations to provide the additional charge required for band bending.\cite{Chen2022EffectBarrier,Zervos2024AHEMTs} However, unintentional doping would also be present in other layers of the heterostructure, and the doping concentration would vary depending on Al composition and other factors. The introduction of doping would inevitably alter the carrier concentrations. To simplify the analysis and avoid unnecessary complexity, we refrain from including doping in our HEMT structures. 

%%%%%%%%%%%%%%%%%%%%%%%%%%%%%%%%%%%%%%%%%%%%%%%%%%%%%%%%%%%%%%%%%%%%%%%%%%%%%%
\section{\label{sec:secS3}Material and model parameters}
Table~\ref{tab:tableS1} presents the material parameters used for the Nextnano++ simulations, mobility, and figure-of-merit calculations in this article. The parameters for the (Al,Ga)N alloy ($P$) are derived from the corresponding binary materials, GaN and AlN, using the following formula:
\begin{equation}
    P_{\text{Al}_x \text{Ga}_{1-x} \text{N}} = x\,P_{\text{AlN}} + (1-x)\,P_{\text{GaN}} - b_{_P}\,x\,(1-x)  \label{eq:eq31}
\end{equation}
where $x$ is the Al mole fraction in Al$_x$Ga$_{1-x}$N, and $b_{_P}$ is the bowing parameter. 

The parameters for the 6-band \textbf{k}$\cdot$\textbf{p} model are listed in Table~\ref{tab:tableS2}. As mentioned in Sec.~\ref{sec:secS1}, strain-induced energy shifts are incorporated into the Nextnano++ simulations through deformation potentials. In wurtzite structure, crystal anisotropy leads to two distinct conduction band deformation potentials at the $\Gamma$ point: one parallel ($D_\parallel$) and one perpendicular ($D_\perp$) to the c-axis. For the valence bands, six deformation potentials ($D_1, D_2, D_3, D_4, D_5,$ and $D_6$) are used, reflecting a full treatment of the effect of strain effects on the 6-band Hamiltonian.\cite{Chuang1996} The deformation potential values used in this study are provided in Table~\ref{tab:tableS3}.

\begin{table}[!ht]
\caption{\label{tab:tableS1}Material parameters for AlN, GaN, and Al$_x$Ga$_{1-x}$N alloy used in this paper for 2DEG density, low-field mobility, and figure-of-merit calculations.}
\begin{ruledtabular}
\begin{tabular}{llcccc}
Parameters & Symbol (Unit) & GaN ($x=0$) & AlN ($x=1$)	& Bowing ($b_{_P}$) & Ref. \\
\hline
Schottky barrier height & $-$ (eV) & 1.11	& 3.40 & 0 & \citenum{Armstrong2016Polarization-inducedAlloys,Arehart2006EffectCharacteristics} \\
Mass density & $\rho$ (kg m$^{-3}$) & 6150 & 3230	& 0	& \citenum{NSMSemiconductors} \\
Lattice constants & $a_0$ (\r{A}) at 300 K & 3.189 & 3.112 & 0 & \citenum{Vurgaftman2003BandSemiconductors} \\
 & $c_0$ (\r{A}) at 300 K & 5.185 & 4.982 & 0 & \citenum{Vurgaftman2003BandSemiconductors} \\
Thermal expansion coefficient\footnotemark[1] & $\alpha_{a_0}$ ($10^{-5}$ \r{A} K$^{-1}$) & 1.783 & 1.291 & 0 & \cite{NSMSemiconductors} \\
 & $\alpha_{c_0}$ ($10^{-5}$ \r{A} K$^{-1}$) & 1.644 & 2.626 & 0 & \citenum{NSMSemiconductors} \\
Bandgap energy & E$_\text{g}$ (eV) at 0 K & 3.51 & 6.25	& 0.7 & \citenum{Vurgaftman2003BandSemiconductors} \\ 
Varshni's parameter & $\alpha$ (meV/K)\footnotemark[2] & 0.909 & 1.799 & 0 & \citenum{Vurgaftman2003BandSemiconductors} \\
 & $\beta$ (K) & 830 & 1462 & 0 & \citenum{Vurgaftman2003BandSemiconductors} \\
Valence band offset & $-$ (eV) & $-0.726$ & $-1.526$ & 0 & \citenum{Wei1998CalculatedOrbitals}\\
Alloy-disordered scattering potential\footnotemark[3] & $U_0$ (eV) & 1.0 & 1.8 & $-1.6$ & \citenum{Pant2020HighStructure}\\
% \end{tabular}
% \end{ruledtabular}
% \end{table}

% \begin{table}
% \begin{ruledtabular}
% \begin{tabular}{llcccc}
Perpendicular to c-axis electron effective mass & $m_\perp \,(m_0)$ & 0.20 & 0.30 & 0 & \citenum{Vurgaftman2003BandSemiconductors}\\
Parallel to c-axis electron effective mass & $m_\parallel \,(m_0)$ & 0.20 & 0.32 & 0 & \citenum{Vurgaftman2003BandSemiconductors} \\
Isotropic electron effective mass & $m^* \,(m_0)$ & 0.20 & 0.31 & 0 & \citenum{Vurgaftman2003BandSemiconductors} \\
Static dielectric constant & $\varepsilon_s \,(\varepsilon_0)$ & 8.90 & 8.50 & 0 & \citenum{Bassaler2024AlRichMobility} \\
High frequency dielectric constant & $\varepsilon_h \,(\varepsilon_0)$ & 5.35 & 4.60 & 0 & \citenum{Bassaler2024AlRichMobility} \\
Longitudinal acoustic phonon velocity & $v_{LA}$ (ms$^{-1}$) & 6560 & 9060 & 0 & \citenum{Bassaler2024AlRichMobility} \\
Transversal acoustic phonon velocity & $v_{TA}$ (ms$^{-1}$) & 2680 & 3700 & 0 & \citenum{Bassaler2024AlRichMobility} \\
Deformation potential\footnotemark[4] & $E_D$ (eV) & 8.3 & 9.5 & 0 & \citenum{Bassaler2024AlRichMobility} \\
Polar optical phonon energy\footnotemark[4] & $E_{pop}$ (meV) & 91.2 & 99.0 & 0 & \citenum{Bassaler2024AlRichMobility} \\
Electromechanical coupling coefficient\footnotemark[4] & $K^2$ & 0.045 & 0.106 & 0 & \citenum{Bassaler2024AlRichMobility} \\
Spontaneous polarization constant & $P_{sp}$ (C m$^{-2}$) & $-0.034$ & $-0.090$ & $-0.021$ & \citenum{Vurgaftman2003BandSemiconductors} \\
Elastic constants & $C_{11}$ (GPa) & 390 & 396 & 0 & \citenum{Vurgaftman2003BandSemiconductors} \\
 & $C_{12}$ (GPa) & 145 & 137 & 0 & \citenum{Vurgaftman2003BandSemiconductors} \\
 & $C_{13}$ (GPa) & 106 & 108 & 0 & \citenum{Vurgaftman2003BandSemiconductors} \\
 & $C_{33}$ (GPa) & 398 & 373 & 0 & \citenum{Vurgaftman2003BandSemiconductors} \\
 & $C_{44}$ (GPa) & 105 & 116 & 0 & \citenum{Vurgaftman2003BandSemiconductors} \\
Piezoelectric constants & $e_{31}$ (C m$^{-2}$) & $-0.35$ & $-0.50$ & 0 & \citenum{Vurgaftman2001BandAlloys} \\
 & $e_{33}$ (C m$^{-2}$) & 1.27 & 1.79 & 0 & \citenum{Vurgaftman2001BandAlloys} \\
 & $e_{15}$ (C m$^{-2}$) & $-0.30$ & $-0.48$ & 0 & \citenum{Bassaler2024AlRichMobility} 
\end{tabular}
\end{ruledtabular}
\footnotetext[1]{In Nextnano++, the thermal expansion of the lattice parameter $a$(T) at temperature T is given by:\[ a (\text{T}) = a (\text{300 K}) + \alpha_{exp} \times  (\text{T}-300)\]
where $\alpha_{exp}$ is the linear thermal expansion coefficient. The definition of $\alpha_{exp}$ used here differs slightly from the conventional definitions found in the literature. The relationship between the two can be expressed as: $\alpha_{exp} = a$(300 K) $\times \alpha_{Literature}$; where $\alpha_{Literature}$ is the conventional linear thermal expansion coefficient, expressed in K$^{-1}$.}

\footnotetext[2]{In Nextnano++, the temperature correction to bandgap (E$_{\text{g}}$) is included using Varshni's formula: \[ \text{E}_{\text{g}} (\text{T}) = \text{E}_{\text{g}} (0) + \delta \text{E}_{\text{g}} (\text{T})\]
where the temperature correction $\delta$E$_{\text{g}}$ (T) for alloy is interpolated in a slightly different manner than that of Eq.~\ref{eq:eq31}. The interpolation is expressed as: 
\[ \delta \text{E}_{\text{g, Al}_x\text{Ga}_{1-x}N} (\text{T},\,x) = x \,\frac{-\alpha_{\text{AlN}}\,\text{T}^2}{\text{T} + \beta_{\text{AlN}}} + (1-x)\,\frac{-\alpha_{\text{GaN}}\,\text{T}^2}{\text{T} + \beta_{\text{GaN}}} - x\,(1-x)\,\frac{-\alpha_{\text{(Al,Ga)N}}\,\text{T}^2}{\text{T} + \beta_{\text{(Al,Ga)N}}}\]
where $\alpha_{\text{AlN}}$, $\alpha_{\text{GaN}}$, $\beta_{\text{AlN}}$, and $\beta_{\text{GaN}}$ are the Varshni's parameters of binary compounds and  $\alpha_{\text{(Al,Ga)N}}$, $\beta_{\text{(Al,Ga)N}}$ are the related alloy bowing. Note that the temperature variation only impacts the conduction bands, since the valence bands act as reference energies for the band offsets.}

\footnotetext[3]{The alloy-disordered scattering potential is obtained from a quadratic fit of the data presented in Table 1 of the referenced paper.}

\footnotetext[4]{These parameters are used for the mobility calculation in Sec.~\ref{sec:secS6} below.}

\end{table}

\begin{table}[!ht]
\raggedleft
\caption{\label{tab:tableS2} Valence band effective mass parameters for the 6-band \textbf{k}$\cdot$\textbf{p} Hamiltonian used in this study.}
\begin{ruledtabular}
\begin{tabular}{lccccccc}
Material & $A_1$ & $A_2$ & $A_3$ & $A_4$ & $A_5$ & $A_6$ & Ref.\\
\hline
AlN	& $-3.86$ & $-0.25$ & 3.58 & $-1.32$ & $-1.47$ & $-1.64$ & \citenum{Vurgaftman2003BandSemiconductors} \\
GaN & $-7.21$ & $-0.44$ & 6.68 & $-3.46$ & $-3.40$ & $-4.90$ & \citenum{Vurgaftman2003BandSemiconductors} 
\end{tabular}
\end{ruledtabular}
\end{table}

\begin{table}[!ht]
\caption{\label{tab:tableS3} Valence and conduction band deformation potentials to account for the strain-induced energy shifts in Nextnano++ simulations.}
\begin{ruledtabular}
\begin{tabular}{lccccccccc}
Material & $D_\parallel$\footnotemark[5] &	$D_\perp$\footnotemark[5] & $D_1$ & $D_2$ & $D_3$ & $D_4$ & $D_5$ & $D_6$ & Ref. \\
\hline
AlN & $-20.5$ & $-3.9$ & $-17.1$ & 7.9 & 8.8 & $-3.9$ & $-3.4$ & $-3.4$ & \citenum{Vurgaftman2003BandSemiconductors} \\
GaN	& $-8.6$ & $-6.8$ & $-3.7$ & 4.5 & 8.2 & $-4.1$ & $-4.0$ & $-5.5$ & \citenum{Vurgaftman2003BandSemiconductors}
\end{tabular}
\end{ruledtabular}
\footnotetext[5]{Note that the $a_1$ and $a_2$ deformation potential values listed in Ref.~\citenum{Vurgaftman2003BandSemiconductors}, refer to the interband deformation potentials, which describe the energetic shift of bandgaps with strain. Here, we have added the valence band deformation potential to those values to obtain the deformation potential for conduction bands, i.e., $D_\parallel = a_1 + D_1$ and $D_\perp = a_2+D_2$ .}
\end{table}

%%%%%%%%%%%%%%%%%%%%%%%%%%%%%%%%%%%%%%%%%%%%%%%%%%%%%%%%%%%%%%%%%%%%%%%%%%%%%%
\section{\label{sec:secS4}2DEG and 2DHG density maps for varying barrier thickness}
\begin{figure}[!htbp]
    \centering
    \subfloat[L$_{\text{B}}$ = 5 nm	]{\label{fig:figS4a}\includegraphics[width=0.42\textwidth]{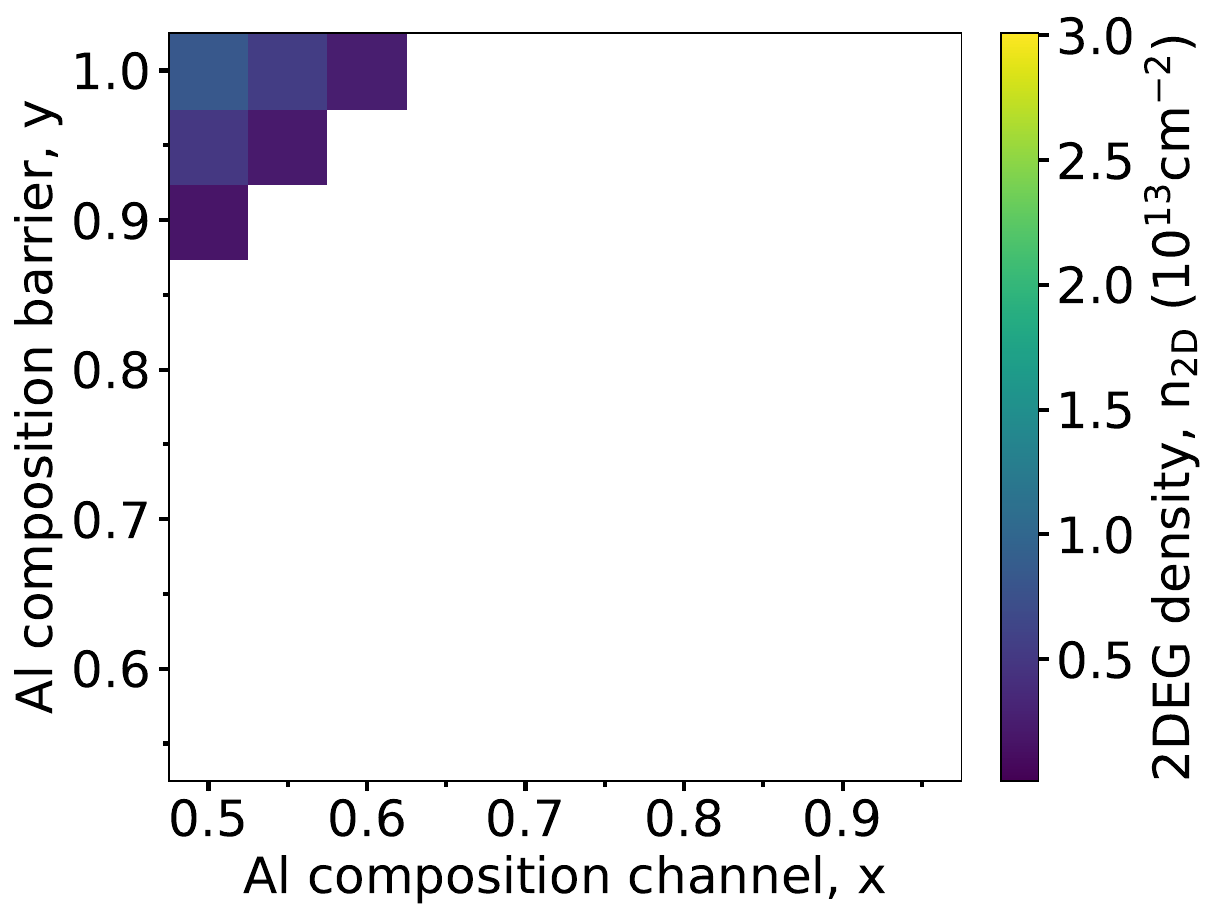}}
    \hspace{1.5cm}
    \subfloat[L$_{\text{B}}$ = 15 nm]{\label{fig:figS4b}\includegraphics[width=0.42\textwidth]{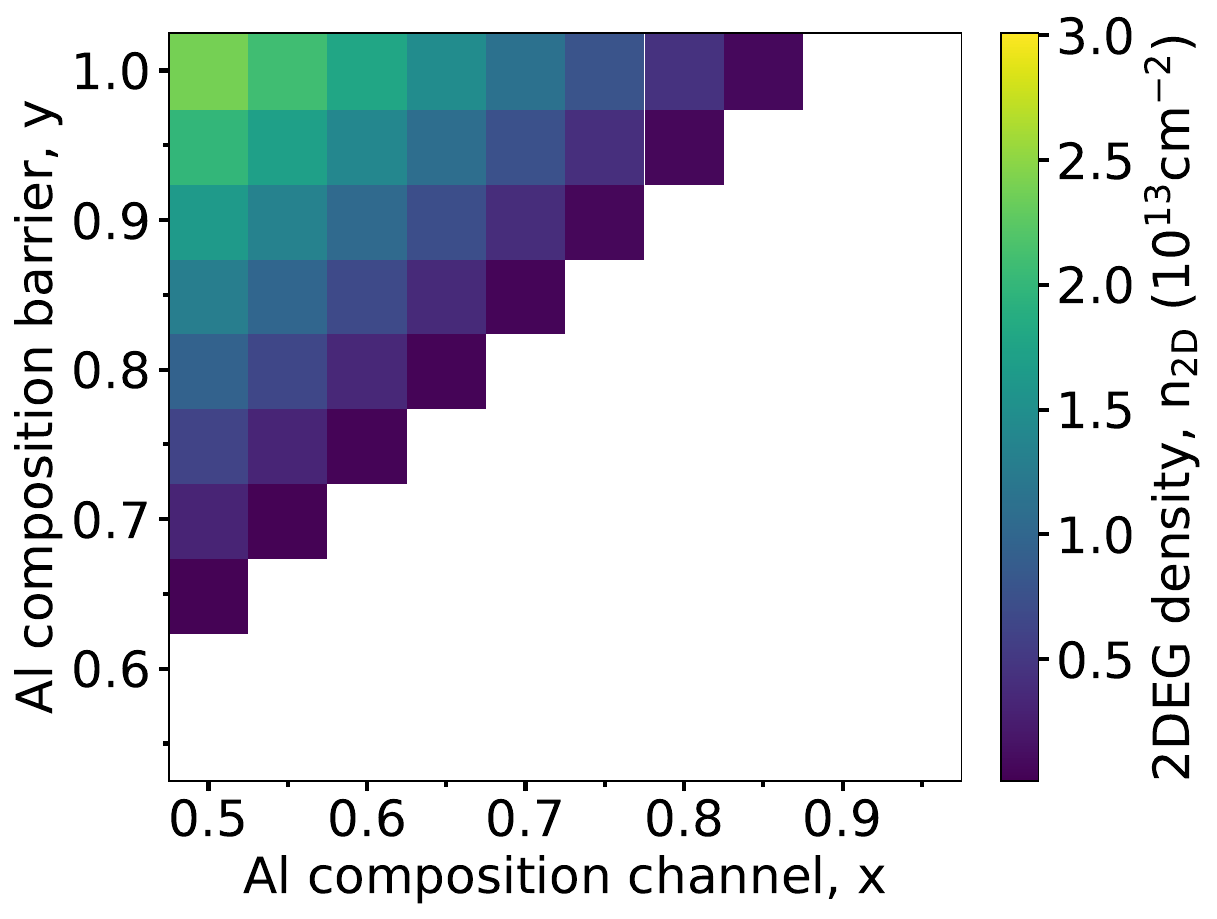}}\\
    \subfloat[L$_{\text{B}}$ = 30 nm]{\label{fig:figS4c}\includegraphics[width=0.42\textwidth]{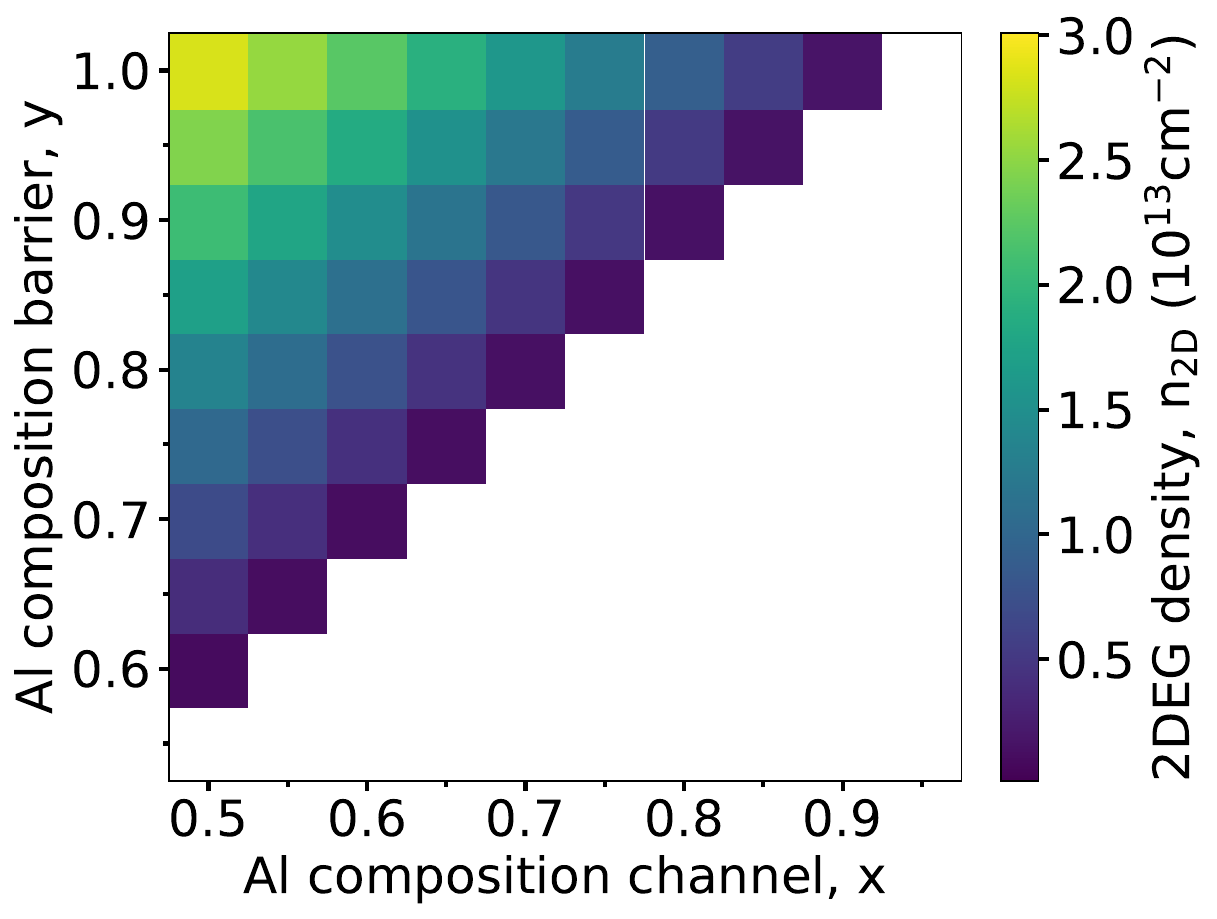}}
    \hspace{1.5cm}
    \subfloat[L$_{\text{B}}$ = 50 nm]{\label{fig:figS4d}\includegraphics[width=0.42\textwidth]{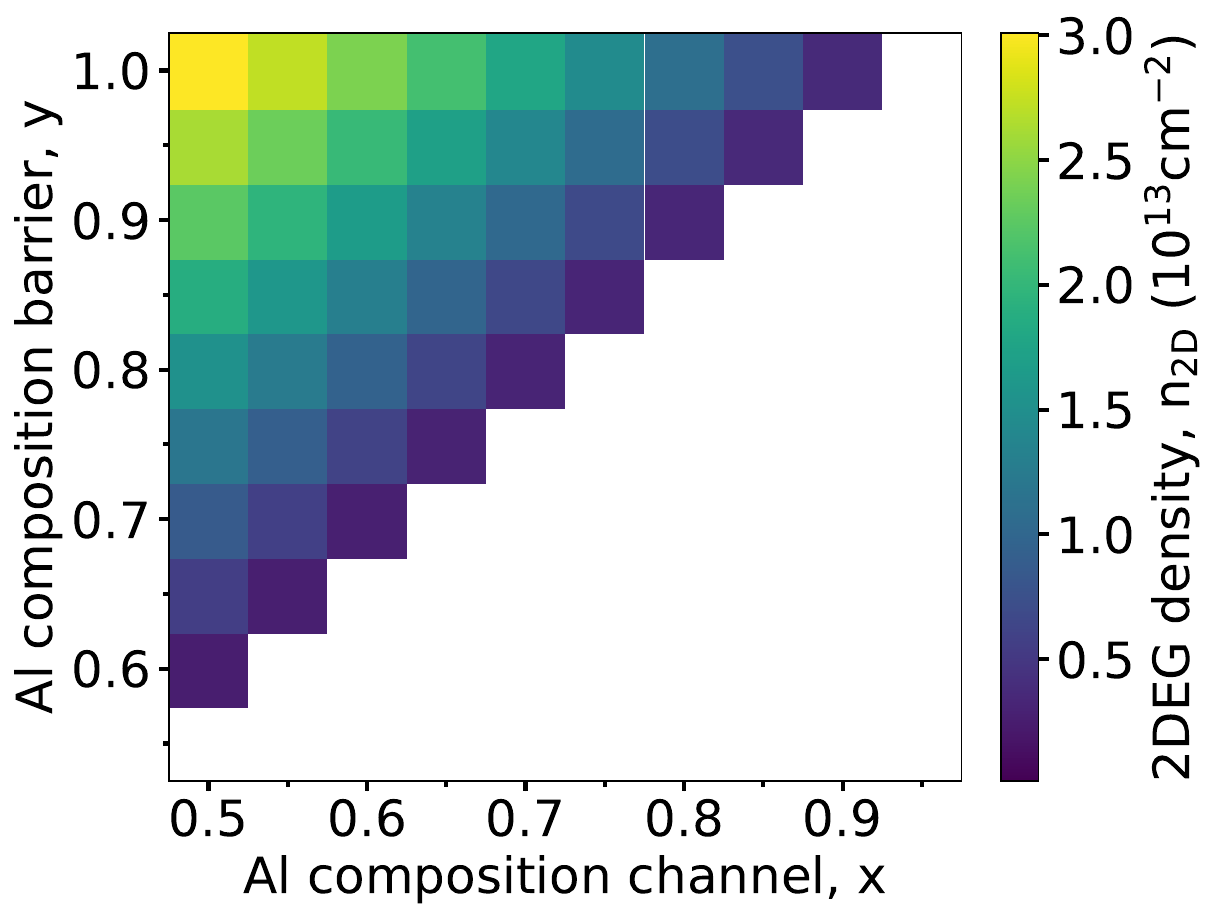}}
    \caption{2DEG density, $n_{\text{2D}}$, as a function of Al composition in barrier, $y$, and channel, $x$, for various barrier thicknesses, L$_{\text{B}}$. For clarity, results are shown for (a) L$_{\text{B}}$ = 5 nm, (b) L$_{\text{B}}$ = 15 nm, (c) L$_{\text{B}}$ = 30 nm, and (d) L$_{\text{B}}$ = 50 nm; the trends are consistent across all other L$_{\text{B}}$ values. The complete dataset is provided in the SI attachment. In all cases a channel length of L$_{\text{C}}$ = 300 nm is used. }
    \label{fig:figS4}
\end{figure}

Figure~\ref{fig:figS4} reveals that the 2DEG density, $n_{\text{2D}}$, increases with increasing barrier thickness. The highest $n_{\text{2D}}$ is achieved when the Al composition contrast between the barrier and channel is maximized. However, as discussed in the main text, $n_{\text{2D}}$ is not solely a function of the Al composition contrast. The formation of the 2DEG is linked to the polarization field discontinuity at the interface between barrier and channel, which can vary slightly depending on the absolute Al content in the individual layers. This is illustrated in detail in Fig.~2(b) of the main text. 

Furthermore, generating a 2DEG requires not only sufficient composition contrast but also a minimum barrier thickness.\cite{Coltrin2017AnalysisAlloys} This is evident from Fig.~\ref{fig:figS4}, where one finds a significant composition contrast region lacking a 2DEG for thin barriers, e.g., L$_{\text{B}}$ = 5 nm, which diminishes as the barrier thickness increases.

\begin{figure}[!htbp]
    \centering
    \subfloat[L$_{\text{B}}$ = 5 nm	]{\label{fig:figS5a}\includegraphics[width=0.42\textwidth]{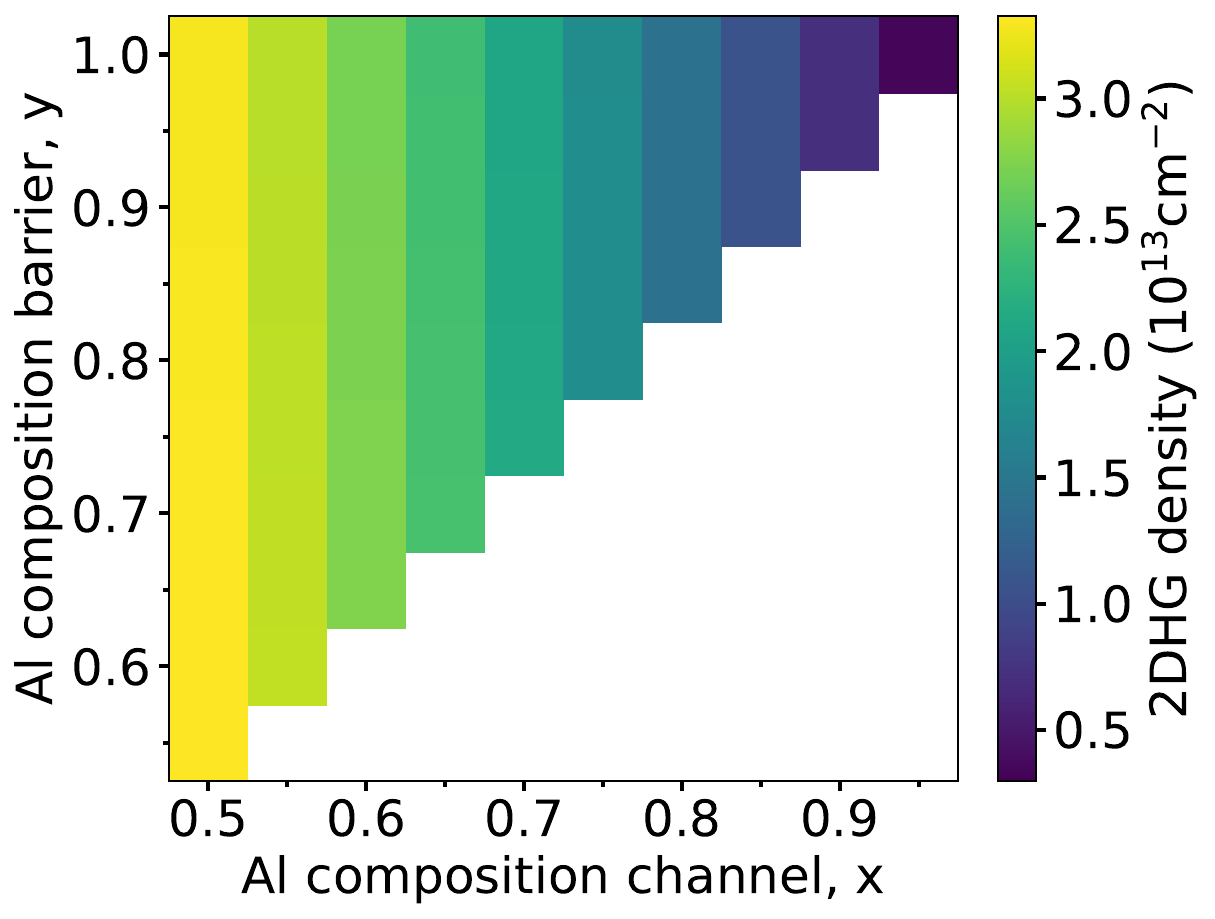}}
    \hspace{1.5cm}
    \subfloat[L$_{\text{B}}$ = 50 nm]{\label{fig:figS5b}\includegraphics[width=0.42\textwidth]{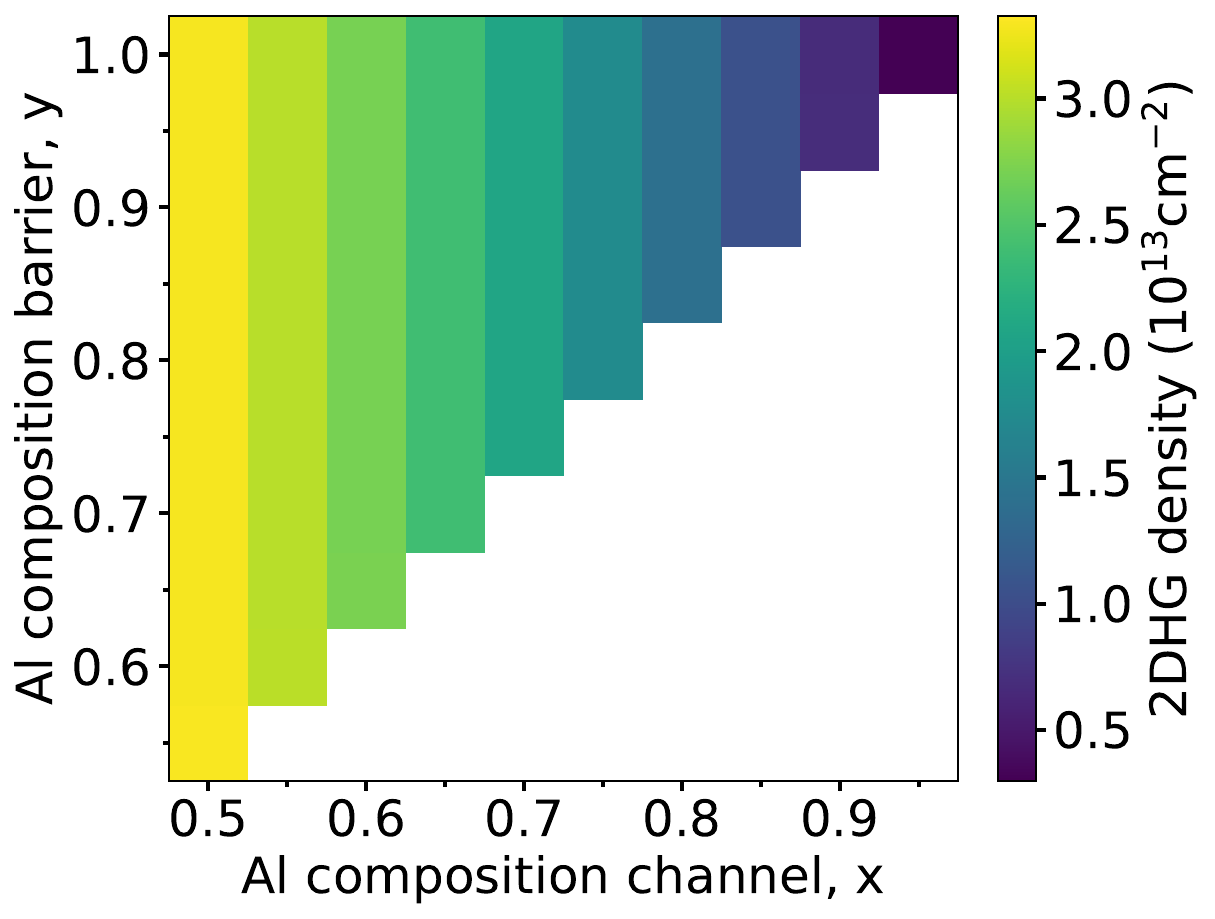}}
    \caption{2DHG density as a function of Al composition in barrier, $y$, and channel, $x$, for various barrier thicknesses, L$_{\text{B}}$. For clarity, results are shown for (a) L$_{\text{B}}$ = 5 nm and (b) L$_{\text{B}}$ = 50 nm; the trends are consistent across all other L$_{\text{B}}$ values. Complete dataset is provided in the SI attachment. In all cases the channel length L$_{\text{C}}$ = 300 nm is used.}
    \label{fig:figS5}
\end{figure} 

In contrast to the 2DEG density, maps of the 2DHG density shown in Fig.~\ref{fig:figS5} indicate that this quantity remains largely independent of the barrier thickness. This independence is because the 2DHG is generated at the channel-buffer interface, which is separated from the barrier region by a 300 nm thick channel, minimizing thus the impact of barrier thickness variations. Additionally, the 2DHG density remains unaffected by the Al compositions in the barrier and decreases with increasing Al content in the channel. This occurs because the buffer is always pure AlN, and as the Al composition in channel increases, the composition contrast between the AlN buffer and (Al,Ga)N channel diminishes, which leads to a reduction in 2DHG density. This is evident from Fig.~\ref{fig:figS5}, where the same colours are seen to be parallel to the y-axis.  

%%%%%%%%%%%%%%%%%%%%%%%%%%%%%%%%%%%%%%%%%%%%%%%%%%%%%%%%%%%%%%%%%%%%%%%%%%%%%%
\section{\label{sec:secS5}Variation of 2DEG and 2DHG densities with layer thicknesses} 
\begin{figure}[!htbp]
    \centering
    \subfloat[]{\label{fig:figS6a}\includegraphics[width=0.43\textwidth]{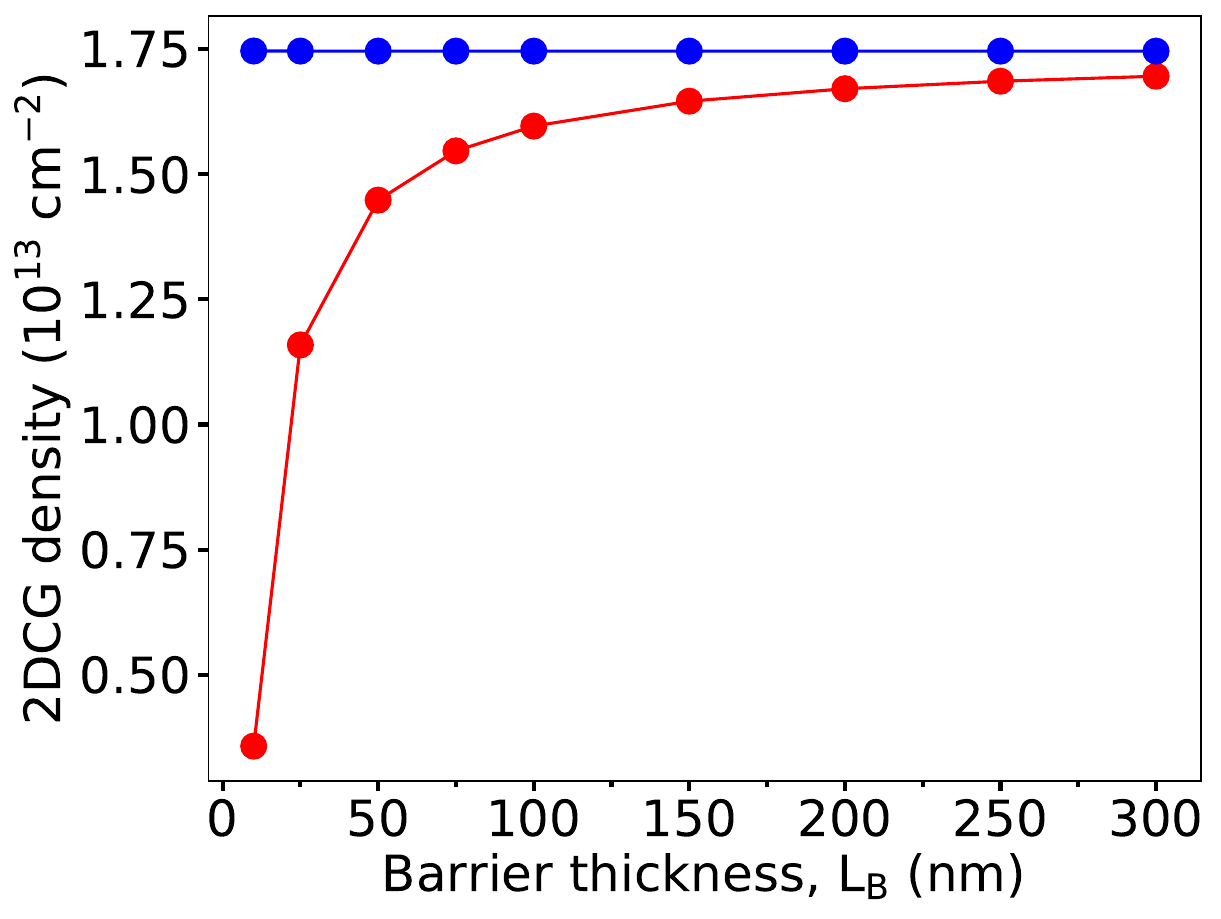}}
    \hspace{.7cm}
    \subfloat[]{\label{fig:figS6b}\includegraphics[width=0.43\textwidth]{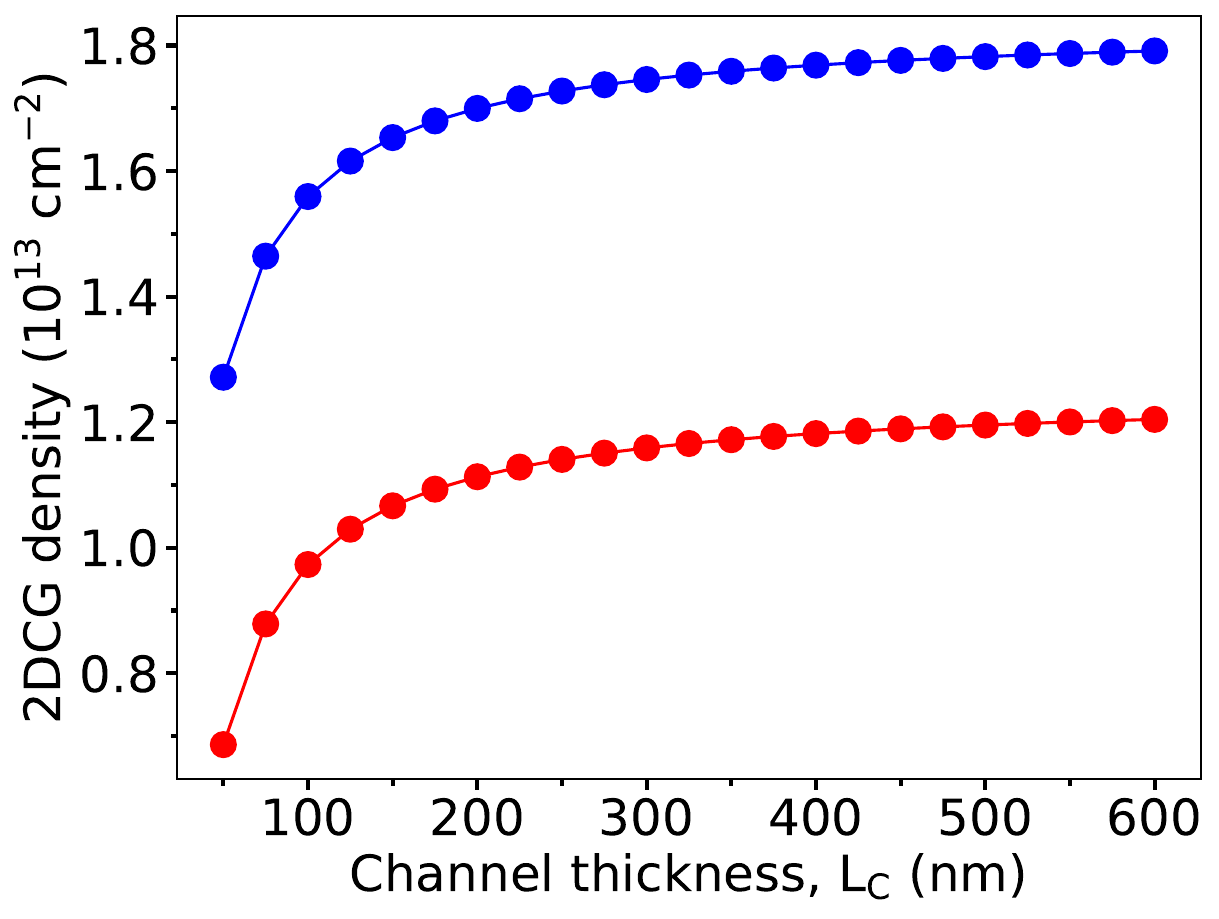}}
    \caption{Variation of  2D carrier gas (2DCG) densities with (a) barrier thickness, L$_{\text{B}}$, for an AlN(L$_{\text{B}}$)/Al$_{0.75}$Ga$_{0.25}$N(300nm) HEMT and (b) channel thickness, L$_{\text{C}}$, for an AlN(25nm)/Al$_{0.75}$Ga$_{0.25}$N(L$_{\text{C}}$) HEMT. The 2D electron gas (2DEG) is shown in red, and the 2D hole gas (2DHG) is shown in blue.}
    \label{fig:figS6}
\end{figure}
While the barrier thickness, L$_{\text{B}}$, influences the 2DEG densities, a thick 300 nm channel places the channel-buffer interface sufficiently far away to have minimal impact on the 2DHG densities [Fig.~\ref{fig:figS6a}]. Reducing the channel width, L$_{\text{C}}$, leads to interactions between the 2DEG and 2DHG, which results in a decrease in both densities [Fig.~\ref{fig:figS6b}].

\begin{figure}[!htbp]
    \centering
    \subfloat[]{\label{fig:figS7a}\includegraphics[width=0.43\textwidth]{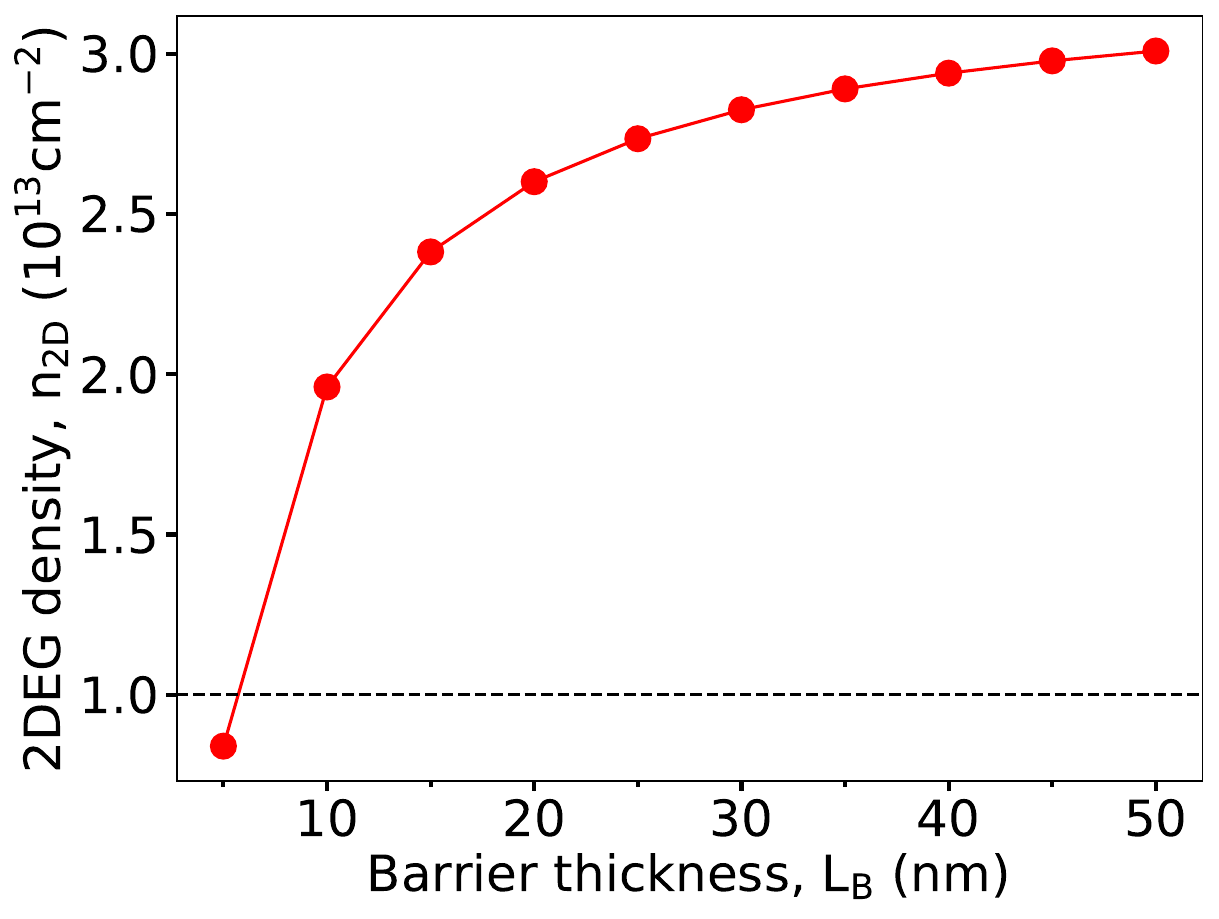}}
    \hspace{.7cm}
    \subfloat[]{\label{fig:figS7b}\includegraphics[width=0.43\textwidth]{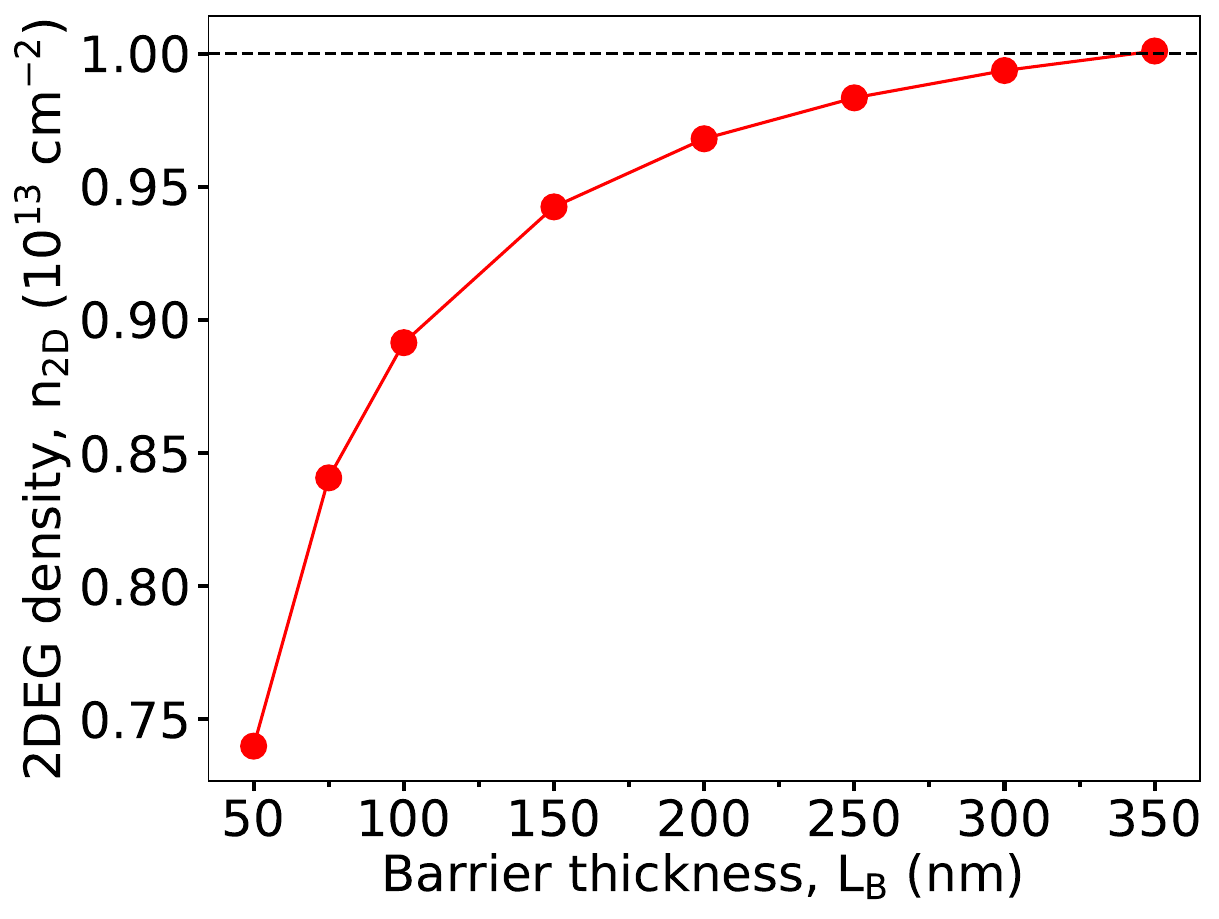}}
    \caption{2DEG density, $n_{\text{2D}}$, as a function of barrier thickness, L$_{\text{B}}$, for (a) AlN(L$_{\text{B}}$)/Al$_{0.50}$Ga$_{0.50}$N(300nm), and (b) AlN(L$_{\text{B}}$)/Al$_{0.85}$Ga$_{0.15}$N(300nm) HEMT structures. The horizontal black dashed line represents $n_{\text{2D}}$ = $1 \times 10^{13}$ cm$^{-2}$.}
    \label{fig:figS7}
\end{figure}

Figure~\ref{fig:figS7} presents the variation of 2DEG density, $n_{\text{2D}}$, as a function of barrier thickness, L$_{\text{B}}$, for two different HEMT structures:  Al$_y$Ga$_{1-y}$N/Al$_x$Ga$_{1-x}$N HEMT, with $(y, x) = (1.00, 0.50)$ and $(1.00, 0.85)$. The constant $n_{\text{2D}}$ = $1 \times 10^{13}$ cm$^{-2}$ are indicated by the black horizontal dashed lines. The figures clearly show that while a $n_{\text{2D}} = 1 \times 10^{13}$ cm$^{-2}$ can be achieved with a relatively thin 10 nm AlN barrier in AlN/Al$_{0.50}$Ga$_{0.50}$N structures, achieving the same $n_{\text{2D}}$ in an AlN/Al$_{0.85}$Ga$_{0.15}$N structure requires an experimentally impractically thick barrier of over 300 nm. 

%%%%%%%%%%%%%%%%%%%%%%%%%%%%%%%%%%%%%%%%%%%%%%%%%%%%%%%%%%%%%%%%%%%%%%%%%%%%%%
\section{\label{sec:secS6}Low-field 2DEG mobility models}
We calculate the low-field electron mobility following Matthiessen's rule, which relates the mobility to various scattering mechanisms as follows:\cite{Bassaler2024AlRichMobility,Zhang2008TheContent}
\begin{equation}
    \mu ^{-1} = \sum_i \mu_i^{-1}
\end{equation}

Within the momentum relaxation approximation, each mobility component $\mu_i$ can be described by: 
\begin{equation}
    \mu_i = \frac{e\,\tau_i}{m^*}
\end{equation}
where $e$ is the elementary electron charge, $\tau_i$ is the momentum relaxation time associated with the specific scattering mechanism and $m^*$ is the isotropic effective mass. Since the electron effective masses in in-plane and out-of-plane directions are very similar for both GaN and AlN (Table~\ref{tab:tableS1}), using an isotropic effective mass is justified.

We consider two primary scattering origins:\cite{Bassaler2024AlRichMobility,Zhang2008TheContent} (a) temperature-independent carrier scattering resulting from material and crystal quality properties, and (b) temperature-dependent electron-phonon interaction-mediated scattering. For the temperature-independent scattering mechanisms, we consider alloy disorder (AD), interface roughness (IRF), and dislocation (DIS) scattering, which are related to microstructure of the materials. The electron-phonon interactions are described by the acoustic phonon (AP) effects, the deformation-potential (DP), piezoelectric (PE) contributions, and polar optical phonon (POP) effects. In this study, we have neglected the carrier-carrier scattering effect. The alloy materials parameters used in our mobility calculations are detailed in Table~\ref{tab:tableS1}.  

In mobility models for (Al,Ga)N/(Al,Ga)N heterostructures, scattering contributions should ideally be considered from both the barrier and the channel.\cite{Miyoshi2015NumericalHeterostructures} However, we find that the 2DEG density distributions are primarily confined within the channel region (Sec.~\ref{sec:secS8}). Consequently, in our mobility models, we disregard the small spill-over of 2DEG into the barrier and apply the materials parameters --- such as compositions, effective mass, and scattering potentials --- corresponding to the channel (Al,Ga)N alloy. Due to the minimal penetration of the 2DEG into the large bandgap barrier, the error introduced by this assumption is expected to be of secondary importance. 

The momentum relation rates $1/\tau_i$ for various scattering models are summarized below. The equations are taken from Refs.~\citenum{Bassaler2024AlRichMobility} and \citenum{Zhang2008TheContent}. For detailed references to individual equations, we refer to the extensive reference list provided in those two publications. We have implemented the mobility equations in an open-source Python package, ``mobilitypy,'' which is available on GitHub.\cite{Mondal2024Mobilitypy}

A short summary of the parameters used in the momentum relaxation rate equations in our calculations for different scattering mechanisms is presented in Table~\ref{tab:tableS4}. Required unit adjustments have been made in the implementation of the mobility equations within the ``mobilitypy'' package.

\begin{table}[!ht]
\caption{\label{tab:tableS4} Summary of parameters used in the momentum relaxation rate equations below.}
\begin{ruledtabular}
\begin{tabular}{lccc}
Parameters name	& Symbol & Value & Unit\\
\hline
Electron elementary charge & $e$ & $1.602 \times 10^{-19}$ & C \\
Electron rest mass & $m_0$ & $9.11 \times 10^{-31}$ & kg \\
Vacuum permittivity & $\varepsilon_0$ & $8.85\times 10^{-12}$ & C V$^{-1}$ m$^{-1}$ \\
Boltzmann constant & $k_B$ & $1.38 \times 10^{-23}$ & JK$^{-1}$ \\
Temperature	& T	 & $-$ & K \\
Unit cell volume & $\Omega_0$ & $-$ & \r{A}$^3$ \\
Root mean square (RMS) interface roughness & $\Delta$ & 0.3 & nm \\
Interface roughness correlation length & $L$ & 3.0 & nm \\
(Threading) Dislocation density	& $N_{DIS}$ & $10^{10}$ & cm$^{-2}$\\
Occupancy of dislocation-introduced defect states inside bandgap & $f_{DIS}$ & 0.3 & unitless \\
Scattering wave vector & $k$ & Eq.~\ref{eq:eqS611} & m$^{-1}$\\
Fermi wave-vector & $k_{_F}$ & Eq.~\ref{eq:eqS612} & m$^{-1}$\\
Thomas-Fermi wave vector & $q_{_{TF}}$ & Eq.~\ref{eq:eqS613} & m$^{-1}$\\
Fang-Howard variational wave function parameter & $b$ & Eq.~\ref{eq:eqS614} & m$^{-1}$\\
Fang-Howard wave function form factor & $G,\,F$ & Eqs.~\ref{eq:eqS622}, ~\ref{eq:eqS633} & unitless\\
Polar optical phonon wave vector & $k_{pop}$ & Eq.~\ref{eq:eqS637} & m$^{-1}$\\
\end{tabular}
\end{ruledtabular}
\end{table}

In Table~\ref{tab:tableS4} the values for interface root mean square roughness ($\Delta$), roughness correlation length ($L$), dislocation density ($N_{DIS}$), and occupancy of dislocation-introduced defect states inside the bandgap ($f_{DIS}$) are taken from Refs. \citenum{Bassaler2024AlRichMobility} and \citenum{Zhang2008TheContent}, which produced properties, such as mobility, in good agreement with the experimental data. However, we emphasize that these quantities can be strongly influenced by factors such as growth conditions, strain relaxation, Al content, and layer thickness in a HEMT heterostructure.\cite{Singhal2022TowardHeterostructures,Chaudhuri2022InSituTransistors} Moreover, the RMS roughness at the heterostructure interfaces is not directly measurable; typically, only the surface roughness of the final heterostructure is accessible. Since stress transfer to multiple layers tends to reduce interface roughness relative to surface roughness, we expected the interface to be smoother than the measured surface.\cite{Zhang2008TheContent, Bassaler2024AlRichMobility} Nevertheless, following the previous works,\cite{Zhang2008TheContent, Bassaler2024AlRichMobility} for simplicity and consistency in our comparative analysis, we chose the same value of the mentioned parameters for all investigated structures. 

\subsection{\label{sec:secS61}General definitions and Fang Howard approximation}
The following general definitions related to the different scattering mechanisms are used in the momentum relaxation rate equations. The scattering wave vector, $k$, is defined as:
\begin{equation}
    k = 2\,k_{_F} \,\sin(\theta/2) \equiv 2\, k_{_F}\, u \label{eq:eqS611}
\end{equation}
where the angle $\theta$ is the wave vector deviation between the initial and final scattering states and the Fermi wave-vector, $k_{_F}$, is defined in terms of 2DEG density, $n_{\text{2D}}$, as:
\begin{equation}
    k_{_F} = (2\,\pi\,n_{\text{2D}})^{1/2} \label{eq:eqS612}
\end{equation}
The screening length of the 2DEG is characterized by the Thomas-Fermi wave vector, $q_{_{TF}}$, expressed as:
\begin{equation}
    q_{_{TF}} = \frac{e^2\,m^*}{2\,\pi\,\varepsilon_s\,\hbar^2} \label{eq:eqS613}
\end{equation}
The variational wave function parameter, $b$, in the Fang-Howard approximation that quantifies the special repartition of the 2DEG, is given by:
\begin{equation}
    b = \left(\frac{33\,e^2\,m^*\,n_{\text{2D}}}{8\,\varepsilon_s\,\hbar^2}\right)^{1/3} \label{eq:eqS614}
\end{equation}

The exact sub-band Hartree-Fock wave functions are approximated using the Fang-Howard variational wave function.\cite{Fang1966NegativeSurfaces} This simplification allows us to replace the explicit wave-functions dependence in the mobility models with simple 2DEG density, $n_{\text{2D}}$.

\subsection{\label{sec:secS62}Structural related scattering}
\subsubsection{\label{sec:secS621}Interface-roughness mediated scattering}
The interface roughness momentum relaxation rate is expressed as:
\begin{equation}
    \frac{1}{\tau_{IFR}} = \frac{e^4\,m^*\,(\Delta\,L\,n_{\text{2D}})^2}{8\,\varepsilon_s^2\,\hbar^3} \times \int_0^1 \frac{u^4\,\text{e}^{-(L\,k_{_F}\,u)^2}}{\left[u+\frac{q_{_{TF}}\,G(u)}{2\,k_{_F}}\right]^2 \sqrt{1-u^2}}\,du \label{eq:eqS621}
\end{equation}
where $G(u)$ is the Fang-Howard wave function form factor, defined as:
\begin{equation}
    G(u) = \frac{2\,\eta(u)^3 + 3\,\eta(u)^2 + 3\,\eta(u)}{8} \label{eq:eqS622}
\end{equation}
with
\begin{equation}
    \eta(u) = \frac{b}{b+2\,k_{_F}\,u} \label{eq:eqS623}
\end{equation}

\subsubsection{\label{sec:secS622}Dislocation mediated scattering}
In lateral transport, threading dislocation along the growth direction can significantly affect carrier mobility, thereby affecting the carrier transport. In this case, the dislocation momentum relaxation rate is expressed as:
\begin{equation}
        \frac{1}{\tau_{DIS}} = \frac{e^4\,m^*\,N_{DIS}\,f^2_{DIS}}{4\,\pi\,k_{_F}^4\,c_0^2\,\varepsilon_s^2\,\hbar^3} \times \int_0^1 \frac{1}{\left(u+\frac{q_{_{TF}}}{2\,k_{_F}}\right)^2 \sqrt{1-u^2}}\,du \label{eq:eqS624}
\end{equation}

\subsubsection{\label{sec:secS623}Alloy-disordered mediated scattering}
The alloy disorder momentum relaxation rate is expressed as:
\begin{eqnarray}
    \frac{1}{\tau_{AD}} &=& \frac{m^*\,\Omega_0\,U_0^2\,x\,(1-x)}{\hbar^3} \times \int_{-\infty}^{\infty} \Psi(z)^4\,dz \label{eq:eqS625}\\
    &\approx& \frac{m^*\,\Omega_0\,U_0^2\,x\,(1-x)}{\hbar^3} \frac{3\,b}{16} \label{eq:eqS626}
\end{eqnarray}
where $\Omega_0 = \sqrt{3}/2\times  a_0^2 \,c_0$ is the unit cell volume, and $\Psi(z)$ corresponds to the real part of the 2DEG wave function along the growth direction. Following the previous works,\cite{Zhang2008TheContent, Bassaler2024AlRichMobility} we use the simplified 2DEG density dependence through the Fang-Howard approximation\cite{Fang1966NegativeSurfaces} instead of explicitly considering the wave-function dependence. The effects of lattice thermal expansion and (substrate) strain are also not included when calculating $\Omega_0$. 

\subsection{\label{sec:secS63}Phonon related scattering}
\subsubsection{\label{sec:secS631}Deformation-potential momentum relaxation}
The deformation-potential momentum relaxation rate is given by:
\begin{equation}
        \frac{1}{\tau_{DP}} = \frac{6\,m^*\,E_D^2\,k_B\,\text{T}\,k_{_F}^2\,b}{2\,\pi\,\rho\,v_{LA}^2\,\hbar^3} \times \int_0^1 \frac{u^4}{\left[2\,k_{_F}\,u + q_{_{TF}}\, F(u)\right]^2\, \sqrt{1-u^2}}\,du \label{eq:eqS631}
\end{equation}

\subsubsection{\label{sec:secS632}Piezoelectric momentum relaxation }
The piezoelectric momentum relaxation rate is expressed as:
\begin{equation}
        \frac{1}{\tau_{PE}} = \frac{4\,e^2\,m^*\,K^2\,k_B\,\text{T}\,k_{_F}}{\pi\,\varepsilon_s\,\hbar^3} \times \int_0^1 \frac{u^3\,F(u)}{\left[2\,k_{_F}\,u + q_{_{TF}}\, F(u)\right]^2\, \sqrt{1-u^2}}\,du \label{eq:eqS632}
\end{equation}
where $F(u)$ is the Fang-Howard wave function form factor, defined as:
\begin{equation}
    F(u) = \eta(u)^3 \label{eq:eqS633}
\end{equation}
with
\begin{equation}
    \eta(u) = \frac{b}{b+2\,k_{_F}\,u} \label{eq:eqS634}
\end{equation}

\subsubsection{\label{sec:secS633}Acoustic phonon mediated scattering}
The acoustic phonon scattering is calculated from the deformation-potential and piezoelectric effects as:
\begin{equation}
    \frac{1}{\tau_{AP}} = \frac{1}{\tau_{DP}} + \frac{1}{\tau_{PE}} \label{eq:eqS635}
\end{equation}

\subsubsection{\label{sec:secS634}Polar optical phonon mediated scattering}
The polar optical phonon momentum relaxation rate is expressed as:
\begin{equation}
        \frac{1}{\tau_{pop}} = \frac{e^2\,m^*\,E_{pop}\,G(k_{pop})}{2\,\varepsilon^*\,k_{pop}\,\hbar^3} \times \frac{d\,(1+d-\text{e}^{-d})^{-1}}{\text{e}^{\frac{E_{pop}}{k_B\text{T}}}-1} \label{eq:eqS636}
\end{equation}
where $k_{pop}$ is the polar optical phonon wave vector:
\begin{equation}
    k_{pop} = \sqrt{\frac{2\,m^*\,E_{pop}}{\hbar^2}} \label{eq:eqS637}
\end{equation}
and $G(k_{pop})$ is the Fang-Howard form factor applied on the vector $k_{pop}$:
\begin{equation}
    G(k_{pop}) = \frac{2\,\eta(k_{pop})^3 + 3\,\eta(k_{pop})^2 + 3\,\eta(k_{pop})}{8} \label{eq:eqS638}
\end{equation}
with
\begin{equation}
    \eta(k_{pop}) = \frac{b}{b+k_{pop}} \label{eq:eqS639}
\end{equation}
The effective dielectric constant is defined as:
\begin{equation}
    \frac{1}{\varepsilon^*} = \frac{1}{\varepsilon_h} - \frac{1}{\varepsilon_s} \label{eq:eqS640}
\end{equation}
Finally, the dimensionless parameter $d$ is defined as:
\begin{equation}
    d = \frac{\pi\,\hbar^2\,n_{\text{2D}}}{m^*\,k_B\,\text{T}} \label{eq:eqS641}
\end{equation}

It is important to note that due to the large polar optical phonon energy in AlN and GaN ($> 90$ meV), polar optical phonons are expected to have a more significant impact on carrier mobility than acoustic phonons at elevated temperatures. Moreover, in polar systems like wurtzite (Al,Ga)N, non-polar optical phonon effects on mobility are expected to be negligible and are not considered in this study.

\section{\label{sec:secS7}Breakdown of mobility contributions}
\begin{figure}[!htbp]
    \centering
    \includegraphics[width=0.45\textwidth]{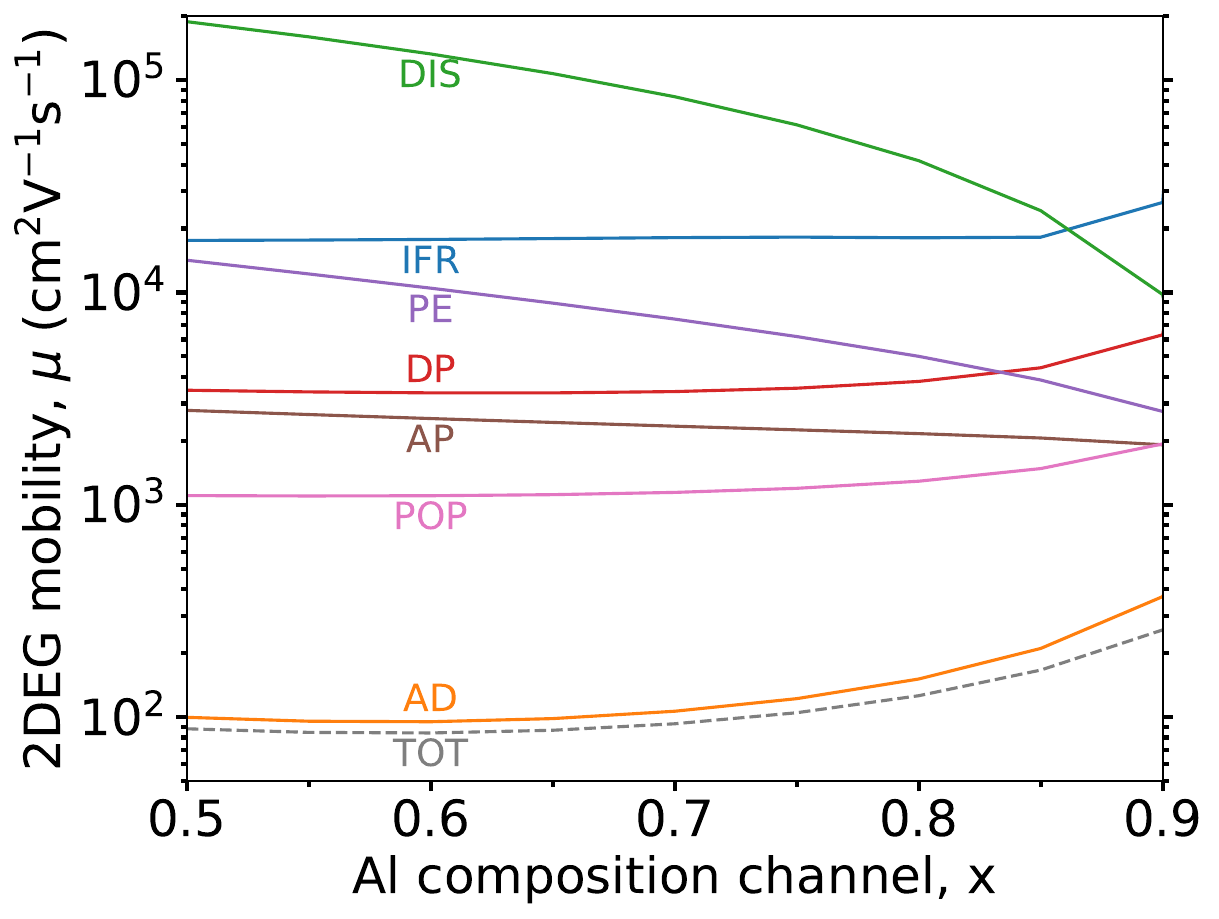}
    \caption{Decomposition of total mobility (TOT) into individual scattering mechanisms as a function of channel Al composition, $x$, for an AlN(50nm)/Al$_x$Ga$_{1-x}$N(300nm) HEMT structure at 300 K. The contributions from various scattering mechanism are shown: alloy disorder (AD), interface roughness (IRF), dislocation (DIS) scattering, acoustic phonon (AP) effects, deformation-potential (DP), piezoelectric (PE) effects, and polar optical phonon (POP) effects.}
    \label{fig:figS8}
\end{figure}

Figure~\ref{fig:figS8} shows the decomposition of total mobility, TOT, into individual scattering mechanisms for AlN(50nm)/Al$_x$Ga$_{1-x}$N(300nm) HEMT structures, with $x = 0.5-0.9$. The figure clearly shows that alloy disorder scattering is the dominant contributor limiting the total mobility for AlN/(Al,Ga)N HEMT structures. 
%The detailed maps for 2DEG density, mobility, the decomposition of mobility contributions, and LFOMs for various HEMT structures as a function of Al composition in barrier and channel - with barrier thickness varied from 5 to 50 nm at 300 K - are provided in the SI attachment as a zip file (Detailed\_FiguresS8.zip). The complete raw data set is available in the SI attachment.

\section{\label{sec:secS8}2DEG confinement}

\begin{figure}[!htbp]
    \centering
    \subfloat[]{\label{fig:figS9a}\includegraphics[width=0.45\textwidth]{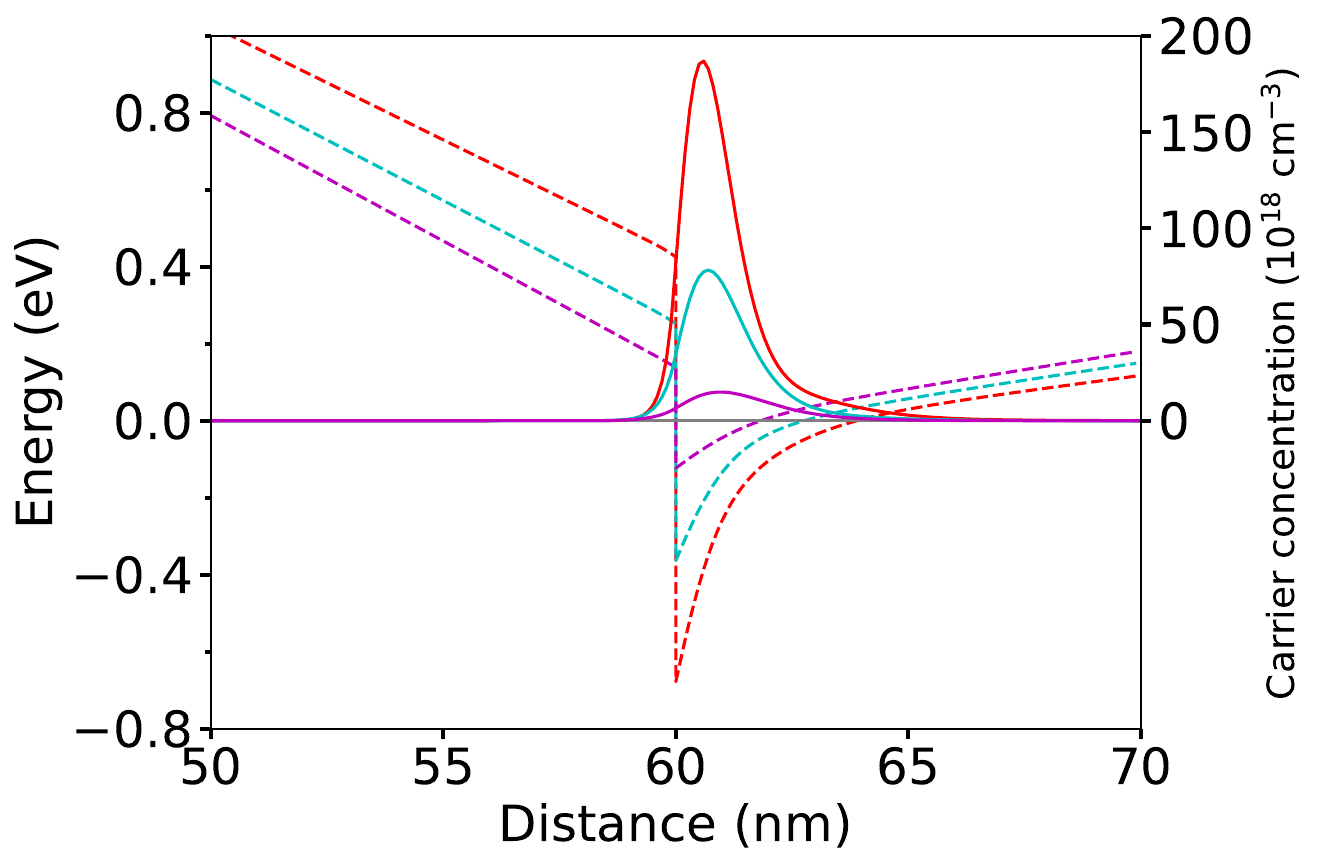}}
    \hspace{0.5cm}
    \subfloat[]{\label{fig:figS9b}\includegraphics[width=0.45\textwidth]{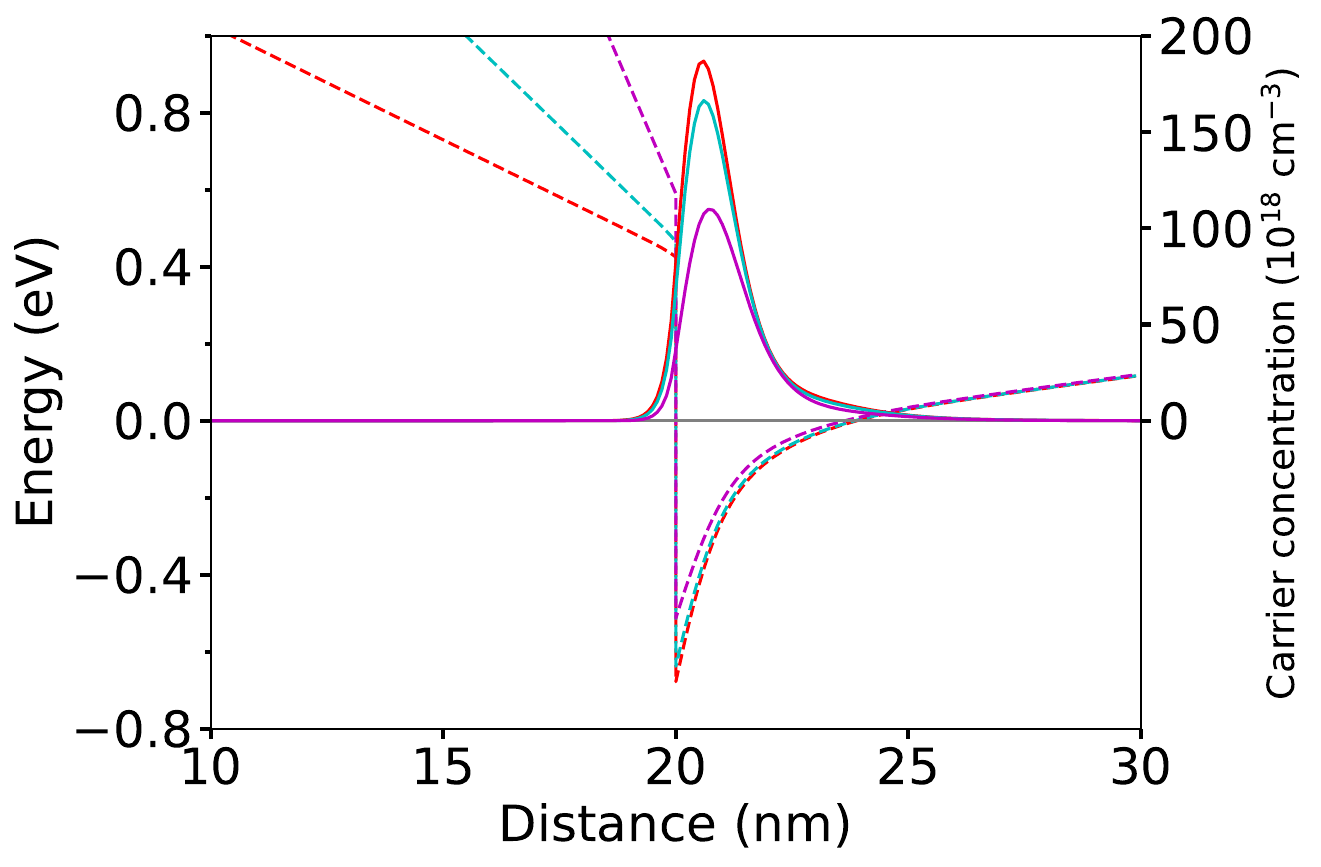}}
    \caption{2DEG density distribution near the barrier-channel interface for (a) varying Al content in the Al$_x$Ga$_{1-x}$N channel of an AlN(50nm)/Al$_x$Ga$_{1-x}$N(300nm) HEMT structure and (b) varying barrier thickness, L$_\text{B}$, in an AlN(L$_\text{B}$)/Al$_{0.5}$Ga$_{0.5}$N(300nm) HEMT structure. Band diagrams in the vicinity of the barrier-channel interface [at Distance = 60 nm for (a), and 20 nm for (b)] show the Fermi level (gray solid line), conduction band edges (dashed lines), and 2DEG distributions (solid lines) for $x = 0.5$ (red), 0.75 (cyan), and 0.9 (purple) in (a), and L$_\text{B}$ = 50 (red), 25 (cyan), and 10 nm (purple) in (b).}
    \label{fig:figS9}
\end{figure} 

Figure~\ref{fig:figS9} demonstrates the 2DEG confinement in cases with a fixed AlN barrier and varying Al composition in the (Al,Ga)N channel [Fig.~\ref{fig:figS9a}], and a fixed Al composition in the Al$_{x}$Ga$_{1-x}$N channel, $x=0.5$, and varying barrier thickness [Fig.~\ref{fig:figS9b}]. In both cases, it is evident that the 2DEG density distributions are primarily confined within the channel region. A similar degree of confinement is also observed in cases where barrier thicknesses and/or barrier (Al,Ga)N compositions are varied. These results are in agreement with previous studies.\cite{Bassaler2024AlRichMobility,Coltrin2017AnalysisAlloys,Miyoshi2015NumericalHeterostructures} 

Given the strong confinement of the 2DEG distributions in the channel region, we apply the materials parameters corresponding to the channel (Al,Ga)N alloy in our mobility models. The minimal penetration of the 2DEG into the large bandgap barrier ensures that any error introduced by this assumption remains negligible. 

\section{\label{sec:secS9}Figure-of-merit and reference GaN-channel HEMT}
For HEMTs, power-switching performance is evaluated using the lateral figure-of-merit (LFOM).\cite{Coltrin2017AnalysisAlloys} The LFOM is primarily determined by two factors: breakdown voltage and transport properties. It is expressed as:\cite{Coltrin2017AnalysisAlloys}
\begin{equation}
    \text{LFOM} = \frac{\text{V}_{\text{br}}}{\text{R}_{\text{(on,sp)}}} = e\,n_{\text{2D}}\,\mu \,\text{E}_{\text{cr}}^2 \label{eq:eqS91}
\end{equation}
where V$_{\text{br}}$ is the breakdown voltage, R$_{\text{(on,sp)}}$ is the so-called specific on-resistance, $e$ is the elementary charge, $n_{\text{2D}}$ is the 2DEG density, $\mu$ is the low-field mobility, and E$_{\text{cr}}$ is the critical electric field. The critical electric field has been shown to scale with the semiconductor bandgap (E$_\text{g}$) according to the relation:\cite{Hudgins2003AnDevices}
\begin{equation}
    \text{E}_{\text{cr}} = 1.73 \times 10^5 \times \text{E}_{\text{g}}^{2.5} \label{eq:eqS92}
\end{equation} 
for direct bandgap materials.

In this study, we neglect the effects of strain, layer thickness, and temperature on the critical electric field (accordingly on bandgap). It is worth noting that the above relationship between E$_{\text{cr}}$ and E$_\text{g}$, (Eq.~\ref{eq:eqS92}) remains controversial within the literature,\cite{Coltrin2017AnalysisAlloys} and a deeper analysis is beyond the scope of this paper. 

Furthermore, we conducted a comparative performance analysis using the LFOM metric to evaluate the advantages or disadvantages of (Al,Ga)N-channel HEMTs against a reference GaN-channel HEMT. Several types of GaN-channel HEMT-based discrete power devices and power integrated circuits are available on the market. Extensive reviews of these devices and technologies can be found in Refs.~\citenum{Udrea2024GaNTechnology} and \citenum{Ma2019ReviewW}. However, detailed structural specifications of HEMT heterostructures for commercial settings are rarely disclosed due to competitive concerns.  While previous studies\cite{Bassaler2024AlRichMobility,Coltrin2017AnalysisAlloys,Zhang2008TheContent,Bajaj2014ModelingVoltage,Ambacher1999Two-dimensionalHeterostructures} have performed similar comparative studies, they typically assumed a constant 2DEG density of $n_{\text{2D}} = 1 \times 10^{13}$ cm$^{-2}$, which eliminated the need for detailed structural specifications for the reference GaN-channel HEMT. In contrast, our study accounts for these structural aspects, requiring detailed specifications of the GaN HEMT. After an extensive literature review, we selected an Al$_{0.25}$Ga$_{0.75}$N(25nm)/GaN(300nm)/GaNsubstrate HEMT as our reference. Based on our literature search this heterostructure is one of the most widely studied GaN-channel HEMTs.\cite{Udrea2024GaNTechnology,Ma2019ReviewW} Note that we use a 300 nm thick GaN channel in our simulations, instead of $\sim \mu m$ thickness found in experimentally realized devices. This channel thickness is sufficient to ensure a negligible substrate impact on the 2DEG density at the far-side barrier-channel interface in our simulations, while also reducing computational cost. For consistency, the 2DEG density in GaN devices is calculated using our same methodology, yielding $n_{\text{2D}} = 6.25 \times 10^{12}$ cm$^{-2}$. This value of $n_{\text{2D}}$ is subsequently used in the mobility and LFOM$_{\text{GaN}}$ calculations ($\mu = 1.65 \times 10^3$ cm$^2$V$^{-1}$s$^{-1}$, LFOM$_{\text{GaN}}$  = $2.64 \times 10^4$ MW/cm$^2$ at 300 K). 

We note that due to this different choice of the reference GaN-channel HEMT, the normalized LFOM in the constant $n_{\text{2D}}$ case from our study [LFOM$^{\text{A}}_{\text{norm}}$, blue curve in Fig.~5(a) in main text] crosses unity at a lower Al composition compared to in Ref.~\citenum{Bassaler2024AlRichMobility}, where the crossover occurred at a channel Al composition of 0.85. Moreover, studies incorporating complex design engineering have achieved better-performing GaN-channel HEMTs than the reference device we used in this study, and such improvements are indeed an active research field.\cite{Udrea2024GaNTechnology,Ma2019ReviewW} We acknowledge that selecting a different reference GaN-HEMT would shift some of the quantitative results, such as those presented in Fig.~5(b) of the main text. Moreover, some of the scattering mechanisms considered for (Al,Ga)N alloys in the mobility calculations in Sec.~\ref{sec:secS6} assume ``worst case" scenarios, such as high defect density and large interface roughness in the heterostructure (Table~\ref{tab:tableS4}). Therefore, overall improvements in both (Al,Ga)N and GaN HEMTs can be expected based on the quality of the grown heterostructure. However, we emphasize that this would not impact the overall conclusions in this article.  

\pagebreak
\section{Figure-of-merit for AlN/(Al,Ga)N HEMTs}
\begin{figure}[!htbp]
    \centering
    \includegraphics[width=0.45\textwidth]{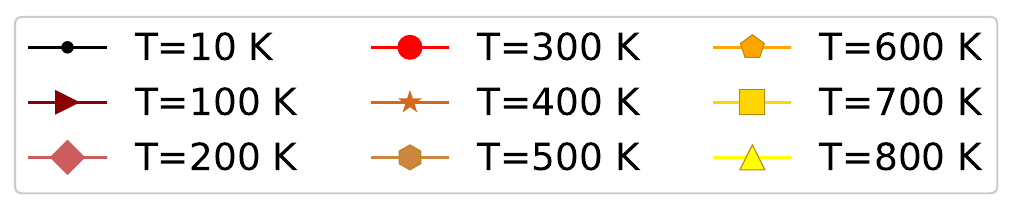}\\
    \subfloat[]{\label{fig:figS10a}\includegraphics[width=0.5\textwidth]{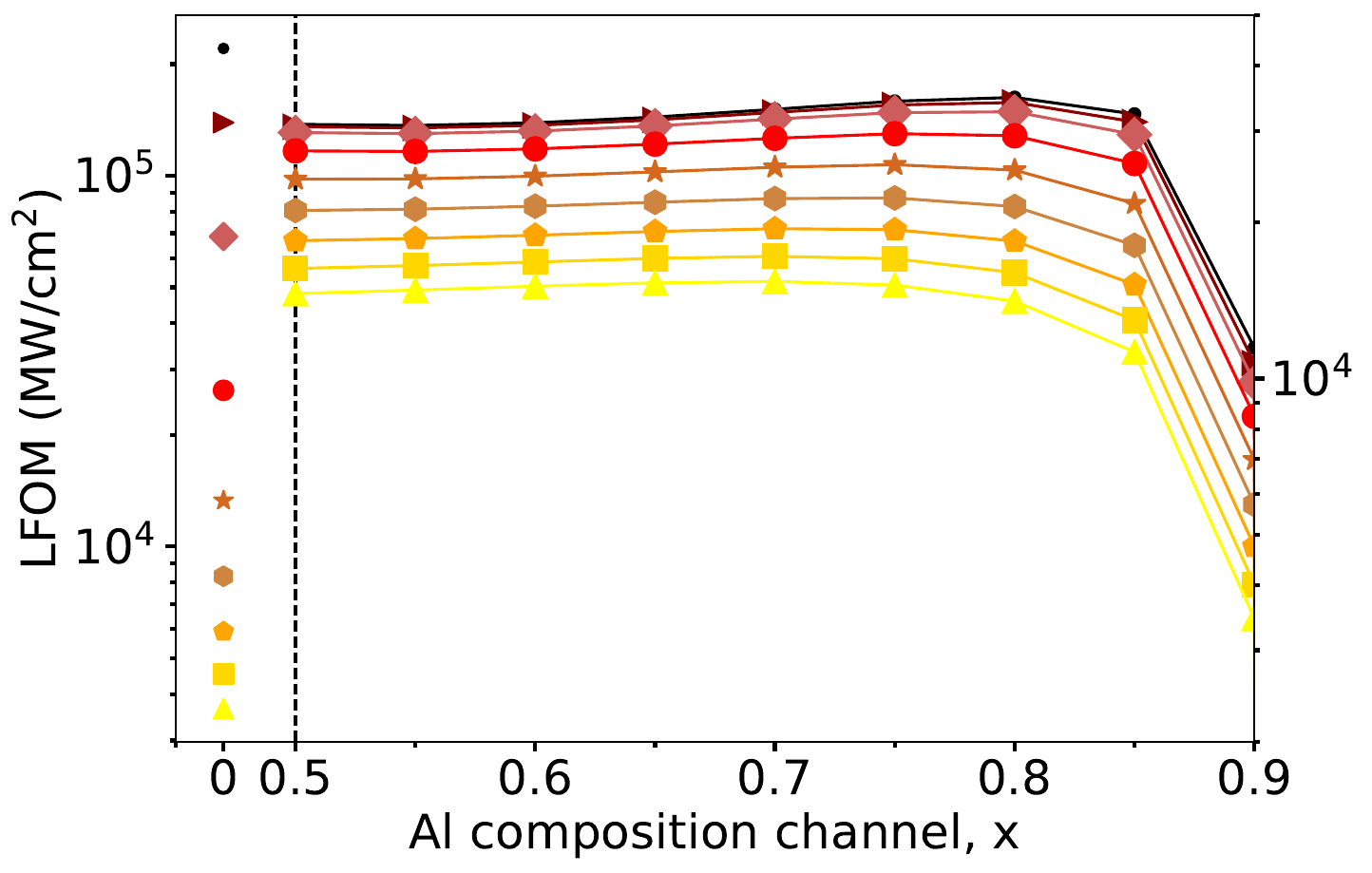}}\\
    %\hspace{0.5cm}
    \subfloat[]{\label{fig:figS10b}\includegraphics[width=0.45\textwidth]{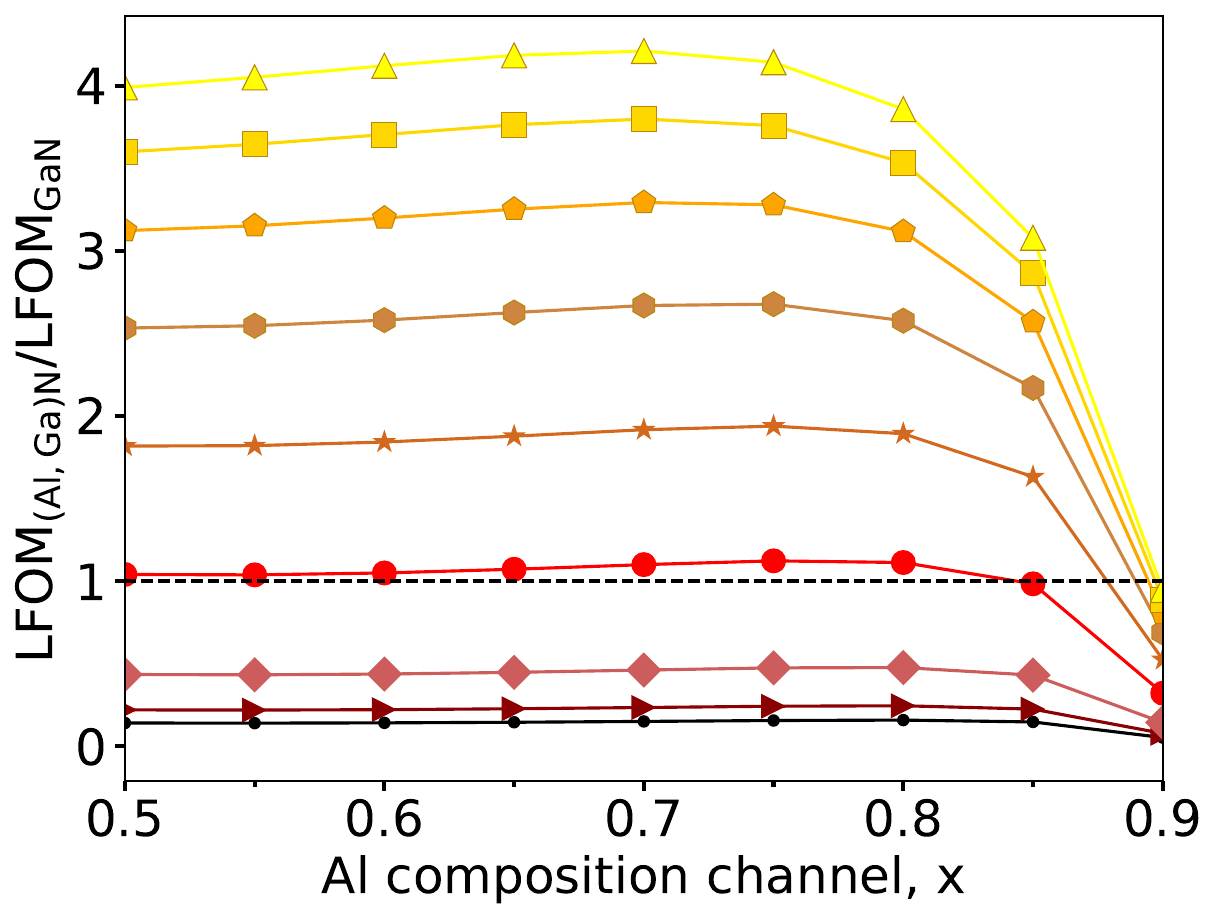}}
    \hspace{1.5cm}
    \subfloat[]{\label{fig:figS10c}\includegraphics[width=0.45\textwidth]{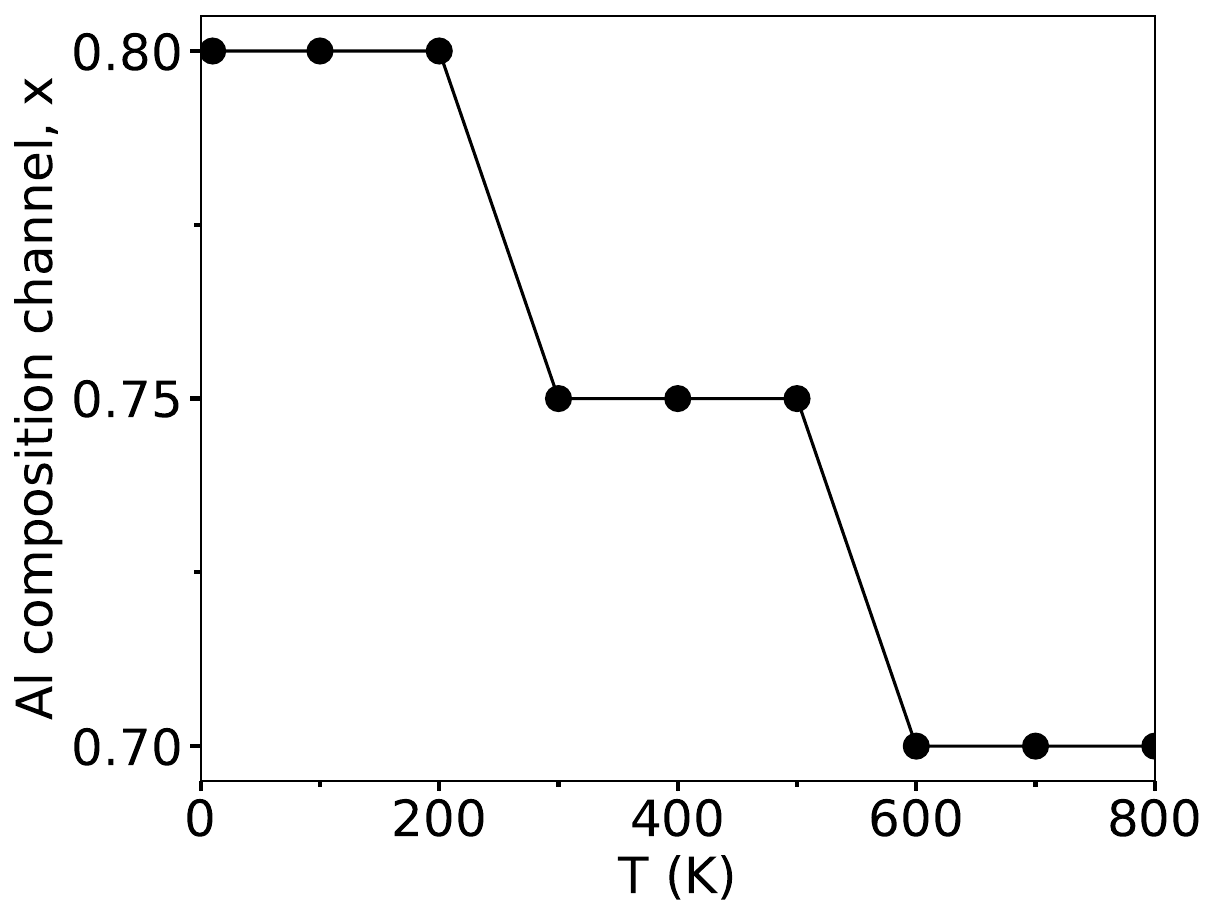}}
    \caption{(a) Absolute lateral figure-of-merit, LFOM, versus Al composition in the channel, $x$, at various temperatures, T. Results for the reference GaN-channel Al$_{0.25}$Ga$_{0.75}$N(25nm)/GaN(300nm)/GaN HEMT are plotted on the left axis, while those for the (Al,Ga)N-channel AlN(25nm)/Al$_x$Ga$_{1-x}$N(300nm)/AlN HEMT are shown on the right axis. The vertical line separates the two datasets. (b) Normalized LFOM (LFOM$_{\text{norm}}$ = LFOM$_{\text{(Al,Ga)N}}$ / LFOM$_{\text{GaN}}$), for AlN(25nm)/Al$_x$Ga$_{1-x}$N(300nm)/AlN HEMT. LFOM$_{\text{(Al,Ga)N}}$ is normalized against LFOM$_{\text{GaN}}$ at each temperature. Temperature (in K) legends are shown at the top. (c) Channel Al composition, $x$, corresponding to the highest LFOM$_{\text{norm}}$s at each temperature obtained from (a).}
    \label{fig:figS10}
\end{figure}

In reference to Fig.~5(b) in the main text, Fig.~\ref{fig:figS10} presents the temperature dependence of the LFOM for  25 nm barrier AlN(25nm)/Al$_x$Ga$_{1-x}$N HEMT structures, across a temperature range of $10 - 800$ K. Similar to the results for the 50 nm barrier AlN(50nm)/Al$_x$Ga$_{1-x}$N HEMTs presented in the main text, here also we observe a comparable decrease in LFOM [Figs.~\ref{fig:figS10a} and \ref{fig:figS10b}], as well as a shift of the highest normalized LFOM peak towards lower Al composition with increasing temperature [Fig.~\ref{fig:figS10c}]. 

\begin{figure}%[!ht]
    \centering
    \includegraphics[width=0.33\textwidth]{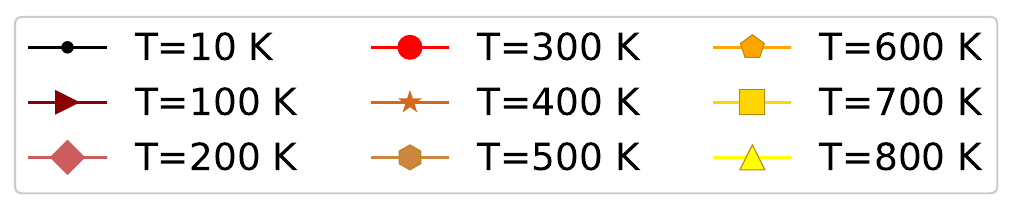}\\
    \includegraphics[width=0.45\textwidth]{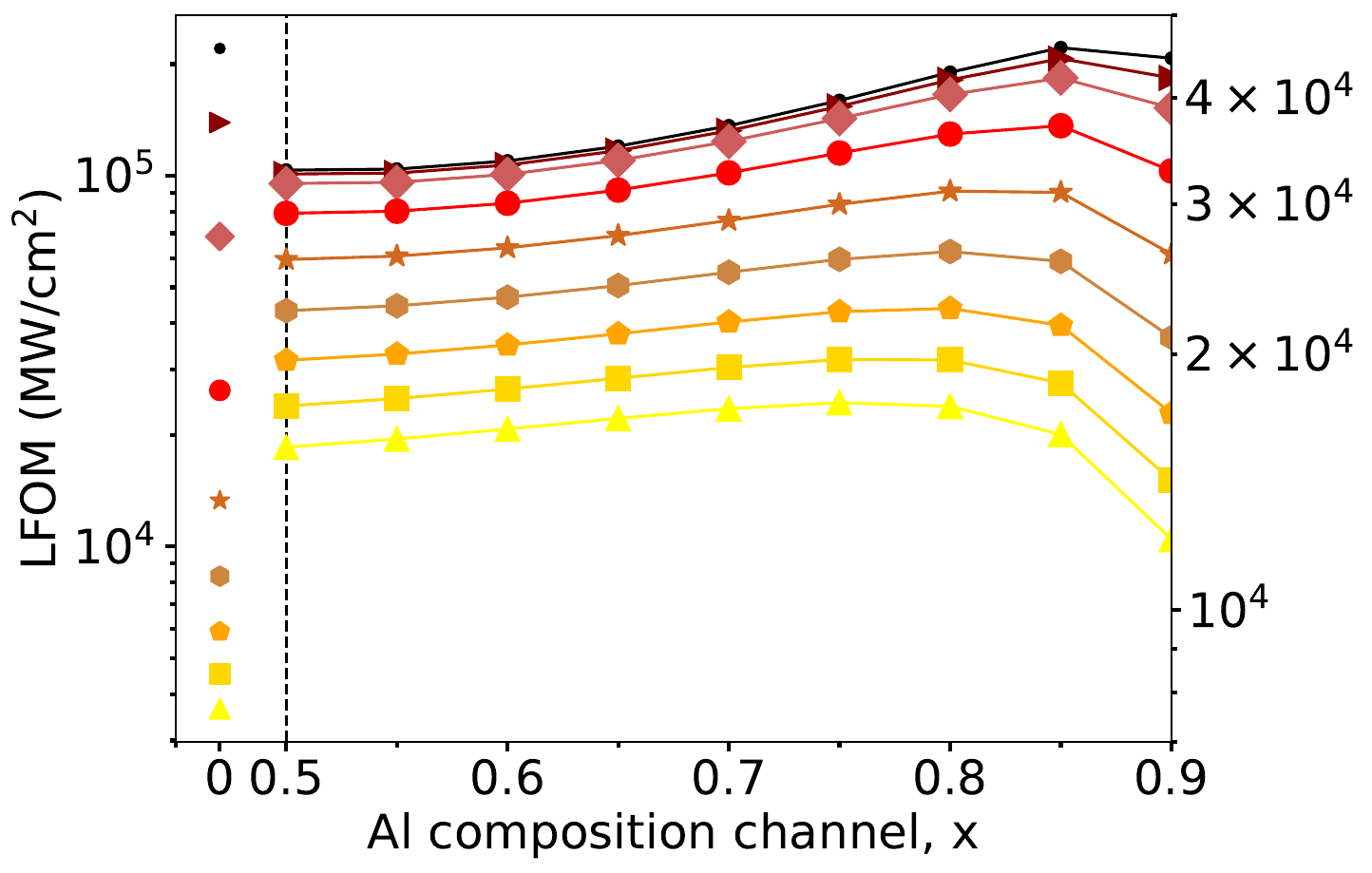}
    \caption{Absolute lateral figure-of-merit, LFOM, as a function of channel Al composition, $x$, at various temperatures, T. Results for the reference GaN-channel Al$_{0.25}$Ga$_{0.75}$N(25nm)/GaN(300nm)/GaN HEMT are plotted on the left axis, while those for the (Al,Ga)N-channel AlN(50nm)/Al$_x$Ga$_{1-x}$N(300nm)/AlN HEMT are shown on the right axis. The vertical dashed line separates the two datasets. Temperature (in K) legends are shown at the top.}
    \label{fig:figS11}
\end{figure}

\pagebreak
Figure~\ref{fig:figS11} compares the effect of temperature on the absolute LFOM for GaN- and (Al,Ga)N-channel HEMTs. It is evident from the figure that the absolute LFOMs for both (Al,Ga)N- and GaN-channel HEMTs decrease with increasing temperature. However, due to the predominant temperature-independent alloy-disorder-limited mobility contributions, (Al,Ga)N-channel HEMTs exhibit a relatively lower temperature dependence in LFOM compared to GaN-channel devices. This stems from the fact that in the latter case, temperature-dependent phonon scattering processes become the dominant mobility-limiting mechanism,\cite{Coltrin2017AnalysisAlloys,Singhal2022TowardHeterostructures} resulting in a greater temperature dependence of LFOM (LFOM $\propto$ mobility, Eq.~\ref{eq:eqS91}).

\begin{figure}[!ht]
    \centering
    \includegraphics[width=0.45\textwidth]{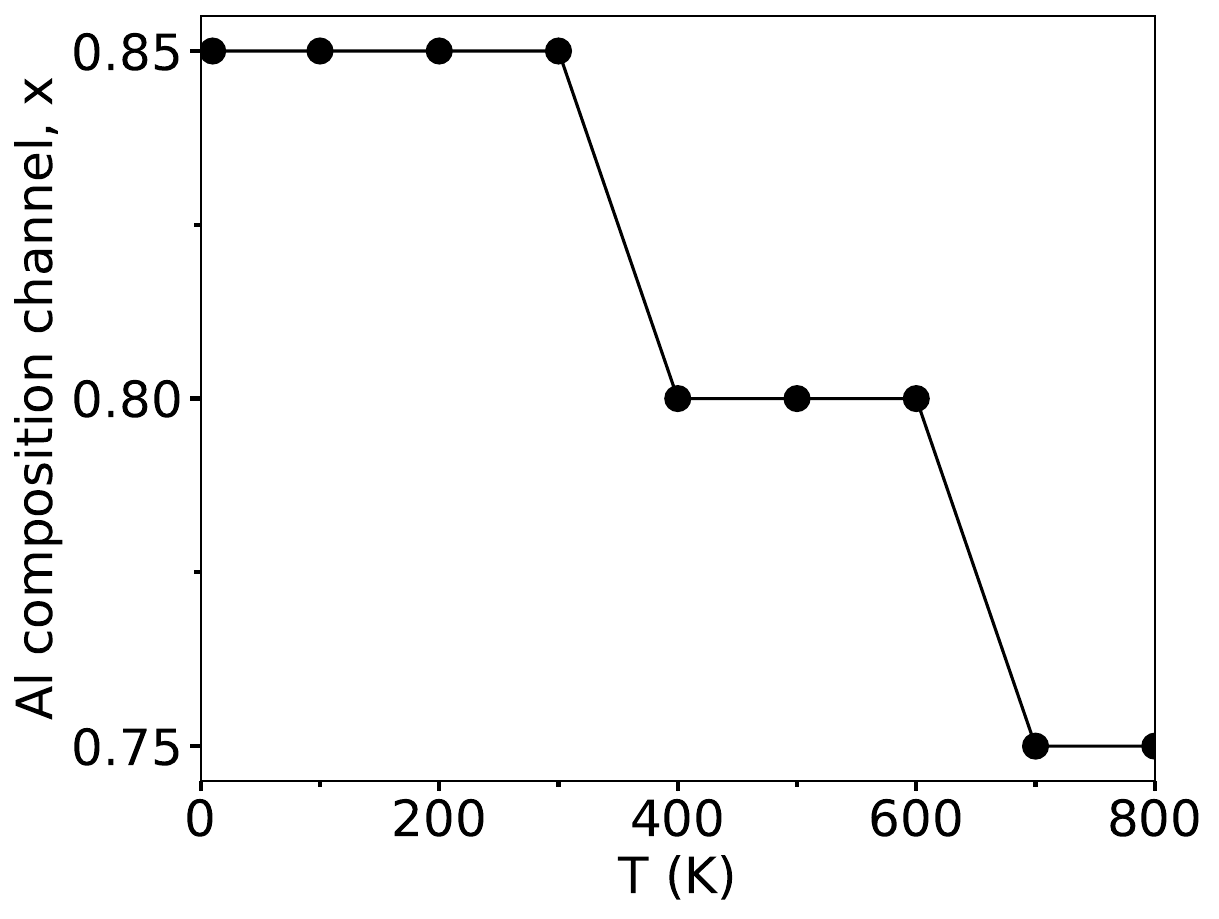}
    \caption{Channel Al composition, $x$, for AlN(50nm)/Al$_x$Ga$_{1-x}$N HEMTs corresponding to the highest normalized LFOMs at each temperature from Fig.~5(b) in the main text. }
    \label{fig:figS12}
\end{figure}

Figure~\ref{fig:figS12} presents the highest normalized LFOM values for AlN(50 nm)/Al$_x$Ga$_{1-x}$N HEMTs at various temperatures, as depicted in Fig.~5(b) of the main text. We find that the Al$_{0.85}$Ga$_{0.15}$N-channel exhibits the highest LFOM$_{\text{norm}}$ up to 300 K, and the peak shifts to a lower Al content, reaching the Al$_{0.75}$Ga$_{0.25}$N-channel at 800 K.

\clearpage
%===============================================================================

\bibliography{supplement}% Produces the bibliography via BibTeX.